\documentclass[a4paper,11pt]{article}

\usepackage{JiaHep} 

\usepackage[numbers,sort&compress]{natbib} 
\usepackage{url}
\usepackage{amsmath}
\usepackage{rotating}

\usepackage{multirow}

\usepackage{bbm} 

\usepackage{subcaption}

\usepackage[T1]{fontenc} 

\usepackage{float, array, xspace, amscd, amsthm}
\usepackage{fancyhdr}
\usepackage{longtable}

\newtheorem{theorem}{Theorem}[section]
\newtheorem{lemma}[theorem]{Lemma}
\newtheorem{proposition}[theorem]{Proposition}
\newtheorem{definition}[theorem]{Definition}
\newtheorem{corollary}[theorem]{Corollary}

\allowdisplaybreaks

\usepackage[usenames,dvipsnames,svgnames,table,x11names]{xcolor}
\definecolor{MagentaXD}{RGB}{204, 48, 152}
\definecolor{MagentaXDdetail}{RGB}{150, 79, 126}
\definecolor{GreenMAF}{RGB}{28, 112, 46}
\definecolor{GreenMAFdetail}{RGB}{80, 117, 88}
\definecolor{detail}{RGB}{110,110,110}
\definecolor{quantumviolet}{HTML}{53257F} 
\definecolor{quantumgray}{HTML}{555555} 
\definecolor{quantumgreen}{HTML}{007474} 
\definecolor{quantumblue}{HTML}{002366} 
\definecolor{quantumpurple}{HTML}{66023C} 
\definecolor{quantumdarkviolet}{HTML}{5D3954} 

\newcommand{\Zbb}{\mathbb{Z}}

\newcommand{\identity}{\mathds{1}}

\newcommand{\be}{\begin{equation}}
	\newcommand{\ee}{\end{equation}}
\def\bea#1\eea{\begin{align}#1\end{align}}

\usepackage{tikz}
\usetikzlibrary{positioning,intersections}
\usetikzlibrary{calc}
\usetikzlibrary{shapes.geometric}
\usepackage{braids}
\usepackage{tikz-cd}

\usepackage{longtable}
\usepackage{booktabs}

\usepackage{upgreek}
\usepackage[normalem]{ulem}
\usepackage{mathtools}
\usepackage{datetime}
\usepackage{cancel}

\usepackage[all,cmtip,knot]{xy}
\usepackage{xypic}

\newcommand{\orcid}[1]{\href{https://orcid.org/#1}{\includegraphics[width=8pt]{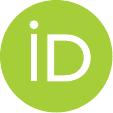}}}

\usepackage{amsmath,amsfonts,amsthm}

\theoremstyle{definition}

\theoremstyle{remark}
\newtheorem{remark}{Remark}[section]

\theoremstyle{remark}

\usepackage{bm,latexsym,galois,euscript,dsfont}

\newcommand\Rep {\mathsf{Rep}}

\newcommand\Irr {\mathrm{Irr}}

\newcommand\Hom {\mathrm{Hom}}
\newcommand{\one}{\mathbbm{1}}

\newcommand{\CC}{\mathsf{CC}_{\Zbb_2}}

\settimeformat{ampmtime}

\begin{document}

\title{Topological quantum color code model on infinite lattice}

\author[a, \ddagger]{Shiyu Cao,}

\author[b,c,d,\dagger, \ddagger]{Zhian Jia\orcid{0000-0001-8588-173X},}

\author[e,\dagger, \ddagger]{Sheng Tan\orcid{0009-0008-3318-9942}}

\affiliation[a]{Chern Institute of Mathematics and LMPC, Nankai University, Tianjin 300071, China}

\affiliation[b]{Institute of Quantum Physics, School of Physics, Central South University,
Changsha 410012, China}
\affiliation[c]{Centre for Quantum Technologies, National University of Singapore, Singapore 117543, Singapore}
\affiliation[d]{Department of Physics, National University of Singapore, SG 117543, Singapore}
\affiliation[e]{School of Mathematical Sciences, Capital Normal University, Beijing 100048, China}

\affiliation[\dagger]{Corresponding authors: \href{mailto:giannjia@foxmail.com}{giannjia@foxmail.com}, \href{mailto:tansheng2018@outlook.com}{tansheng2018@outlook.com}}
\affiliation[\ddagger]{The authors are listed in alphabetical order.}

\abstract{
The color code model is a crucial instance of a Calderbank-Shor-Steane (CSS)-type topological quantum error-correcting code, which notably supports transversal implementation of the full Clifford group.  
Its robustness against local noise is rooted in the structure of its topological excitations. From the perspective of quantum phases of matter, it is essential to understand these excitations in the thermodynamic limit.  
In this work, we analyze the color code model on an infinite lattice within the quasi-local $C^{*}$-algebra framework, using a cone-localized Doplicher-Haag-Roberts (DHR) analysis.  
We classify its irreducible anyon superselection sectors and construct explicit string operators that generate anyonic excitations from the ground state.  
We further examine the fusion and braiding properties of these excitations.  
Our results show that the topological order of the color code is described by $\mathsf{Rep}(D(\mathbb{Z}_2 \times \mathbb{Z}_2)) \simeq \mathsf{Rep}(D(\mathbb{Z}_2)) \boxtimes \mathsf{Rep}(D(\mathbb{Z}_2))$, which is equivalent to a double layer of the toric code and consistent with established analyses on finite lattices. We also establish the existence and uniqueness of the KMS state at every finite inverse temperature and derive an explicit description of the corresponding thermal distribution of stabilizer excitations.
}


\maketitle

\section{Introduction}

The interplay between topological quantum matter and topological quantum information processing, encompassing topological quantum error correction and topological quantum computation, has attracted significant attention over the past several decades~\cite{freedman2002modular,Kitaev2003}. Two-dimensional topological lattice models, such as the quantum double model and its various generalizations~\cite{Kitaev2003,Buerschaper2013a,jia2023boundary,Jia2023weak,jia2018efficient,chang2014kitaev,meusburger2017kitaev,Beigi2011the} and string-net models and their generalizations~\cite{Levin2005,Kitaev2012a,Hu2018full}, have found fruitful applications in both condensed matter physics and quantum information and computation theory.

The color code is one of the simplest and most prominent topological quantum codes for implementing quantum error correction and enabling universal topological quantum computation~\cite{Bombin2006colorcode,bombin2012universal,Kubica2015color,Kubica2015ColorCode}.  
The toric code model (the $\mathbb{Z}_2$ quantum double model, also called the surface code) cannot implement the full Clifford group of gates transversally; lattice surgery is required to realize some Clifford gates, which increases circuit depth and overall overhead.  
In contrast, the 2d color code supports the full Clifford group of transversal gates, making it easier to achieve universality~\cite{Bombin2006colorcode,bombin2012universal,Kubica2015color}.  
Its ability to implement fault-tolerant logical gates depends strongly on the topological quantum phases of the underlying Hamiltonian model~\cite{kesselring2024anyon,Kesselring2018boundaries}.  
Therefore, developing a deep understanding of the topological order of the color code model is a crucial problem.

\begin{figure}[t]
    \centering
    \begin{subfigure}{0.38\textwidth}
        \centering
        \includegraphics[width=\textwidth]{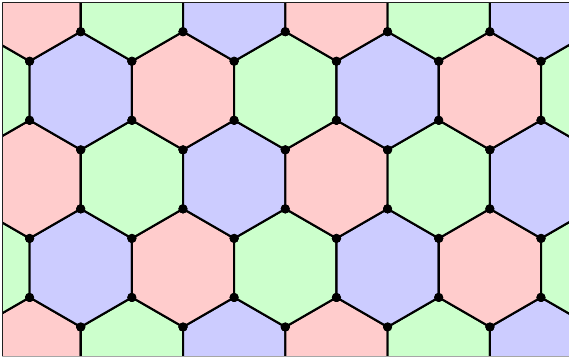}
    \end{subfigure}
    \hspace{1cm}
    \begin{subfigure}{0.38\textwidth}
        \centering
        \includegraphics[width=\textwidth]{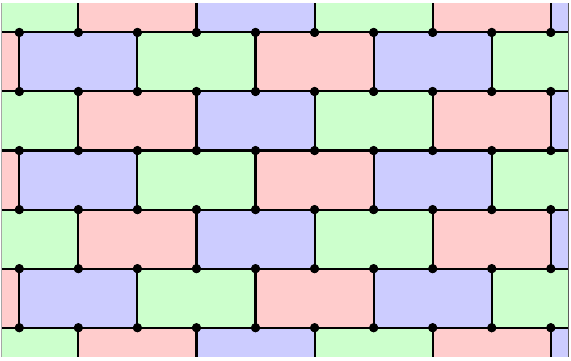}
    \end{subfigure}
    \caption{Two equivalent colored trivalent lattices used in the color code model. Each vertex (represented by a black dot) hosts a physical qubit. The lattice is composed of hexagonal (left) or square (right) faces, each face colored red, green, or blue, in such a way that no two adjacent faces share the same color. Each vertex is trivalent, meaning exactly three edges meet at each vertex.   \label{fig:color_code}}
    \label{fig:both}
\end{figure}

The color code model is defined on a specific type of lattice, which satisfies the following conditions:  
(1) it is trivalent, meaning exactly three edges meet at each vertex;  
(2) its faces are colored with three distinct colors such that no two adjacent faces share the same color.  
We label the faces as ``red'', ``green'', and ``blue''. An edge connecting two red faces is called a red edge, and similarly for the other colors. Consequently, a face coloring naturally induces a corresponding edge coloring.
In this work, we focus on the trivalent lattice with square faces, denoted by $\Sigma$; see Figure~\ref{fig:color_code} (right) for an illustration.  
The sets of vertices, edges, and faces of $\Sigma$ are denoted by $V(\Sigma)$, $E(\Sigma)$, and $F(\Sigma)$, respectively.

The Hamiltonian for the color code model is defined as the negative sum of two types of stabilizer operators, with each operator acting on the six vertices of a face on the lattice.
Specifically, for each face $f \in F(\Sigma)$, define 
\begin{equation*}
    K_f:= \bigotimes_{v\in  \partial f} \sigma_v^{\mathtt{x}},\quad J_f:= \bigotimes_{v\in  \partial f} \sigma_v^{\mathtt{z}},
\end{equation*}
where $v\in\partial f$ means that $v$ is one of the vertices of the face $f$, and
we use the Pauli matrices $\sigma^{\mathtt{x}}$, $\sigma^{\mathtt{y}}$, $\sigma^{\mathtt{z}}$:
\begin{equation*}
    \sigma^{\mathtt{x}}= \begin{bmatrix}
        0 & 1 \\
        1 & 0
    \end{bmatrix}, \quad 
    \sigma^{\mathtt{y}}= \begin{bmatrix}
        0 & -i \\
        i & 0
    \end{bmatrix}, \quad 
    \sigma^{\mathtt{z}}= \begin{bmatrix}
        1 & 0 \\
        0 & -1
    \end{bmatrix}. 
\end{equation*}
Note that the Pauli matrices satisfy the following relations:
\begin{equation*}
   \sigma^i \sigma^j = \delta_{ij} \, \mathds{1} + i \, \varepsilon_{ijk} \, \sigma^k,
\end{equation*}
where $i = 1, 2, 3$ correspond to $\mathtt{x}, \mathtt{y}, \mathtt{z}$, respectively, and $\varepsilon_{ijk}$ denotes the Levi-Civita symbol.  
It follows immediately that any two Pauli matrices either commute or anticommute with each other.

It is straightforward to verify that 
\begin{equation*}
    [K_f, J_g] = 0, \quad  (K_f)^2 = (J_f)^2 = \mathds{1}, \quad  (K_f)^{\dagger} = K_f, \quad (J_f)^{\dagger} = J_f,
\end{equation*}
for any $f, g \in F(\Sigma)$. Indeed, any two faces share an even number of common vertices, ensuring commutativity through the relation $\sigma^{\mathtt{x}} \sigma^{\mathtt{z}} = -\sigma^{\mathtt{z}} \sigma^{\mathtt{x}}$. The remaining identities follow directly from $(\sigma^i)^2 = \mathds{1}$ and $(\sigma^i)^{\dagger} = \sigma^i$ for $i = {\mathtt{x}}, {\mathtt{z}}$.
Let $\Lambda \subset V(\Sigma)$ be a finite subset. The associated local Hamiltonian is defined as 
\begin{equation*}
    H_\Lambda = -\sum_{f:f \subset \Lambda} K_f - \sum_{f:f \subset \Lambda} J_f.
\end{equation*}
The sums run over all faces whose vertices lie entirely within $\Lambda$. When $\Sigma$ is a finite lattice, we obtain the color code Hamiltonian $H_{\Sigma}$. 
For a lattice $\Sigma_g$ on a two-dimensional surface of genus $g$, the ground-state degeneracy of $H_{\Sigma_g}$ is \cite{Bombin2006colorcode,bombin2013introduction}
\begin{equation*}
    \operatorname{GSD} = 2^{4g} = 2^{2(2 - \chi)},
\end{equation*}
where $\chi = 2 - 2g$ is the Euler characteristic of the surface. 
Since the ground-state degeneracy depends only on the genus—a topological invariant—it is robust to local perturbations, reflecting the topological nature of the color code. The ground state space serves as the code space, safeguarding logical quantum information against noise-induced errors.

The aim of this work is to carry out a systematic investigation of the color code model in the thermodynamic limit, namely on an infinite planar lattice $\Sigma$.  
Notice that in this setting, the summation appearing in the Hamiltonian may fail to converge and should therefore be understood only as a formal expression.  
A suitable framework for treating such systems is the quasi-local algebra~\cite{bratteli1987operator,bratteli1997operator}, an approximately finite-dimensional (AF) $C^*$-algebra that captures the underlying locality structure.  
The analysis of superselection sectors is based on the theory of localized and transportable sectors developed by Doplicher, Haag, and Roberts~\cite{doplicher1971local,doplicher1974local}, together with the auxiliary algebra approach originating in the work of Buchholz and Fredenhagen~\cite{Buchholz1982locality}.  
Recently, there has been substantial progress along this direction for various quantum spin models; see, for example, \cite{naaijkens2011localized,fiedler2015haag,wallick2023algebraic,bols2025classification,Cha2018infinite,bols2025sector,bols2026sectorII,bols2026anyon}.
In this work, we develop the corresponding theory for the color code model on an infinite lattice.
For finite-lattice color codes, prior studies have firmly established that topological excitations are categorically characterized by $\Rep(D(\mathbb{Z}_2 \times \mathbb{Z}_2))$, the representation category of the quantum double $D(\mathbb{Z}_2 \times \mathbb{Z}_2)$~\cite{Bombin2006colorcode,bombin2012universal,Kubica2015color,Kubica2015ColorCode,kesselring2024anyon,Kesselring2018boundaries}.  
Extending this result to the infinite-lattice setting, we construct the superselection sector theory for the thermodynamic-limit color code and prove that its category of anyon sectors is isomorphic to $\Rep(D(\mathbb{Z}_2 \times \mathbb{Z}_2))$. We also show that the ground state of the color code model satisfies Haag duality and discuss the Kubo--Martin--Schwinger (KMS) state \cite{kubo1957statistical,martin1959KMStheory,Haag1967KMS} for the model.

The remainder of the paper is organized as follows.  
Section~\ref{sec:GS} establishes the existence and purity of the ground state.  
Section~\ref{sec:TO} develops the superselection sector theory for topological excitations of the color code, together with their fusion rules and braiding; we then show that these excitations form a theory equivalent to the category $\Rep(D(\Zbb_2\times \Zbb_2))$.  
Section~\ref{sec:haag_duality} establishes Haag duality for the color code.  
Section~\ref{sec:KMS} constructs the unique KMS state of the infinite color code model at every finite inverse temperature.
For completeness, the appendix provides details on the representation theory of the quantum double $D(\Zbb_2\times \Zbb_2)$.

\section{Ground state of color code model}
\label{sec:GS}

In this section, we establish the existence and essential properties of the ground state within the $C^*$-algebraic framework for quantum spin systems \cite{bratteli1987operator,bratteli1997operator}. We begin by reviewing necessary tools from the theory of operator algebras and constructing the quasi-local algebra associated with the infinite colored lattice. We then prove that the color code Hamiltonian admits a unique translationally invariant ground state, which is a pure state satisfying the ground state condition. This result provides the foundational vacuum sector upon which the theory of topological excitations, developed in the subsequent sections, will be built.

A state on a $C^*$-algebra $\mathcal{A}$ is a positive linear functional $\omega:\mathcal{A} \to \mathbb{C}$ of norm one. 
If $\mathcal{A}$ is unital with unit element $\identity$, the norm one condition reads $\omega(\identity)=1$. 
The quasi-local algebra we consider here is unital. 
The positive condition means that $\omega(a^{\dagger}a)\geq 0$ for all $a\in \mathcal{A}$. 
The following provides a fundamental tool for evaluating a state on a general element. 
The proof follows immediately by virtue of the Cauchy--Schwarz inequality for states, cf. \cite{alicki2007statistical}. 

\begin{lemma}[{\cite[Section~2.1.1]{alicki2007statistical}}] \label{lem:evaluate_state}
    Let $\omega$ be a state on a unital $C^*$-algebra $\mathcal{A}$. Let $a\in \mathcal{A}$ be a self-adjoint element such that $a\leq \identity$ and $\omega(a)=1$. Then for any $b\in \mathcal{A}$, one has 
    \begin{equation*}
        \omega(b) = \omega(ab) = \omega(ba). 
    \end{equation*}
\end{lemma}

\begin{lemma} \label{lem:state_abelian}
    Let $\mathcal{A}$ be a $C^*$-algebra, and define the \emph{stabilizer subalgebra} $\mathcal{A}_{\rm stab}$ as the subalgebra generated by a set of commuting involutionary elements $\{g_i\}_{i\in I}$, 
    \[
    \mathcal{A}_{\rm stab} = \big\langle g_i \in \mathcal{A} \,:\, g_i g_j = g_j g_i,  \, g_i^2 = \identity, g_i^{\dagger}=g_i, \forall \,i,j \in I \big\rangle.
    \]
Then there is a state on $\mathcal{A}_{\rm stab}$ such that
    \begin{equation*}
        \omega(g_i) = 1, \quad \forall\, i \in I.
    \end{equation*}
\end{lemma}

\begin{proof}

Consider the index set configuration space over the stabilizer algebra's generator index set $I$ defined by 
\[
\mathcal{X} = \{ f : I \to \{\pm 1\} \},
\]
which is equivalently the product space $\{\pm 1\}^I$.  
Equipping $\{\pm 1\}$ with the discrete topology, the product space $\mathcal{X}$ becomes a compact Hausdorff space under the product topology, by Tychonoff’s theorem.
Then we obtain a commutative $C^*$ algebra $C(\mathcal{X})$ consisting of complex-valued continuous functions on $\mathcal{X}$ (the multiplication is multiplication of functions and $*$-operation is complex conjugation).

The evaluation map $\tilde{g}_i:\mathcal{X} \to \{\pm 1\}\subset \mathbb{C}$ is defined as
\[
\tilde{g}_i(f):=f(i).
\]
It is clear that $\tilde{g}_i^\dagger(f)=\overline{f(i)} =f(i) =\tilde{g}_i(f)$; thus, $\tilde{g}_i^\dagger=\tilde{g}_i$. Also, $\tilde{g}_i^2(f)= \tilde{g}_i(f) \tilde{g}_i(f)=f(i)^2 = 1$, and therefore $\tilde{g}_i^2 = 1$.
Let $\tilde{\mathcal A}\subset C(\mathcal{X})$ be the subalgebra generated by the functions $\tilde g_i$. Then we have the isomorphism
\[
\mathcal{A}_{\rm stab}  \cong \tilde{\mathcal{A}}.
\]

Let $1_I\in \mathcal{X}$ denote the configuration that takes the value $1$ at every index. The evaluation functional
\[
    \omega_0(\tilde h):=\tilde h(1_I),\quad \tilde h\in\tilde{\mathcal A},
\]
is a state on the unital $C^*$-subalgebra $\tilde{\mathcal A}\subset C(X)$. Moreover,
\[
    \omega_0(\tilde g_i)=\tilde g_i(1_I)=1
\]
for every $i\in I$. Using the isomorphism $\mathcal A_{\rm stab}\cong\tilde{\mathcal A}$, the claim follows.
\end{proof}

\subsection{Quasi-local algebra}

We now construct the quasi-local algebra $\mathcal{A}$ associated with a colored trivalent lattice $\Sigma$ embedded in the plane, which is equal to the inductive limit of a local net of finite algebras. 
In a color code model, a qubit is placed on each vertex of the lattice. 
Recall that $V(\Sigma)$ denotes the set of vertices of $\Sigma$. 
For each site $v\in V(\Sigma)$, the local state space is $\mathcal{H}_{\{v\}} = \mathbb{C}[\mathbb{Z}_2]\simeq \mathbb{C}^2$, where the observables are $\mathcal{A}(\{v\}) = \mathbf{B}(\mathbb{C}^2) = M_2(\mathbb{C})$. Here $M_2(\mathbb{C})$ is the matrix algebra of complex $2\times 2$ matrices.

For a finite subset $\Lambda\subset V(\Sigma)$, $\mathcal{A}(\Lambda)$ is defined to be the algebra consisting of observables living on the vertices of $\Lambda$, i.e., $\mathcal{A}(\Lambda) = \otimes_{v\in \Lambda} M_2(\mathbb{C})$. This algebra acts on the local Hilbert space supported on $\Lambda$: $\mathcal{H}_\Lambda = \otimes_{v\in \Lambda} \mathbb{C}^2$. For an inclusion $\Lambda_1\subset \Lambda_2$ of finite subsets, there is a natural inclusion of algebras 
\begin{equation} \label{eq:inclusion}
    \mathcal{A}(\Lambda_1) \hookrightarrow \mathcal{A}(\Lambda_2), \quad A\mapsto A\otimes \mathds{1}_{\Lambda_2\setminus \Lambda_1},
\end{equation}
where $\mathds{1}_{\Lambda_2\setminus \Lambda_1} = \otimes_{v\in {\Lambda_2\setminus \Lambda_1}}\mathds{1}_{\{v\}}$. 
Here the identification $\mathcal{H}_{\Lambda_2} = \mathcal{H}_{\Lambda_1} \otimes \mathcal{H}_{\Lambda_2 \setminus \Lambda_1}$ induces the algebra isomorphism $\mathcal{A}({\Lambda_2}) \simeq \mathcal{A}({\Lambda_1}) \otimes \mathcal{A}({\Lambda_2 \setminus \Lambda_1})$.
The algebra of local observables is given by 
\begin{equation*}
    \mathcal{A}_{\rm loc} = \bigcup_{\Lambda\subset_f V(\Sigma)} \mathcal{A}(\Lambda),
\end{equation*}
where $\Lambda\subset_fV(\Sigma)$ means the union is taken over all finite subsets $\Lambda$ of $V(\Sigma)$ under the natural inclusion~\eqref{eq:inclusion}. 
This algebra is equipped with a norm $\|\cdot \|$ induced from the norm from the matrix algebra $M_2(\mathbb{C})$. 
Hence the quasi-local algebra $\mathcal{A}$ is defined to be the norm completion of $\mathcal{A}_{\rm loc}$:
\begin{equation*}
    \mathcal{A} = \overline{\mathcal{A}_{\rm loc}}^{\|\cdot\|} = \overline{\bigcup_{\Lambda\subset_f V(\Sigma)} \mathcal{A}(\Lambda)}^{\|\cdot\|}. 
\end{equation*}
For an arbitrary subset $\Lambda\subset V(\Sigma)$, we can perform an analogous construction to obtain the algebra localized in $\Lambda$: 
\begin{equation*}
    \mathcal{A}(\Lambda) = \overline{\bigcup_{\Lambda'\subset_f \Lambda} \mathcal{A}(\Lambda')}^{\|\cdot\|},
\end{equation*}
where $\Lambda'\subset_f\Lambda$ means the union is taken over all finite subsets $\Lambda'$ of $\Lambda$.  If $A\in \mathcal{A}(\Lambda)$, we say that $A$ is localized in $\Lambda$; the support $\operatorname{supp}(A)$ of $A$ is the smallest $\Lambda\subset V(\Sigma)$ in which $A$ is localized. 

Any state on a $C^*$ algebra gives rise to a GNS representation. A pure state corresponds to an irreducible representation.
We will use the following to determine if two GNS representations are equivalent. 

\begin{proposition}[{\cite[Proposition~3.2.8]{naaijkens2017quantum}}] \label{prop:equiv_GNS}
    The GNS representations of two pure states $\omega$, $\omega'$ on the quasi-local algebra $\mathcal{A}$ are equivalent if and only if for any $\varepsilon>0$, there is a finite subset $S_\varepsilon\subset V(\Sigma)$ such that $|\omega(A) - \omega'(A)| <\varepsilon\|A\|$ for all $A\in \mathcal{A}(S)$ with $S$ any finite subset of $S_{\varepsilon}^c := V(\Sigma)\setminus S_\varepsilon$. 
\end{proposition}

\begin{remark}
    This is actually a special case of \cite[Corollary 2.6.11]{bratteli1987operator}. 
\end{remark}

\subsection{Ground states}

For a finite subset $\Lambda\subset V(\Sigma)$, define its one-step enlargement by
\[
    \Lambda(1) := \bigcup_{f:\,\partial f\cap\Lambda\neq\emptyset}\partial f.
\]
This enlarged region contains the supports of all stabilizers that can fail to commute with an observable supported in $\Lambda$. For $A\in\mathcal A(\Lambda)$, define 
\[
    \delta(A):=-i\sum_{f:\,\partial f\cap\Lambda\neq\emptyset}[K_f,A]-i\sum_{f:\,\partial f\cap\Lambda\neq\emptyset}[J_f,A]. 
\]
The sum is finite and is independent of the choice of the finite region containing the support of $A$. Equivalently, 
\[
    \delta(A)=i[H_{\Lambda(1)},A]. 
\]
This defines a derivation on $\mathcal A_{\rm loc}$ whose closure generates a strongly continuous one-parameter group of automorphisms $\alpha_t$ of $\mathcal A$; see \cite[Theorem~6.2.4]{bratteli1997operator}. For a finite region $\Lambda$, the truncated Hamiltonian $H_\Lambda$ defines the finite-volume dynamics 
\[
    \alpha_t^\Lambda(A):=e^{itH_\Lambda}Ae^{-itH_\Lambda}
\]
on $\mathcal A(\Lambda)$. For the infinite-volume dynamics, Proposition~\ref{prop:exact_locality} below shows that $\alpha_t(A)=\alpha_t^{\Lambda(1)}(A)$ for every $A\in\mathcal A(\Lambda)$.

\begin{definition} \label{def:groud_state}
    A ground state for the system $(\mathcal{A},\alpha)$ is a state $\omega$ on the quasi-local algebra $\mathcal{A}$ satisfying the condition 
    \begin{equation*}
        -i\omega(A^\dagger\delta(A))\geq 0,\quad \text{for~all}~A\in \mathcal{A}_{\rm loc}. 
    \end{equation*}
\end{definition}

We next recall the local topological order condition that will be used to construct the ground state. For a finite region $\Lambda\subset V(\Sigma)$, let $p_\Lambda$ denote the projection onto the local ground state space of $H_\Lambda$. Since the stabilizers commute, this projection is given by
\[
    p_\Lambda = \prod_{f:\,\partial f\subset\Lambda}
    \frac{\mathds{1}+K_f}{2}
    \frac{\mathds{1}+J_f}{2}.
\]
Equivalently, $p_\Lambda$ is the ground state projection of the frustration-free Hamiltonian
\begin{equation} \label{eq:Hamiltonian-1}
    \widetilde H_\Lambda = \sum_{f:\,\partial f\subset\Lambda} \frac{\mathds{1}-K_f}{2} + \sum_{f:\,\partial f\subset\Lambda} \frac{\mathds{1}-J_f}{2},
\end{equation}
which has the same ground state space as $H_\Lambda$.

We recall the local topological order condition introduced in \cite{jones2025local}. This condition is closely related to the LTQO condition of \cite{Michalakis2013stability}, which in turn builds on the TQO conditions introduced in \cite{Bravyi2010topological}.

\begin{definition}
    The local ground state projections $\{p_\Lambda\}_\Lambda$ are said to satisfy the local topological order condition {\rm (LTO1)} if there exists a fixed constant $s>0$ such that, whenever a finite region $\Delta$ completely surrounds $\Lambda$ by at least $s$,\footnote{This means that the distance from $\Lambda$ to $\Delta^c$ is at least $s$, with respect to the graph distance on the lattice.}
    \begin{equation}
        p_\Delta \mathcal A(\Lambda) p_\Delta
        =
        \mathbb C p_\Delta.
    \end{equation}
\end{definition}

Condition~{\rm (LTO1)} expresses local indistinguishability: every observable supported sufficiently far from the boundary of $\Delta$ acts as a scalar on the local ground state space of $H_\Delta$. We refer to \cite{jones2025local} for the general formulation of {\rm (LTO1)} and its relation to earlier notions of topological quantum order.

\begin{theorem}  \label{prop:GS}
    The local ground state projections of the color code model satisfy the condition {\rm (LTO1)}. Consequently, they determine a canonical state $\omega_0$ on the quasi-local algebra $\mathcal A$. Moreover, $\omega_0$ is pure and is the unique translation-invariant ground state of the color code Hamiltonian. This is the unique state $\omega$ on $\mathcal A$ satisfying $\omega(K_f)=\omega(J_f)=1$ for every face $f$.
\end{theorem}

\begin{proof}
We first prove {\rm (LTO1)}. Let $\Lambda\subset\Delta$ be finite regions such that $\Delta$ completely surrounds $\Lambda$ by a sufficiently large fixed width. By linearity, it suffices to consider a Pauli monomial $A\in\mathcal A(\Lambda)$.

Suppose first that $A$ anticommutes with some stabilizer $Q_f\in\{K_f,J_f\}$ supported in $\Delta$. Since $Q_f p_\Delta=p_\Delta Q_f=p_\Delta$, we obtain
\[
    p_\Delta A p_\Delta = p_\Delta Q_f A p_\Delta = -p_\Delta A Q_f p_\Delta = -p_\Delta A p_\Delta.
\]
Hence $p_\Delta A p_\Delta=0$. 

It remains to consider the case in which $A$ commutes with every stabilizer supported in $\Delta$. We now adapt the stabilizer elimination procedure of \cite[Algorithm~3.10]{jones2025local} to the color code setting. Consider the northernmost, and then easternmost, vertex $v\in\operatorname{supp}(A)$. Then $v$ cannot be the vertex $v_3$ in the following configuration:
\begin{equation} \label{eq:north-east-fig}
    \begin{aligned}
        \begin{tikzpicture}[scale=0.7]
            \draw[line width=0.9pt] (-0.5,1) -- (6.5,1); 
            \draw[line width=0.9pt] (-0.5,2) -- (6.5,2); 
            \draw[line width=0.9pt] (-0.5,3) -- (6.5,3); 
            \draw[line width=0.9pt] (0,1) -- (0,2);
            \draw[line width=0.9pt] (2,1) -- (2,2);
            \draw[line width=0.9pt] (4,1) -- (4,2);
            \draw[line width=0.9pt] (6,1) -- (6,2);
            \draw[line width=0.9pt] (1,2) -- (1,3);
            \draw[line width=0.9pt] (3,2) -- (3,3);
            \draw[line width=0.9pt] (5,2) -- (5,3);
            \node[ line width=0.6pt, dashed, draw opacity=0.5] (a) at (3,1.8){$\scriptstyle v_1$};
            \node[ line width=0.6pt, dashed, draw opacity=0.5] (a) at (4,2.2){$\scriptstyle v_2$};
            \node[ line width=0.6pt, dashed, draw opacity=0.5] (a) at (5,1.8){$\scriptstyle v_3$};
            \node[ line width=0.6pt, dashed, draw opacity=0.5] (a) at (4,2.6){$\scriptstyle f_1$};
            \node[ line width=0.6pt, dashed, draw opacity=0.5] (a) at (6,2.6){$\scriptstyle f_2$};
            \node[ line width=0.6pt, dashed, draw opacity=0.5] (a) at (5,1.4){$\scriptstyle f_3$};
            \node[ line width=0.6pt, dashed, draw opacity=0.5] (a) at (3,1.4){$\scriptstyle f_4$};
        \end{tikzpicture}
    \end{aligned}
\end{equation}
since otherwise $A$ would anticommute with either $K_{f_2}$ or $J_{f_2}$. Consequently, $v$ must be the vertex $v_2$. The neighboring vertex $v_1$ must also belong to $\operatorname{supp}(A)$; otherwise $A$ would anticommute with either $K_{f_1}$ or $J_{f_1}$. Moreover, the Pauli operators acting at $v_1$ and $v_2$ must be of the same type, since different Pauli types would again produce an anticommutation with one of the adjacent stabilizers. We may therefore multiply $A$ by an appropriate stabilizer supported on the adjacent face $f_4$: by $K_{f_4}$ for Pauli type $\mathtt x$, by $J_{f_4}$ for Pauli type $\mathtt z$, and by $K_{f_4}J_{f_4}$ for Pauli type $\mathtt y$. This removes the Pauli operators at $v_1$ and $v_2$ without changing the commutation of $A$ with the stabilizers. Repeating this procedure successively removes the support of $A$. 

Since the procedure is local, there is a fixed constant $s>0$ depending only on the lattice such that, whenever $\Delta$ completely surrounds $\Lambda$ by at least $s$, all stabilizers used in the elimination are supported in $\Delta$. It follows that there exist stabilizers $Q_1,\ldots,Q_m\in\{K_f,J_f\}$ supported in $\Delta$ and a scalar
$\lambda_A\in\mathbb C$ such that $A=\lambda_A Q_1\cdots Q_m$. Since $Q_i p_\Delta=p_\Delta$ for every $i$, we obtain $p_\Delta A p_\Delta = \lambda_A p_\Delta$. Together with the anticommuting case, this proves $p_\Delta\mathcal A(\Lambda)p_\Delta = \mathbb{C} p_\Delta$, and hence the color code model satisfies {\rm (LTO1)}.

By \cite[Corollary 2.19, Definition~2.20]{jones2025local}, {\rm (LTO1)} determines a canonical state $\omega_0$ on the quasi-local algebra $\mathcal A$. By \cite[Corollary~2.24]{jones2025local}, this state is pure. Finally, since the color code Hamiltonian is translation invariant and frustration free, \cite[Remark~2.25]{jones2025local} implies that $\omega_0$ is the unique translation-invariant ground state.

Finally, let $\omega$ be a state satisfying $\omega(K_f)=\omega(J_f)=1$ for every face $f$. These conditions imply $\omega(p_\Delta)=1$ for every finite region $\Delta$. Given $A\in\mathcal A(\Lambda)$, choose $\Delta$ sufficiently large that {\rm (LTO1)} applies. Since $p_\Delta A p_\Delta=\omega_0(A)p_\Delta$, Lemma~\ref{lem:evaluate_state} gives $\omega(A)=\omega(p_\Delta A p_\Delta)=\omega_0(A)$. The norm density of $\mathcal A_{\rm loc}$ in $\mathcal A$ then implies $\omega=\omega_0$.
\end{proof}

\begin{remark}
    Lemma~\ref{lem:state_abelian} provides an alternative direct construction of the ground state. Applied to the stabilizer subalgebra generated by $K_f, J_f$, it gives a state satisfying
    \[
        \omega(K_f)=\omega(J_f)=1
    \]
    for every face $f$. After extending this state to $\mathcal A$, one can verify directly that it is a ground state and use the stabilizer elimination argument to prove uniqueness and purity. We instead use the stronger {\rm (LTO1)} formulation, which allows these properties to be obtained from the general results of \cite{jones2025local}.
\end{remark}

In the following, we denote by $(\pi_0,\mathcal{H},\Omega)$ a GNS representation for $\omega_0$, which exists by the Gelfand--Naimark--Segal construction.  
In other words, $\mathcal{H}$ is a Hilbert space, $\Omega\in \mathcal{H}$ is a vector, and $\pi_0:\mathcal{A}\to \mathbf{B}(\mathcal{H})$ is a homomorphism such that $\{\pi_0(A)\Omega:A\in\mathcal{A}\}$ is dense in $\mathcal{H}$ and $\omega_0(A) = \langle \Omega,\pi_0(A)\Omega\rangle$ for all $A\in \mathcal{A}$. 
Such a triple exists uniquely up to unitary equivalence. 
The stabilizer operators $K_f$, $J_f$ satisfy $\pi_0(K_f)\Omega=\Omega=\pi_0(J_f)\Omega$ for any $f\in F(\Sigma)$. In fact,  
\begin{equation*}
    \|\pi_0(K_f-\mathds{1})\Omega\|^2 = \omega_0((K_f-\mathds{1})^\dagger (K_f-\mathds{1})) =0, 
\end{equation*}
implying the identity for $K_f$. The identity for $J_f$ is shown similarly.

\section{Topological excitations of color code}
\label{sec:TO}

In this section, we develop the superselection sector theory for the topological excitations of the quantum color code within the Doplicher-Haag-Roberts (DHR) framework. Our goal is to construct and classify the irreducible anyon sectors, derive their associated fusion rules and braiding statistics, and demonstrate that the resulting braided tensor category is equivalent to $\Rep(D(\Zbb_2 \times \Zbb_2))$. This provides a rigorous characterization of the topological order of the color code model in the thermodynamic limit, consistent with known results for finite lattices~\cite{Bombin2006colorcode,bombin2012universal,Kubica2015color,Kubica2015ColorCode,kesselring2024anyon,Kesselring2018boundaries}. We begin by recalling the known modular tensor category structure, then explicitly construct the superselection sectors via string operators, and finally establish their fusion and braiding properties.

\subsection{Modular tensor category of topological excitations in the color code}

The topological excitations in color code models are point-like quasiparticles that cannot be created or annihilated by local operators. 
It is known that, on a finite lattice, the modular tensor category $\mathsf{CC}_{\Zbb_2}$ describing these topological excitations 
is equivalent to that obtained by stacking two layers of the toric code~\cite{bombin2012universal,Kubica2015ColorCode,Kesselring2018boundaries,kesselring2024anyon}:
\begin{equation*}
    \mathsf{CC}_{\Zbb_2} \simeq \mathsf{TC}_{\Zbb_2} \boxtimes  \mathsf{TC}_{\Zbb_2} \simeq \Rep(D(\Zbb_2\times \Zbb_2)),
\end{equation*}
where $D(\Zbb_2\times \Zbb_2)$ is the Drinfeld quantum double of $\Zbb_2\times \Zbb_2$ and ``$\boxtimes$'' denotes the Deligne tensor product.
This correspondence is useful for understanding the topological order of the color code model and plays a crucial role in quantum error correction.

\begin{table}[h!]
\centering
\resizebox{\textwidth}{!}{
\begin{tabular}{c|cccccccccccccccc}
\hline\hline
{Anyon} ($\mathsf{CC}_{\mathbb{Z}_2}$) &
$\mathbbm{1}$ &
$\mathtt{rx}$ & $\mathtt{ry}$ & $\mathtt{rz}$ &
$\mathtt{gx}$ & $\mathtt{gy}$ & $\mathtt{gz}$ &
$\mathtt{bx}$ & $\mathtt{by}$ & $\mathtt{bz}$ &
$\mathtt{f}_1$ & $\mathtt{f}_2$ & $\mathtt{f}_3$ & $\mathtt{f}_4$ & $\mathtt{f}_5$ & $\mathtt{f}_6$ \\
\hline
{Anyon} ($\mathsf{TC}_{\mathbb{Z}_2} \boxtimes \mathsf{TC}_{\mathbb{Z}_2}$) &
$11$ &
$e1$ & $em$ & $1m$ &
$ee$ & $ff$ & $mm$ &
$1e$ & $me$ & $m1$ &
$f1$ & $1f$ & $ef$ & $fe$ & $mf$ & $fm$ \\
\hline
{Topological spin} $\theta_a$ &
$+1$ &
$+1$ & $+1$ & $+1$ &
$+1$ & $+1$ & $+1$ &
$+1$ & $+1$ & $+1$ &
$-1$ & $-1$ & $-1$ & $-1$ & $-1$ & $-1$ \\
\hline\hline
\end{tabular}
}
\caption{Topological excitations of the color code and their topological spins.}
\label{tab:colorcode_anyons_row}
\end{table}

There are 16 simple objects in $\mathsf{CC}_{\mathbb{Z}_2}$, among which 10 are bosons. 
Following the convention in~\cite{kesselring2024anyon}, we denote these bosonic excitations as
\begin{equation*}
 \mathbbm{1};\quad    \mathtt{rx}, \mathtt{ry}, \mathtt{rz}; \quad
    \mathtt{gx}, \mathtt{gy}, \mathtt{gz}; \quad
    \mathtt{bx}, \mathtt{by}, \mathtt{bz}.
\end{equation*}
Here, the labels $\mathtt{r}$, $\mathtt{g}$, and $\mathtt{b}$ stand for ``red'', ``green'', and ``blue'', respectively, while 
$\mathtt{x}$, $\mathtt{y}$, and $\mathtt{z}$ correspond to the three types of Pauli operators.
The significance of this naming convention will become clear later.
There are 6 fermionic charges, which we denote as
\begin{equation*}
    \mathtt{f}_1,\cdots, \mathtt{f}_6.
\end{equation*}
To obtain the fusion rules, braiding, and topological spins, we use the correspondence between these charges and those of the double-layer toric code. 
The four simple objects of $\mathsf{TC}_{\Zbb_2}$ are denoted as $1,e,m,f$, where $f=e\otimes m$. 
Hereinafter, we omit the stacking tensor ``$\boxtimes$'' whenever no ambiguity arises.
For bosonic charges, we have ($\one=11$)
\begin{equation*}
\begin{array}{c|ccc}
    \hline\hline
     & \mathtt{r} & \mathtt{g} & \mathtt{b} \\
    \hline
    \mathtt{x} & e1 & ee & 1e \\
    \mathtt{y} & em & ff & me \\
    \mathtt{z} & 1m & mm & m1 \\
    \hline\hline
\end{array}
\end{equation*}
This kind of table is sometimes called the Mermin--Peres magic square (a terminology from quantum contextuality \cite{Peres1991MerminPeres,mermin1990simple}), since the fusion in each row and each column is cyclic: fusing any two anyons yields the remaining one.
For fermionic charges, we have
\begin{equation*}
    \mathtt{f}_1 = f 1,\,
    \mathtt{f}_2 = 1 f,\,
    \mathtt{f}_3 = e f,\,
    \mathtt{f}_4 = f e,\,
    \mathtt{f}_5 = m f,\,
    \mathtt{f}_6 = f m.
\end{equation*}
See Table~\ref{tab:colorcode_anyons_row} for a summary.
We further emphasize that the correspondence between the charges of the color code and those of the double-layer toric code is not unique. The color code possesses 72 distinct automorphisms, which are essential for the implementation of Floquet quantum error correction; see Ref.~\cite{Davydova2024quantumcomputation}.

\begin{table}[h!]
\centering
\resizebox{\textwidth}{!}{
\begin{tabular}{c|cccccccccccccccc}
\hline\hline
$\otimes$ & $\mathbbm{1}$ & $\mathtt{rx}$ & $\mathtt{ry}$ & $\mathtt{rz}$ & $\mathtt{gx}$ & $\mathtt{gy}$ & $\mathtt{gz}$ & $\mathtt{bx}$ & $\mathtt{by}$ & $\mathtt{bz}$ & $\mathtt{f}_1$ & $\mathtt{f}_2$ & $\mathtt{f}_3$ & $\mathtt{f}_4$ & $\mathtt{f}_5$ & $\mathtt{f}_6$ \\
\hline
$\mathbbm{1}$ & $\mathbbm{1}$ & $\mathtt{rx}$ & $\mathtt{ry}$ & $\mathtt{rz}$ & $\mathtt{gx}$ & $\mathtt{gy}$ & $\mathtt{gz}$ & $\mathtt{bx}$ & $\mathtt{by}$ & $\mathtt{bz}$ & $\mathtt{f}_1$ & $\mathtt{f}_2$ & $\mathtt{f}_3$ & $\mathtt{f}_4$ & $\mathtt{f}_5$ & $\mathtt{f}_6$ \\
$\mathtt{rx}$ & $\mathtt{rx}$ & $\mathbbm{1}$ & $\mathtt{rz}$ & $\mathtt{ry}$ & $\mathtt{bx}$ & $\mathtt{f}_5$ & $\mathtt{f}_6$ & $\mathtt{gx}$ & $\mathtt{f}_4$ & $\mathtt{f}_1$ & $\mathtt{bz}$ & $\mathtt{f}_3$ & $\mathtt{f}_2$ & $\mathtt{by}$ & $\mathtt{gy}$ & $\mathtt{gz}$ \\
$\mathtt{ry}$ & $\mathtt{ry}$ & $\mathtt{rz}$ & $\mathbbm{1}$ & $\mathtt{rx}$ & $\mathtt{f}_2$ & $\mathtt{by}$ & $\mathtt{f}_1$ & $\mathtt{f}_3$ & $\mathtt{gy}$ & $\mathtt{f}_6$ & $\mathtt{gz}$ & $\mathtt{gx}$ & $\mathtt{bx}$ & $\mathtt{f}_5$ & $\mathtt{f}_4$ & $\mathtt{bz}$ \\
$\mathtt{rz}$ & $\mathtt{rz}$ & $\mathtt{ry}$ & $\mathtt{rx}$ & $\mathbbm{1}$ & $\mathtt{f}_3$ & $\mathtt{f}_4$ & $\mathtt{bz}$ & $\mathtt{f}_2$ & $\mathtt{f}_5$ & $\mathtt{gz}$ & $\mathtt{f}_6$ & $\mathtt{bx}$ & $\mathtt{gx}$ & $\mathtt{gy}$ & $\mathtt{by}$ & $\mathtt{f}_1$ \\
$\mathtt{gx}$ & $\mathtt{gx}$ & $\mathtt{bx}$ & $\mathtt{f}_2$ & $\mathtt{f}_3$ & $\mathbbm{1}$ & $\mathtt{gz}$ & $\mathtt{gy}$ & $\mathtt{rx}$ & $\mathtt{f}_1$ & $\mathtt{f}_4$ & $\mathtt{by}$ & $\mathtt{ry}$ & $\mathtt{rz}$ & $\mathtt{bz}$ & $\mathtt{f}_6$ & $\mathtt{f}_5$ \\
$\mathtt{gy}$ & $\mathtt{gy}$ & $\mathtt{f}_5$ & $\mathtt{by}$ & $\mathtt{f}_4$ & $\mathtt{gz}$ & $\mathbbm{1}$ & $\mathtt{gx}$ & $\mathtt{f}_6$ & $\mathtt{ry}$ & $\mathtt{f}_3$ & $\mathtt{f}_2$ & $\mathtt{f}_1$ & $\mathtt{bz}$ & $\mathtt{rz}$ & $\mathtt{rx}$ & $\mathtt{bx}$ \\
$\mathtt{gz}$ & $\mathtt{gz}$ & $\mathtt{f}_6$ & $\mathtt{f}_1$ & $\mathtt{bz}$ & $\mathtt{gy}$ & $\mathtt{gx}$ & $\mathbbm{1}$ & $\mathtt{f}_5$ & $\mathtt{f}_2$ & $\mathtt{rz}$ & $\mathtt{ry}$ & $\mathtt{by}$ & $\mathtt{f}_4$ & $\mathtt{f}_3$ & $\mathtt{bx}$ & $\mathtt{rx}$ \\
$\mathtt{bx}$ & $\mathtt{bx}$ & $\mathtt{gx}$ & $\mathtt{f}_3$ & $\mathtt{f}_2$ & $\mathtt{rx}$ & $\mathtt{f}_6$ & $\mathtt{f}_5$ & $\mathbbm{1}$ & $\mathtt{bz}$ & $\mathtt{by}$ & $\mathtt{f}_4$ & $\mathtt{rz}$ & $\mathtt{ry}$ & $\mathtt{f}_1$ & $\mathtt{gz}$ & $\mathtt{gy}$ \\
$\mathtt{by}$ & $\mathtt{by}$ & $\mathtt{f}_4$ & $\mathtt{gy}$ & $\mathtt{f}_5$ & $\mathtt{f}_1$ & $\mathtt{ry}$ & $\mathtt{f}_2$ & $\mathtt{bz}$ & $\mathbbm{1}$ & $\mathtt{bx}$ & $\mathtt{gx}$ & $\mathtt{gz}$ & $\mathtt{f}_6$ & $\mathtt{rx}$ & $\mathtt{rz}$ & $\mathtt{f}_3$ \\
$\mathtt{bz}$ & $\mathtt{bz}$ & $\mathtt{f}_1$ & $\mathtt{f}_6$ & $\mathtt{gz}$ & $\mathtt{f}_4$ & $\mathtt{f}_3$ & $\mathtt{rz}$ & $\mathtt{by}$ & $\mathtt{bx}$ & $\mathbbm{1}$ & $\mathtt{rx}$ & $\mathtt{f}_5$ & $\mathtt{gy}$ & $\mathtt{gx}$ & $\mathtt{f}_2$ & $\mathtt{ry}$ \\
$\mathtt{f}_1$ & $\mathtt{f}_1$ & $\mathtt{bz}$ & $\mathtt{gz}$ & $\mathtt{f}_6$ & $\mathtt{by}$ & $\mathtt{f}_2$ & $\mathtt{ry}$ & $\mathtt{f}_4$ & $\mathtt{gx}$ & $\mathtt{rx}$ & $\mathbbm{1}$ & $\mathtt{gy}$ & $\mathtt{f}_5$ & $\mathtt{bx}$ & $\mathtt{f}_3$ & $\mathtt{rz}$ \\
$\mathtt{f}_2$ & $\mathtt{f}_2$ & $\mathtt{f}_3$ & $\mathtt{gx}$ & $\mathtt{bx}$ & $\mathtt{ry}$ & $\mathtt{f}_1$ & $\mathtt{by}$ & $\mathtt{rz}$ & $\mathtt{gz}$ & $\mathtt{f}_5$ & $\mathtt{gy}$ & $\mathbbm{1}$ & $\mathtt{rx}$ & $\mathtt{f}_6$ & $\mathtt{bz}$ & $\mathtt{f}_4$ \\
$\mathtt{f}_3$ & $\mathtt{f}_3$ & $\mathtt{f}_2$ & $\mathtt{bx}$ & $\mathtt{gx}$ & $\mathtt{rz}$ & $\mathtt{bz}$ & $\mathtt{f}_4$ & $\mathtt{ry}$ & $\mathtt{f}_6$ & $\mathtt{gy}$ & $\mathtt{f}_5$ & $\mathtt{rx}$ & $\mathbbm{1}$ & $\mathtt{gz}$ & $\mathtt{f}_1$ & $\mathtt{by}$ \\
$\mathtt{f}_4$ & $\mathtt{f}_4$ & $\mathtt{by}$ & $\mathtt{f}_5$ & $\mathtt{gy}$ & $\mathtt{bz}$ & $\mathtt{rz}$ & $\mathtt{f}_3$ & $\mathtt{f}_1$ & $\mathtt{rx}$ & $\mathtt{gx}$ & $\mathtt{bx}$ & $\mathtt{f}_6$ & $\mathtt{gz}$ & $\mathbbm{1}$ & $\mathtt{ry}$ & $\mathtt{f}_2$ \\
$\mathtt{f}_5$ & $\mathtt{f}_5$ & $\mathtt{gy}$ & $\mathtt{f}_4$ & $\mathtt{by}$ & $\mathtt{f}_6$ & $\mathtt{rx}$ & $\mathtt{bx}$ & $\mathtt{gz}$ & $\mathtt{rz}$ & $\mathtt{f}_2$ & $\mathtt{f}_3$ & $\mathtt{bz}$ & $\mathtt{f}_1$ & $\mathtt{ry}$ & $\mathbbm{1}$ & $\mathtt{gx}$ \\
$\mathtt{f}_6$ & $\mathtt{f}_6$ & $\mathtt{gz}$ & $\mathtt{bz}$ & $\mathtt{f}_1$ & $\mathtt{f}_5$ & $\mathtt{bx}$ & $\mathtt{rx}$ & $\mathtt{gy}$ & $\mathtt{f}_3$ & $\mathtt{ry}$ & $\mathtt{rz}$ & $\mathtt{f}_4$ & $\mathtt{by}$ & $\mathtt{f}_2$ & $\mathtt{gx}$ & $\mathbbm{1}$ \\
\hline\hline
\end{tabular}
}
\caption{Fusion rules of the 16 anyons of the color code ($\mathsf{CC}_{\mathbb{Z}_2}$), where ``row label'' $\otimes$ ``column label''.}
\label{tab:colorcode_fusion}
\end{table}

Using this correspondence, we establish that all anyons in the color code model are their own antiparticles and are Abelian, ensuring that their fusion yields a unique outcome. Consequently, for all \( a \in \mathsf{CC}_{\mathbb{Z}_2} \), the fusion rule satisfies \( a \otimes a = \one \). 
The fusion rules can be directly derived from the data of the double-layer toric code, see Table~\ref{tab:colorcode_fusion}.
For the toric code, recall that $\theta_{1}=\theta_{e}=\theta_m=1$ and $\theta_f=-1$. From this, we directly obtain the topological spin $\theta_a$ for $\CC$ (see Table~\ref{tab:colorcode_anyons_row}).
The braiding properties can also be derived from this correspondence. We will focus in particular on mutual statistics (or monodromy) which can be regarded as double braiding. For an Abelian topological phase, the mutual statistics \( M_{a,b} = R_{b,a} R_{a,b} \) of charges \( a \) and \( b \) are given by
\begin{equation*}
    M_{a,b} = \frac{\theta_{a \otimes b}}{\theta_a \theta_b}.
\end{equation*}
In the color code model, this quantity takes values \( \pm 1 \), see Table~\ref{tab:colorcode_monodromy}. 

In the color code model, anyons are associated with the faces of the lattice. A violation of the stabilizer conditions for a face \( f \) indicates the presence of an anyon. Given that faces are assigned distinct colors, the elementary bosonic excitations carry corresponding color labels. Moreover, when anyons are created through the application of Pauli operators (\( \mathtt{x} \), \(\mathtt{y} \), or \( \mathtt{z} \)) in string-like configurations, the resulting bosonic excitations inherit both a color label and a Pauli-type label corresponding to the applied operator.

\begin{table}[t]
\centering
\resizebox{\textwidth}{!}{
\begin{tabular}{c|cccccccccccccccc}
\hline\hline
$M_{a,b}$ & $\mathbbm{1}$ & $\mathtt{rx}$ & $\mathtt{ry}$ & $\mathtt{rz}$ & $\mathtt{gx}$ & $\mathtt{gy}$ & $\mathtt{gz}$ & $\mathtt{bx}$ & $\mathtt{by}$ & $\mathtt{bz}$ & $\mathtt{f}_1$ & $\mathtt{f}_2$ & $\mathtt{f}_3$ & $\mathtt{f}_4$ & $\mathtt{f}_5$ & $\mathtt{f}_6$ \\
\hline
$\mathbbm{1}$ & $+1$ & $+1$ & $+1$ & $+1$ & $+1$ & $+1$ & $+1$ & $+1$ & $+1$ & $+1$ & $+1$ & $+1$ & $+1$ & $+1$ & $+1$ & $+1$ \\
$\mathtt{rx}$ & $+1$ & $+1$ & $+1$ & $+1$ & $+1$ & $-1$ & $-1$ & $+1$ & $-1$ & $-1$ & $-1$ & $+1$ & $+1$ & $-1$ & $-1$ & $-1$ \\
$\mathtt{ry}$ & $+1$ & $+1$ & $+1$ & $+1$ & $-1$ & $+1$ & $-1$ & $-1$ & $+1$ & $-1$ & $-1$ & $-1$ & $-1$ & $+1$ & $+1$ & $-1$ \\
$\mathtt{rz}$ & $+1$ & $+1$ & $+1$ & $+1$ & $-1$ & $-1$ & $+1$ & $-1$ & $-1$ & $+1$ & $+1$ & $-1$ & $-1$ & $-1$ & $-1$ & $+1$ \\
$\mathtt{gx}$ & $+1$ & $+1$ & $-1$ & $-1$ & $+1$ & $+1$ & $+1$ & $+1$ & $-1$ & $-1$ & $-1$ & $-1$ & $-1$ & $-1$ & $+1$ & $+1$ \\
$\mathtt{gy}$ & $+1$ & $-1$ & $+1$ & $-1$ & $+1$ & $+1$ & $+1$ & $-1$ & $+1$ & $-1$ & $+1$ & $+1$ & $-1$ & $-1$ & $-1$ & $-1$ \\
$\mathtt{gz}$ & $+1$ & $-1$ & $-1$ & $+1$ & $+1$ & $+1$ & $+1$ & $-1$ & $-1$ & $+1$ & $-1$ & $-1$ & $+1$ & $+1$ & $-1$ & $-1$ \\
$\mathtt{bx}$ & $+1$ & $+1$ & $-1$ & $-1$ & $+1$ & $-1$ & $-1$ & $+1$ & $+1$ & $+1$ & $+1$ & $-1$ & $-1$ & $+1$ & $-1$ & $-1$ \\
$\mathtt{by}$ & $+1$ & $-1$ & $+1$ & $-1$ & $-1$ & $+1$ & $-1$ & $+1$ & $+1$ & $+1$ & $-1$ & $-1$ & $+1$ & $-1$ & $-1$ & $+1$ \\
$\mathtt{bz}$ & $+1$ & $-1$ & $-1$ & $+1$ & $-1$ & $-1$ & $+1$ & $+1$ & $+1$ & $+1$ & $-1$ & $+1$ & $-1$ & $-1$ & $+1$ & $-1$ \\
$\mathtt{f}_1$ & $+1$ & $-1$ & $-1$ & $+1$ & $-1$ & $+1$ & $-1$ & $+1$ & $-1$ & $-1$ & $+1$ & $+1$ & $-1$ & $+1$ & $-1$ & $+1$ \\
$\mathtt{f}_2$ & $+1$ & $+1$ & $-1$ & $-1$ & $-1$ & $+1$ & $-1$ & $-1$ & $-1$ & $+1$ & $+1$ & $+1$ & $+1$ & $-1$ & $+1$ & $-1$ \\
$\mathtt{f}_3$ & $+1$ & $+1$ & $-1$ & $-1$ & $-1$ & $-1$ & $+1$ & $-1$ & $+1$ & $-1$ & $-1$ & $+1$ & $+1$ & $+1$ & $-1$ & $+1$ \\
$\mathtt{f}_4$ & $+1$ & $-1$ & $+1$ & $-1$ & $-1$ & $-1$ & $+1$ & $+1$ & $-1$ & $-1$ & $+1$ & $-1$ & $+1$ & $+1$ & $+1$ & $-1$ \\
$\mathtt{f}_5$ & $+1$ & $-1$ & $+1$ & $-1$ & $+1$ & $-1$ & $-1$ & $-1$ & $-1$ & $+1$ & $-1$ & $+1$ & $-1$ & $+1$ & $+1$ & $+1$ \\
$\mathtt{f}_6$ & $+1$ & $-1$ & $-1$ & $+1$ & $+1$ & $-1$ & $-1$ & $-1$ & $+1$ & $-1$ & $+1$ & $-1$ & $+1$ & $-1$ & $+1$ & $+1$ \\
\hline\hline
\end{tabular}
}
\caption{Monodromy $M_{a,b}$ for the 16 anyons of the color code.}
\label{tab:colorcode_monodromy}
\end{table}

\subsection{String operators} \label{sec:string-ope}

\begin{figure}[b]
    \centering
    \includegraphics[width=8cm]{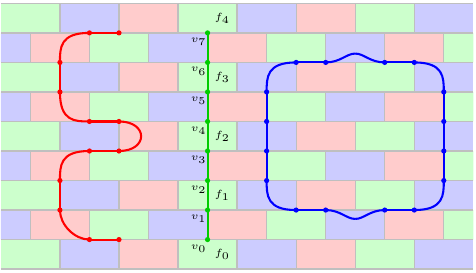}
    \caption{ An illustration of three strings: $\gamma^{\mathtt{r}}$ (red), $\gamma^{\mathtt{g}}$ (green), and $\gamma^{\mathtt{b}}$ (blue), arranged from left to right. The blue string $\gamma^{\mathtt{b}}$ on the right is a closed string.    \label{fig:ribbon}}
\end{figure}

To concretely generate and manipulate the anyonic excitations, we give a precise definition of colored strings on the color code lattice $\Sigma$ and construct the associated string operators.  
These are local operators supported on one-dimensional paths on the lattice, whose endpoints create pairs of topological charges.

Given a coloring of the faces $F(\Sigma)$, we can assign a coloring of the edges $E(\Sigma)$ in the following way: an edge is colored by $c$ if it connects two faces colored by $c$. 
Roughly speaking, a finite string is a path that traverses consecutive edges and faces, all of which share the same color.
The rigorous definition is given as follows. 

\begin{definition}[String]
A null string is a string with no support. 
A single vertex is considered to be a string. 
Let $c\in \{\mathtt{r},\mathtt{g},\mathtt{b}\}$. 
A finite string $\gamma^c$ of type $c$ is a path composed of finitely many consecutive $c$-colored faces and edges, which starts and terminates on faces.
Formally, it is a sequence
\begin{equation*}
\gamma^c = (f_0,v_0, v_1,f_1, v_2, v_3,  \cdots, f_n, v_{2n},v_{2n+1},f_{n+1}),
\end{equation*}
where faces $\{f_i\}_{i=0}^{n+1}$ and edges $\{e_{2i,2i+1}\}_{i=0}^n$ are colored by $c$, such that $v_0\in \partial f_0$, $v_{2i-1},v_{2i} \in \partial f_{i}$ for $1\leq i\leq n$, $v_{2n+1}\in \partial f_{n+1}$, and the edge $e_{2j,2j+1}$ connects the faces $f_{j}$ and $f_{j+1}$, for $0\leq j\leq n$. 
We denote by $\partial_0 \gamma^c = f_0$ the starting face of $\gamma^c$, and by $\partial_1 \gamma^c = f_{n+1}$ the terminating face of $\gamma^c$.
If $\partial_0\gamma^c=\partial_1\gamma^c$, then $\gamma^c$ is called a closed string. 
The support of $\gamma^c$ is the set of vertices $\{v_0,v_1,\cdots,v_{2n+1}\}$.
\end{definition}

As an illustration, Figure~\ref{fig:ribbon} provides examples of three strings: $\gamma^{\mathtt{r}}$, $\gamma^{\mathtt{g}}$, and $\gamma^{\mathtt{b}}$, with their supports labeled by bullets. The green string, for instance, is given by
\begin{equation*}
\gamma^{\mathtt{g}} = (f_0, v_0, v_1, f_1, v_2, v_3, f_2, v_4, v_5, f_3, v_6, v_7, f_4),
\end{equation*}
where $f_0$ and $f_4$ are the starting and terminating faces, respectively, and the support is indicated by the green bullets.
The blue string $\gamma^{\mathtt{b}}$ is a closed string.

\begin{definition}[String operator]
    An identity operator is associated to a null string.
    If $\gamma=\{v\}$ is a string of a single vertex, define its string operators by $S_\gamma^{\mathtt{x}} = \sigma_v^{\mathtt{x}}$, $S_\gamma^{\mathtt{z}} = \sigma_v^{\mathtt{z}}$, and $S_{\gamma}^{\mathtt{y}} = S_{\gamma}^{\mathtt{x}}S_{\gamma}^{\mathtt{z}}$. 
    Let $c\in \{\mathtt{r},\mathtt{g},\mathtt{b}\}$.  Let $\gamma$ be a $c$-colored finite string.  
    The string operators are defined by 
    \begin{equation*}
        S_{\gamma}^{c\mathtt{x}} = \bigotimes_{v_j\in \gamma} \sigma_{v_j}^\mathtt{x},\quad S_{\gamma}^{c\mathtt{z}} = \bigotimes_{v_j\in \gamma} \sigma_{v_j}^z,\; S_{\gamma}^{c\mathtt{y}} = S_{\gamma}^{c\mathtt{x}} S_{\gamma}^{c\mathtt{z}}.
    \end{equation*}
    The string operator $S_{\gamma}^{ck}$ is said to be of type $ck$. 
\end{definition}

For example, the string operators associated to the green string $\gamma=\gamma^{\mathtt{g}}$ in Figure~\ref{fig:ribbon} are given by 
\begin{equation*}
    S_{\gamma}^{\mathtt{g}\mathtt{x}} = \bigotimes_{j=0}^7 \sigma_{v_j}^{\mathtt{x}}, \quad S_{\gamma}^{\mathtt{g}\mathtt{y}} = \bigotimes_{j=0}^7 \sigma_{v_j}^{\mathtt{x}}\sigma_{v_j}^{\mathtt{z}}, \quad S_{\gamma}^{\mathtt{g}\mathtt{z}} = \bigotimes_{j=0}^7 \sigma_{v_j}^{\mathtt{z}}.  
\end{equation*}

\begin{lemma}\label{lem:defom}
    Let $\gamma_1$ and $\gamma_2$ be two different strings of the same type $c\in \{\mathtt{r},\mathtt{g},\mathtt{b}\}$ connecting the same starting and terminating faces. Then for all $k\in\{\mathtt{x},\mathtt{y},\mathtt{z}\}$, $S_{\gamma_1}^{ck}$ can be deformed to $S_{\gamma_2}^{ck}$ by multiplying suitable $K_f$ and $J_f$. 
\end{lemma}

\begin{proof}
    Without loss of generality, we may suppose $c=\mathtt{r}$ and $k=\mathtt{x}$. 
    Let $f_1,\cdots,f_N$ be the faces inside the region bounded by $\gamma_1$ and $\gamma_2$ that are colored either by $\mathtt{g}$ or $\mathtt{b}$. Then one can verify that $S_{\gamma_1}^{\mathtt{r}\mathtt{x}}A = S_{\gamma_2}^{\mathtt{r}\mathtt{x}}$, where $A = \prod_{j=1}^N K_{f_j}$. 
    For $k=\mathtt{y}$, one needs to use both $K$ and $J$ stabilizers in the definition of $A$. 
\end{proof}

The string operator $S_\gamma^{ck}$ associated with an open string $\gamma$ creates excitations at its starting and terminating faces $\partial_0\gamma$, $\partial_1\gamma$. 
This occurs because at each endpoint face, there is exactly one $c$-colored face $\partial_i\gamma$ that shares only a single vertex with $\gamma$; the string operator anticommutes with the corresponding stabilizer ($K_f$ or $J_f$) at this face. 
In contrast, all other faces share an even number of vertices with $\gamma$ and thus commute with the string operator.
An excitation created by $S_\gamma^{ck}$ is labeled by $\pi^{ck}$. All nine of these excitations are bosons. There are six fermions obtained by fusing two bosons that differ in both $c$ and $k$ labels, cf.~\cite{kesselring2024anyon}.  
If $\gamma$ is a closed string, then $S_{\gamma}^{ck}$ is a product of stabilizer operators by the proof of the above lemma.
Since $\pi_0(K_f)\Omega = \pi_0(J_f)\Omega = \Omega$ for all $f$, the same holds for the closed string operator.

\subsection{Superselection sectors}

We adapt the method of \cite{naaijkens2011localized}, originally developed for the toric code, to study excitations in the color code. The idea is to create a pair of excitations using a string operator and then move one of them to infinity. We will construct a braided tensor category and show that its full subcategory of dualizable objects is braided tensor equivalent to $\Rep(D(\mathbb{Z}_2\times \mathbb{Z}_2))$, which characterizes the excitations of the color code model. The objects are the so-called localized and transportable endomorphisms. 

\begin{figure}[t]
    \centering
    \includegraphics[width=6cm]{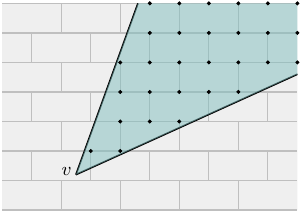}
    \caption{ A cone with apex $v$. The black bullet vertices are considered to be inside the cone.  \label{fig:cone}}
\end{figure}

We will mainly use cone regions in this paper, but the precise shape is not essential in most cases. 
On the plane $\mathbb{R}^2$, a cone with an apex at the origin can be expressed by
\begin{equation*}
    \Lambda = \left\{x\in \mathbb{R}^2\,|\,x\cdot a > \|x\|\cos\theta\right\},
\end{equation*}
for some $\theta\in(0,\pi)$ (the half-opening angle) and unit vector $a\in \mathbb{R}^2$ (the axis).  
Other cones are obtained by translating cones based at the origin.
A cone in the vertex set $V(\Sigma)$ is by convention the intersection of $V(\Sigma)$ with a cone $\Lambda$ in $\mathbb{R}^2$.
By abuse of notation, we also denote this cone by $\Lambda$.
For example, a cone $\Lambda$ in $V(\Sigma)$ is depicted in Figure~\ref{fig:cone}, where $\Lambda$ consists of the black bullet points. 
We denote by $\Lambda^c$ the complement of a cone $\Lambda$ in $V(\Sigma)$. 

Let $(\pi_0,\mathcal{H},\Omega)$ be a GNS representation of the unique ground state $\omega_0$ in Theorem~\ref{prop:GS}.

\begin{definition}
    1. A $*$-homomorphism $\pi:\mathcal{A}\to \mathbf{B}(\mathcal{H})$ is said to be localized in a cone $\Lambda$ if it acts as $\pi_0$ outside $\Lambda$; that is, $\pi(A) = \pi_0(A)$ for all $A \in \mathcal{A}(\Lambda^c)$. 

    2. A $*$-homomorphism $\pi:\mathcal{A}\to \mathbf{B}(\mathcal{H})$ localized in a cone $\Lambda$ is said to be transportable if for any cone $\Lambda'$, there is a $*$-homomorphism $\pi':\mathcal{A}\to \mathbf{B}(\mathcal{H})$ localized in $\Lambda'$ that is unitarily equivalent to $\pi$, namely, there is a unitary $U$ such that $\pi'(A) = U^\dagger\pi(A)U$ for all $A\in \mathcal{A}$. 
\end{definition}

\begin{definition}
    A representation $\pi$ of $\mathcal{A}$ is said to satisfy the superselection criterion if for any cone $\Lambda$, it holds that 
    \begin{equation*}
        \pi|_{\mathcal{A}(\Lambda^c)} \simeq \pi_0|_{\mathcal{A}(\Lambda^c)},
    \end{equation*}
    where $\simeq$ means unitary equivalence. Such a representation is called a superselection sector. 
\end{definition}

Let $\Lambda$ be a cone. We consider a half-infinite string $\gamma$ which is contained in $\Lambda$, meaning the support of $\gamma$ is contained in $\Lambda$. Denote by $\gamma_n$ the finite string formed by the first $n$ faces and all intermediate vertices of $\gamma$. For a local operator $A\in \mathcal{A}_{\rm loc}$, $c\in \{\mathtt{r},\mathtt{g},\mathtt{b}\}$, and $k\in \{\mathtt{x},\mathtt{y},\mathtt{z}\}$, define 
\begin{equation*}
    \rho_\gamma^{ck}(A) := \lim_{n\to\infty} (S_{\gamma_n}^{ck})^\dagger A S_{\gamma_n}^{ck},
\end{equation*}
where the convergence is taken in norm topology. 
This is well-defined, as the support of $A$ is finite, so there exists an $N>0$ such that $\gamma\setminus \gamma_n$ does not intersect the support of $A$ for $n\geq N$, which results in $(S_{\gamma_n}^{ck})^\dagger A S_{\gamma_n}^{ck} = (S_{\gamma_N}^{ck})^\dagger A S_{\gamma_N}^{ck}$. 
Extending by continuity, we obtain an endomorphism $\rho_\gamma^{ck}:\mathcal{A}\to \mathcal{A}$ (cf.~\cite{naaijkens2011localized}).  

\begin{proposition} \label{prop:local-transp}
    Let $\Lambda$ be a cone, $c\in \{\mathtt{r},\mathtt{g},\mathtt{b}\}$, and $k\in \{\mathtt{x},\mathtt{y},\mathtt{z}\}$. Let $\gamma$ be a half-infinite string of type $c$ contained in $\Lambda$, with starting face $f=\partial_0\gamma$ contained in $\Lambda$. Then the state $\omega_f^{ck}:=\omega_0\comp\rho_\gamma^{ck}$ depends only on the starting face $f$, but not on the choice of the half-infinite string $\gamma$. Moreover, the representation $\pi_0\comp\rho_\gamma^{ck}:\mathcal{A}\to\mathbf{B}(\mathcal{H})$ is localized in $\Lambda$ and is transportable.
\end{proposition}

\begin{proof}
    We first show that $\omega_0\comp\rho_\gamma^{ck}$ depends only on the starting face of $\gamma$. Without loss of generality, suppose $c=\mathtt{r}$ and $k=\mathtt{x}$. Let $\gamma^1$ and $\gamma^2$ be two half-infinite strings of type $\mathtt{r}$ with the same starting face $f$, and let $A\in\mathcal{A}_{\rm loc}$. There exists $n_0>0$ such that $(\gamma^1\setminus\gamma_n^1)\cap\operatorname{supp}(A) =  (\gamma^2\setminus\gamma_n^2)\cap\operatorname{supp}(A) = \emptyset$ for all $n\geq n_0$. For sufficiently large $n$, choose a finite string $\eta_n$ of type $\mathtt{r}$ connecting the terminating faces of $\gamma_n^1$ and $\gamma_n^2$, with support disjoint from $\operatorname{supp}(A)$. By Lemma~\ref{lem:defom}, there is an operator $X=\prod_{j=1}^N K_{f_j}$ such that $S_{\gamma_n^2}^{\mathtt{r}\mathtt{x}} = S_{\gamma_n^1\eta_n}^{\mathtt{r}\mathtt{x}}X$, where $f_1,\ldots,f_N$ are the faces inside the region bounded by $\gamma_n^1$, $\gamma_n^2$, and $\eta_n$ that are colored either by $\mathtt{g}$ or $\mathtt{b}$. Hence, for $n$ sufficiently large,
    \begin{equation*}
        \omega_0((S_{\gamma^2_n}^{\mathtt{r}\mathtt{x}})^\dagger AS_{\gamma^2_n}^{\mathtt{r}\mathtt{x}}) = \omega_0(X^\dagger(S_{\gamma^1_n\eta_n}^{\mathtt{r}\mathtt{x}})^\dagger AS_{\gamma^1_n\eta_n}^{\mathtt{r}\mathtt{x}}X) = \omega_0((S_{\gamma^1_n}^{\mathtt{r}\mathtt{x}})^\dagger AS_{\gamma^1_n}^{\mathtt{r}\mathtt{x}}),
    \end{equation*}
    where the last equality follows from Lemma~\ref{lem:evaluate_state} and $[A,S_{\eta_n}^{\mathtt{r}\mathtt{x}}]=0$. Therefore, $\omega_0\comp\rho_{\gamma^1}^{\mathtt{r}\mathtt{x}}(A) = \omega_0\comp\rho_{\gamma^2}^{\mathtt{r}\mathtt{x}}(A)$, for all $A\in\mathcal{A}_{\rm loc}$. The other choices of $c$ and $k$ are analogous. By continuity and the density of $\mathcal{A}_{\rm loc}$ in $\mathcal{A}$, the assertion holds for all $A\in\mathcal{A}$.

    We next prove localization. If $A\in\mathcal{A}(\Lambda^c)$, then $[A,S_{\gamma_n}^{ck}]=0$ for all $n$. Hence $\rho_\gamma^{ck}(A)=A$, and therefore $\pi_0\comp\rho_\gamma^{ck}(A)=\pi_0(A)$. Thus $\pi_0\comp\rho_\gamma^{ck}$ is localized in $\Lambda$.
    Finally, we prove transportability. Let $\Lambda'$ be an arbitrary cone. Choose a $c$-colored face $f'$ contained in $\Lambda'$ and a half-infinite string $\gamma'$ of type $c$ contained in $\Lambda'$ and starting at $f'$. Choose a finite string $\eta$ of type $c$ connecting $f'$ to $f$, and concatenate $\eta$ with $\gamma$ to obtain a half-infinite string $\widehat{\gamma}$ starting at $f'$. Notice that $\widehat{\gamma}$ need not be contained in either $\Lambda$ or $\Lambda'$. This does not affect the argument above, since the string-independence of the state only requires the two half-infinite strings to have the same starting face. Since $\widehat{\gamma}$ and $\gamma'$ both start at $f'$, the first part of the proof gives $\omega_0\comp\rho_{\widehat{\gamma}}^{ck} = \omega_0\comp\rho_{\gamma'}^{ck}$. On the other hand, by construction, $\rho_{\widehat{\gamma}}^{ck} = \rho_\gamma^{ck} \comp \operatorname{Ad}(S_{\eta}^{ck})$, and therefore $\omega_{f'}^{ck} = \omega_f^{ck}\comp\operatorname{Ad}(S_{\eta}^{ck})$. Since $\operatorname{Ad}(S_{\eta}^{ck})$ is an inner automorphism, the GNS representations of $\omega_f^{ck}$ and $\omega_{f'}^{ck}$ are unitarily equivalent. Since $\pi_0\comp\rho_\gamma^{ck}$ and $\pi_0\comp\rho_{\gamma'}^{ck}$ are GNS representations of $\omega_f^{ck}$ and $\omega_{f'}^{ck}$, respectively, it follows that $\pi_0\comp\rho_\gamma^{ck} \simeq \pi_0\comp\rho_{\gamma'}^{ck}$. The latter is localized in $\Lambda'$ by the localization result above. Since $\Lambda'$ was arbitrary, $\pi_0\comp\rho_\gamma^{ck}$ is transportable.
\end{proof}

Let $\pi_f^{ck}$ be a GNS representation of the state $\omega_f^{ck}$. By Proposition~\ref{prop:local-transp}, its unitary equivalence class is independent of the choice of the starting face $f$. We now distinguish the representations corresponding to different labels $ck$.

\begin{theorem}\label{thm:sectors}
    Let $c,c'\in\{\mathtt{r},\mathtt{g},\mathtt{b}\}$ and $k,k'\in\{\mathtt{x},\mathtt{y},\mathtt{z}\}$. Let $f$ and $f'$ be $c$- and $c'$-colored faces, respectively. Then $\pi_f^{ck}\simeq \pi_{f'}^{c'k'}$ if and only if $ck=c'k'$. 
\end{theorem}

\begin{proof}
    If $ck=c'k'$, the equivalence follows from Proposition~\ref{prop:local-transp}.

    Conversely, suppose that $ck\neq c'k'$. Since $\omega_0$ is a pure state and $\rho_\gamma^{ck}$ is an automorphism of $\mathcal{A}$, the states $\omega_f^{ck}$ and $\omega_{f'}^{c'k'}$ are pure. We show that they violate the criterion in Proposition~\ref{prop:equiv_GNS}. Choose $\varepsilon_0=1$. Let $S\subset V(\Sigma)$ be an arbitrary finite subset. We will construct a finite subset
    $R\subset S^c$ and an operator $A\in\mathcal{A}(R)$ such that $|\omega_f^{ck}(A)-\omega_{f'}^{c'k'}(A)| > \varepsilon_0\|A\|$. We distinguish two cases.

    First, suppose that $c\neq c'$. Choose a closed and non-self-intersecting string $\tilde{\gamma}$ of type $c'$ enclosing $S$ as well as the starting faces $f$ and $f'$, such that $R:=\operatorname{supp}(\tilde{\gamma})\subset S^c$. Choose $k''\neq k$ and set $A:=S_{\tilde{\gamma}}^{c'k''}$. Since $\tilde{\gamma}$ is closed, $A$ is a product of stabilizer operators. Hence $\omega_0(A)=1$, $\|A\|=1$. Let $\gamma$ and $\gamma'$ be half-infinite strings defining $\omega_f^{ck}$ and $\omega_{f'}^{c'k'}$, respectively. Since $c\neq c'$, the strings $\gamma$ and $\tilde{\gamma}$ intersect
    in an odd number of vertices. Since also $k\neq k''$, their corresponding string operators anticommute. Thus, for all sufficiently large $n$, $(S_{\gamma_n}^{ck})^\dagger A S_{\gamma_n}^{ck} = -A$, and therefore $\omega_f^{ck}(A) = \lim_{n\to\infty}\omega_0((S_{\gamma_n}^{ck})^\dagger A S_{\gamma_n}^{ck}) = -\omega_0(A) = -1$. On the other hand, $\gamma'$ and $\tilde{\gamma}$ are both of type $c'$, and hence they intersect in an even number of vertices. Their corresponding string operators therefore commute. Thus, for all sufficiently large $n$, $(S_{\gamma'_n}^{c'k'})^\dagger A S_{\gamma'_n}^{c'k'} = A$, which gives $\omega_{f'}^{c'k'}(A) = \omega_0(A) = 1$. 

    Now suppose that $c=c'$ and $k\neq k'$. Choose $c''\neq c$ and a closed and non-self-intersecting string $\tilde{\gamma}$ of type $c''$ enclosing $S$, $f$, and $f'$, such that $R:=\operatorname{supp}(\tilde{\gamma})\subset S^c$. Set $A:=S_{\tilde{\gamma}}^{c''k'}$. Again, since $\tilde{\gamma}$ is closed, $\omega_0(A)=1$, $\|A\|=1$. Since $c''\neq c$, the strings $\gamma$ and $\tilde{\gamma}$ intersect in an odd number of vertices. Moreover, $k\neq k'$, so the corresponding string operators anticommute. Hence $ \omega_f^{ck}(A)=-1$. In contrast, the string operators $S_{\gamma'_n}^{c'k'}$ and $S_{\tilde{\gamma}}^{c''k'}$ are built from the same type $k'$ of Pauli matrices and therefore commute. Consequently, $\omega_{f'}^{c'k'}(A)=1$. 

    Thus, in either case, we have constructed a finite set $R\subset S^c$ and an operator $A\in\mathcal{A}(R)$ such that $|\omega_f^{ck}(A)-\omega_{f'}^{c'k'}(A)| = 2 > \varepsilon_0\|A\|$. Since $S$ is arbitrary, the criterion in Proposition~\ref{prop:equiv_GNS} implies that the GNS representations $\pi_f^{ck}$ and $\pi_{f'}^{c'k'}$ are not unitarily equivalent. This completes the proof.
\end{proof}

\begin{remark}
    The same argument as in the proof of the theorem above shows that $\pi_f^{ck}$ is not equivalent to $\pi_0$. Again, the precise shape of the region $\Lambda$ is not essential; one only requires that the region contains half-infinite strings. 
\end{remark}

From now on, we will denote the GNS representation of $\omega_f^{ck}$ by $\pi^{ck}$, omitting explicit mention of the starting face $f$.    

\begin{corollary}
    The representation $\pi^{ck}$ is a superselection sector of $\pi_0$.
\end{corollary}

\begin{proof}
    Let $\Lambda$ be any cone and choose a half-infinite string $\gamma$ of type $c$ contained in $\Lambda$. By Proposition~\ref{prop:local-transp}, $\pi_0\comp\rho_\gamma^{ck}$ is localized in $\Lambda$, and it is a GNS representation of $\omega_f^{ck}$. Hence $\pi^{ck}\simeq\pi_0\comp\rho_\gamma^{ck}$, and therefore $\pi^{ck}|_{\mathcal{A}(\Lambda^c)} \simeq \pi_0|_{\mathcal{A}(\Lambda^c)}$. Since $\Lambda$ is arbitrary, $\pi^{ck}$ satisfies the superselection criterion.
\end{proof}

\begin{figure}[t]
    \centering
    \begin{subfigure}{0.45\textwidth}
        \centering
        \includegraphics[width=\textwidth]{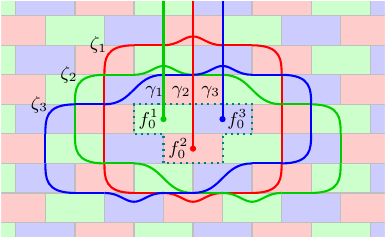}
    \end{subfigure}
    \hspace{1.5cm}
    \begin{subfigure}{0.176\textwidth}
        \centering
        \includegraphics[width=\textwidth]{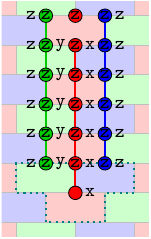}
    \end{subfigure}
    \caption{(Left) Three vertical half-infinite strings $\gamma_1$, $\gamma_2$, and $\gamma_3$ are shown, along with the closed winding strings $\zeta_1$, $\zeta_2$, and $\zeta_3$ that encircle their respective initial faces. The distinct colors of the strings are also indicated. (Right) Deformation of string operators from $S^{\mathtt{rx}}S^{\mathtt{bz}}$ to $S^{\mathtt{ry}}S^{\mathtt{gz}}$. The product of the matrix on the right of the bullet with the matrix at the bullet yields the matrix on the left. Note that the new string $\tilde{\gamma}_2^N$ has a new starting face $f_1^2$. \label{fig:winding_strings}}
\end{figure}

Consider three adjacent faces $f_0^1$, $f_0^2$ and $f_0^3$, colored by $c_1=\mathtt{g}$, $c_2=\mathtt{r}$ and $c_3=\mathtt{b}$, respectively.  Let $\gamma_i$ be the vertical half-infinite string starting from $f_0^i$ with each $\gamma_i$ colored by $c_i$ for $i=1,2,3$; see Figure~\ref{fig:winding_strings} (Left). 
For $k_i, k_j\in \{\mathtt{x},\mathtt{y},\mathtt{z}\}$ with $k_i \neq k_j$, consider the state 
\begin{equation*}
    \omega^{c_ik_i,c_jk_j} = \omega_0\comp \rho_{\gamma_i}^{c_ik_i}\comp \rho_{\gamma_j}^{c_jk_j}.
\end{equation*}
This state is pure because it is obtained by composing the pure state $\omega_0$ with automorphisms of $\mathcal{A}$; hence its GNS representation, denoted as $\pi^{c_ik_i,c_jk_j}$, is irreducible. 
One can also show that, for fixed starting faces, this state does not depend on the choices of the strings. Different choices of starting faces generally give different states, but their corresponding GNS representations are unitarily equivalent.

\begin{theorem} \label{thm:sectors_fermion}
    Each $\pi^{c_ik_i,c_jk_j}$ is a superselection sector of $\omega_0$, and is not equivalent to $\pi_0$ and any $\pi^{ck}$ above. Moreover, we have the following equivalences 
    \begin{align*}
        &\pi^{\mathtt{rx},\mathtt{bz}} \simeq \pi^{\mathtt{ry},\mathtt{gz}} \simeq \pi^{\mathtt{gx},\mathtt{by}},\quad 
        \pi^{\mathtt{rz},\mathtt{bx}} \simeq \pi^{\mathtt{ry},\mathtt{gx}} \simeq \pi^{\mathtt{gz},\mathtt{by}}, \\
        & \pi^{\mathtt{bz},\mathtt{gy}} \simeq \pi^{\mathtt{gx},\mathtt{rz}} \simeq \pi^{\mathtt{bx},\mathtt{ry}}, \quad \pi^{\mathtt{rz},\mathtt{gy}} \simeq \pi^{\mathtt{gx},\mathtt{bz}} \simeq \pi^{\mathtt{rx},\mathtt{by}}, \\
        & \pi^{\mathtt{rx},\mathtt{gy}} \simeq \pi^{\mathtt{bx},\mathtt{gz}} \simeq \pi^{\mathtt{by},\mathtt{rz}}, \quad
        \pi^{\mathtt{bx},\mathtt{gy}} \simeq \pi^{\mathtt{rx},\mathtt{gz}} \simeq \pi^{\mathtt{ry},\mathtt{bz}}. 
    \end{align*}
\end{theorem}

\begin{proof}
    The first statement follows from the preceding results.
    The proof that these representations are not equivalent to $\pi_0$ and $\pi^{ck}$ is similar to the proof of Theorem~\ref{thm:sectors}. 
    Furthermore, by a similar argument, the representations $\pi^{ck,c'k'}$ are pairwise inequivalent unless they are listed as equivalent above.
    Let us show the equivalences; we will show $\pi^{\mathtt{rx},\mathtt{bz}} \simeq \pi^{\mathtt{ry},\mathtt{gz}}$ as an example, the remaining equivalences being proved similarly.
    The idea is to deform their defining string operators using stabilizer operators. 
    We will use the criterion in Proposition~\ref{prop:equiv_GNS}. Fix an $\varepsilon>0$. 
    For $n\geq 1$, let $\gamma_i^n$ be the finite string comprising the first $n+1$ faces, represented as
    \begin{equation*}
        \gamma_i^n = (f_0^i,v_0^i, v_1^i,f_1^i, v_2^i, v_3^i, \cdots, f_n^i).
    \end{equation*}
    Now let $S_\varepsilon$ be a disk such that all the faces $f_0^i$ are in its interior.  
    Let $S\subset S_{\varepsilon}^c$ be any finite subset and $A\in \mathcal{A}(S) \subset \mathcal{A}_{\rm loc}$. 
    Choose $N\geq 1$ such that the support of $A$ does not intersect $\gamma_i\setminus\gamma_i^n$ for all $n\geq N$ and all $i=1,2,3$.
    Hence, 
    \begin{equation*}
        \rho_{\gamma_i}^{c_ik_i}\comp \rho_{\gamma_j}^{c_jk_j}(A) = (S_{\gamma_i^N}^{c_ik_i}S_{\gamma_j^N}^{c_jk_j})^\dagger AS_{\gamma_i^N}^{c_ik_i}S_{\gamma_j^N}^{c_jk_j},
    \end{equation*}
    which results in 
    \begin{equation*}
        \omega^{c_ik_i,c_jk_j}(A) = \langle S_{\gamma_i^N}^{c_ik_i}S_{\gamma_j^N}^{c_jk_j} \Omega, AS_{\gamma_i^N}^{c_ik_i}S_{\gamma_j^N}^{c_jk_j}\Omega \rangle. 
    \end{equation*}
    Consider $c_ik_i=\mathtt{rx}, \mathtt{ry}$ and $c_jk_j=\mathtt{bz}, \mathtt{gz}$. 
    Denote $J:=\prod_{s=1}^NJ_{f_s^2}$. 
    One verifies that 
    \begin{equation*}
        S_{\gamma_2^N}^{\mathtt{rx}}S_{\gamma_3^N}^{\mathtt{bz}} J = \sigma  S^{\mathtt{ry}}_{\tilde{\gamma}_2^N}S^{\mathtt{gz}}_{\gamma_1^N},
    \end{equation*}
    where $\sigma = \sigma_{v_0^2}^{\mathtt{x}}\sigma_{v_1^2}^{\mathtt{x}}\sigma_{v_1^2}^{\mathtt{z}}\sigma_{v_{N+1}^2}^{\mathtt{z}}$, and $\tilde{\gamma}_2^N = \gamma_2^N\setminus\{f_0^2,v_0^2,v_1^2\}$; see Figure~\ref{fig:winding_strings} (Right) for an illustration. 
    Note that the support of $\sigma$ has no overlap with the support of $A$; so $[\sigma,A]=0$. 
    Since $J\Omega = \Omega$, we have 
    \begin{align*}
        \omega^{\mathtt{rx},\mathtt{bz}}(A) 
        & = \langle S_{\gamma_2}^{\mathtt{rx}}S_{\gamma_3}^{\mathtt{bz}} J\Omega, AS_{\gamma_2}^{\mathtt{rx}}S_{\gamma_3}^{\mathtt{bz}}J\Omega\rangle \\  
        & = \langle \sigma S^{\mathtt{ry}}_{\tilde{\gamma}_2^N}S^{\mathtt{gz}}_{\gamma_1^N} \Omega, A\sigma S^{\mathtt{ry}}_{\tilde{\gamma}_2^N}S^{\mathtt{gz}}_{\gamma_1^N} \Omega\rangle \\ 
        & = \langle S^{\mathtt{ry}}_{\tilde{\gamma}_2^N}S^{\mathtt{gz}}_{\gamma_1^N} \Omega, A\sigma ^\dagger\sigma S^{\mathtt{ry}}_{\tilde{\gamma}_2^N}S^{\mathtt{gz}}_{\gamma_1^N} \Omega\rangle \\ 
        & =  \langle S^{\mathtt{ry}}_{\tilde{\gamma}_2^N}S^{\mathtt{gz}}_{\gamma_1^N} \Omega, A S^{\mathtt{ry}}_{\tilde{\gamma}_2^N}S^{\mathtt{gz}}_{\gamma_1^N} \Omega\rangle \\
        & = \omega^{\mathtt{ry},\mathtt{gz}}(A). 
    \end{align*}
    Thus the condition $|\omega^{\mathtt{rx},\mathtt{bz}}(A) - \omega^{\mathtt{ry},\mathtt{gz}}(A)|<\varepsilon\|A\|$ trivially holds for $A\in \mathcal{A}(S)$; this implies that $\pi^{\mathtt{rx},\mathtt{bz}}\simeq \pi^{\mathtt{ry},\mathtt{gz}}$. This completes the proof.
\end{proof}

\begin{definition}
    We denote the pairwise inequivalent superselection sectors above as 
    \begin{align*}
        & \pi^{\mathtt{f}_1} = \pi^{\mathtt{rx},\mathtt{bz}}, \quad \pi^{\mathtt{f}_2} = \pi^{\mathtt{rz},\mathtt{bx}}, \quad \pi^{\mathtt{f}_3} = \pi^{\mathtt{bz},\mathtt{gy}}, \\
        & \pi^{\mathtt{f_4}} = \pi^{\mathtt{rz},\mathtt{gy}}, \quad \pi^{\mathtt{f}_5} = \pi^{\mathtt{rx},\mathtt{gy}}, \quad \pi^{\mathtt{f}_6} = \pi^{\mathtt{bx},\mathtt{gy}}. 
    \end{align*}
    We also denote by $\rho^{\mathtt{f}_i}$ any endomorphism corresponding to $\pi^{\mathtt{f}_i}$, such as $\rho^{\mathtt{f}_1} = \rho^{\mathtt{rx}}\rho^{\mathtt{bz}}$ corresponding to strings $\gamma_2$ and $\gamma_3$ in Figure~\ref{fig:winding_strings} (Left).  
\end{definition}

\subsection{Fusion rules}  \label{subsec:fusion}

We now organize the superselection sectors constructed above into the category of localized and transportable endomorphisms and define their fusion rules. Since the quasi-local algebra $\mathcal A$ is simple, we identify $\mathcal A$ with its image $\pi_0(\mathcal A)\subset \mathbf B(\mathcal H)$ whenever no confusion
arises. Recall that an intertwiner from an endomorphism $\rho$ to an endomorphism $\rho'$ is an operator $T$ satisfying 
\[
    T\rho(A)=\rho'(A)T, \quad A\in\mathcal A. 
\]
If $T$ is unitary, it is also called a charge transporter.

Fix a cone $\Lambda_0$. As in the standard construction of superselection sectors with cone localization, we introduce the auxiliary algebra
\begin{equation*}
    \mathcal{A}^{au} = \overline{\bigcup_{x\in \mathbb{Z}^2} \pi_0(\mathcal{A}(\Lambda_0+x))''}^{\|\cdot\|}, 
\end{equation*}
where the union is taken over the translates of $\Lambda_0$, ordered by inclusion. Clearly, $\mathcal A\subset\mathcal A^{au}$. We will use Haag duality, which is proved in Section~\ref{sec:haag_duality}. In particular, if $\rho$ and $\rho'$ are localized in the same cone $\Lambda$ and $T$ is an intertwiner from $\rho$ to $\rho'$, then
\[
    TA = AT, \quad \forall~A\in\mathcal A(\Lambda^c).
\]
Hence Haag duality gives
\[
    T\in \pi_0\bigl(\mathcal A(\Lambda^c)\bigr)' = \pi_0\bigl(\mathcal A(\Lambda)\bigr)''.
\]
Thus, after enlarging $\Lambda$ to a suitable translate of $\Lambda_0$ if necessary, such intertwiners belong to $\mathcal A^{au}$.

The extension of localized endomorphisms to the auxiliary algebra is standard. More precisely, an endomorphism localized in a translate $\Lambda_0+x$ extends canonically to the corresponding auxiliary algebra; see \cite[Lemma~4.1]{Buchholz1982locality} for the original model-independent argument and \cite{Ogata2022braided} for the quantum spin system setting. In particular, the localized and transportable endomorphisms constructed above admit such extensions. We use the same notation for an endomorphism and its extension whenever no confusion can arise.

\begin{definition}\label{def:Dloc}
    Let $\mathcal D_{\mathrm{loc,tr}}$ be the category whose objects are endomorphisms of $\mathcal A^{au}$ whose restrictions to $\mathcal A$ are localized in $\Lambda_0$ and transportable. For two objects $\rho,\rho'\in\mathcal D_{\mathrm{loc,tr}}$, define
    \[
        \Hom(\rho,\rho'):=\left\{T\in\mathcal A^{au}\,\middle|\,T\rho(A)=\rho'(A)T\text{ for all }A\in\mathcal A\right\}.
    \]
    Composition of morphisms is given by multiplication in $\mathcal A^{au}$, and the identity morphism of every object is $\mathds{1}$.
\end{definition}

The tensor product is given by composition of endomorphisms,
\[
    \rho\otimes\sigma:=\rho\comp\sigma,
\]
while for
\[
    S\in\operatorname{Hom}(\rho_1,\rho_2), \quad T\in\operatorname{Hom}(\sigma_1,\sigma_2),
\]
the tensor product of morphisms is
\begin{equation}\label{eq:tensor_intertwiner}
    S\otimes T
    :=
    S\,\rho_1(T).
\end{equation}
The fact that these operations are well defined follows from the standard extension theory for localized and transportable endomorphisms; see \cite{Buchholz1982locality,Ogata2022braided}.

For every cone $\Lambda$, the cone algebra
\[
    \mathcal R_\Lambda
    :=
    \pi_0\bigl(\mathcal A(\Lambda)\bigr)''
\]
is an infinite factor by the standard argument of \cite[Theorem~5.1]{naaijkens2011localized} and \cite[Proposition~5.3]{keyl2006entanglement}. Together with Haag duality, the general theory of localized and transportable endomorphisms therefore equips $\mathcal D_{\mathrm{loc,tr}}$ with the structure of a braided $C^*$-tensor category; see \cite{Ogata2022braided}.

We now determine explicitly the fusion rules of the superselection sectors constructed above. Since the corresponding localized endomorphisms are obtained from half-infinite Pauli string operators, their compositions can be computed directly from the string construction.

\begin{proposition}\label{prop:fusion_rules}
    The sixteen distinct superselection sectors described in Theorems~\ref{thm:sectors} and~\ref{thm:sectors_fermion} obey the fusion rules given in Table~\ref{tab:colorcode_fusion}.
\end{proposition}

\begin{proof}
    Some of the fusion rules follow directly from Theorem~\ref{thm:sectors_fermion}, together with the identification of the tensor product with composition of localized endomorphisms. The remaining fusion rules are obtained by the same string deformation argument.
\end{proof}

\begin{remark}
This is consistent with the finite-volume color code model. 
In this case, each $\gamma_i$ is a finite string.  
The string operator $S_{\gamma_i}^{c_ik_i}$ creates a boson $c_ik_i$ at the initial face $f_0^i$ of $\gamma_i$, where $c_i$ is the color of $\gamma_i$ and $k_i\in \{\mathtt{x},\mathtt{y},\mathtt{z}\}$. 
Fusion information at these initial faces can be detected using the closed string operators surrounding $f_0^i$.
Consider closed strings $\zeta_1$, $\zeta_2$ and $\zeta_3$ colored by $\mathtt{r}$, $\mathtt{g}$ and $\mathtt{b}$ respectively, which encircle the starting faces of the vertical strings above; see Figure~\ref{fig:winding_strings} for an illustration. 
Let $|\Omega\rangle$ be a ground state.  
Clearly the associated closed string operators satisfy
\begin{equation*}
S_{\zeta_i}^{c_ik_i}|\Omega\rangle = |\Omega\rangle,\quad i=1,2,3,\; k_i\in \{\mathtt{x},\mathtt{y},\mathtt{z}\}.
\end{equation*}
The closed string $\zeta_1$ intersects each of $\gamma_1$ and $\gamma_3$ at a single vertex; hence it is straightforward to verify the following: 
\begin{align*}
    &
    \begin{cases}
        S_{\zeta_1}^{\mathtt{rx}} (S_{\gamma_2}^{\mathtt{rx}}S_{\gamma_3}^{\mathtt{bz}}|\Omega\rangle)  =  S_{\gamma_2}^{\mathtt{rx}} S_{\zeta_1}^{\mathtt{rx}} S_{\gamma_3}^{\mathtt{bz}}|\Omega\rangle = -S_{\gamma_2}^{\mathtt{rx}}S_{\gamma_3}^{\mathtt{bz}} S_{\zeta_1}^{\mathtt{rx}} |\Omega\rangle = -S_{\gamma_2}^{\mathtt{rx}}S_{\gamma_3}^{\mathtt{bz}} |\Omega\rangle, \\
        S_{\zeta_1}^{\mathtt{rx}} (S_{\gamma_2}^{\mathtt{ry}}S_{\gamma_1}^{\mathtt{gz}}|\Omega\rangle)  =  S_{\gamma_2}^{\mathtt{ry}}S_{\zeta_1}^{\mathtt{rx}}S_{\gamma_1}^{\mathtt{gz}}|\Omega\rangle = -S_{\gamma_2}^{\mathtt{ry}}S_{\gamma_1}^{\mathtt{gz}}S_{\zeta_1}^{\mathtt{rx}}|\Omega\rangle = -S_{\gamma_2}^{\mathtt{ry}}S_{\gamma_1}^{\mathtt{gz}}|\Omega\rangle, \\
        S_{\zeta_1}^{\mathtt{rx}} (S_{\gamma_1}^{\mathtt{gx}}S_{\gamma_3}^{\mathtt{by}}|\Omega\rangle)  =   S_{\gamma_1}^{\mathtt{gx}}S_{\zeta_1}^{\mathtt{rx}}S_{\gamma_3}^{\mathtt{by}}|\Omega\rangle = -S_{\gamma_1}^{\mathtt{gx}}S_{\gamma_3}^{\mathtt{by}}S_{\zeta_1}^{\mathtt{rx}}|\Omega\rangle= -S_{\gamma_1}^{\mathtt{gx}}S_{\gamma_3}^{\mathtt{by}}|\Omega\rangle, \\
    \end{cases} \\
    &
    \begin{cases}
        S_{\zeta_1}^{\mathtt{ry}} (S_{\gamma_2}^{\mathtt{rx}}S_{\gamma_3}^{\mathtt{bz}}|\Omega\rangle)  =  S_{\gamma_2}^{\mathtt{rx}} S_{\zeta_1}^{\mathtt{ry}} S_{\gamma_3}^{\mathtt{bz}}|\Omega\rangle = -S_{\gamma_2}^{\mathtt{rx}}S_{\gamma_3}^{\mathtt{bz}} S_{\zeta_1}^{\mathtt{ry}} |\Omega\rangle = -S_{\gamma_2}^{\mathtt{rx}}S_{\gamma_3}^{\mathtt{bz}} |\Omega\rangle, \\
        S_{\zeta_1}^{\mathtt{ry}} (S_{\gamma_2}^{\mathtt{ry}}S_{\gamma_1}^{\mathtt{gz}}|\Omega\rangle)  =  S_{\gamma_2}^{\mathtt{ry}}S_{\zeta_1}^{\mathtt{ry}}S_{\gamma_1}^{\mathtt{gz}}|\Omega\rangle = -S_{\gamma_2}^{\mathtt{ry}}S_{\gamma_1}^{\mathtt{gz}}S_{\zeta_1}^{\mathtt{ry}}|\Omega\rangle = -S_{\gamma_2}^{\mathtt{ry}}S_{\gamma_1}^{\mathtt{gz}}|\Omega\rangle, \\
        S_{\zeta_1}^{\mathtt{ry}} (S_{\gamma_1}^{\mathtt{gx}}S_{\gamma_3}^{\mathtt{by}}|\Omega\rangle)  =  - S_{\gamma_1}^{\mathtt{gx}}S_{\zeta_1}^{\mathtt{ry}}S_{\gamma_3}^{\mathtt{by}}|\Omega\rangle = -S_{\gamma_1}^{\mathtt{gx}}S_{\gamma_3}^{\mathtt{by}}S_{\zeta_1}^{\mathtt{ry}}|\Omega\rangle= -S_{\gamma_1}^{\mathtt{gx}}S_{\gamma_3}^{\mathtt{by}}|\Omega\rangle, 
    \end{cases} \\
    &
    \begin{cases}
        S_{\zeta_1}^{\mathtt{rz}} (S_{\gamma_2}^{\mathtt{rx}}S_{\gamma_3}^{\mathtt{bz}}|\Omega\rangle)  =  S_{\gamma_2}^{\mathtt{rx}} S_{\zeta_1}^{\mathtt{rz}} S_{\gamma_3}^{\mathtt{bz}}|\Omega\rangle = S_{\gamma_2}^{\mathtt{rx}}S_{\gamma_3}^{\mathtt{bz}} S_{\zeta_1}^{\mathtt{rz}} |\Omega\rangle = S_{\gamma_2}^{\mathtt{rx}}S_{\gamma_3}^{\mathtt{bz}} |\Omega\rangle, \\
        S_{\zeta_1}^{\mathtt{rz}} (S_{\gamma_2}^{\mathtt{ry}}S_{\gamma_1}^{\mathtt{gz}}|\Omega\rangle)  =  S_{\gamma_2}^{\mathtt{ry}}S_{\zeta_1}^{\mathtt{rz}}S_{\gamma_1}^{\mathtt{gz}}|\Omega\rangle = S_{\gamma_2}^{\mathtt{ry}}S_{\gamma_1}^{\mathtt{gz}}S_{\zeta_1}^{\mathtt{rz}}|\Omega\rangle = S_{\gamma_2}^{\mathtt{ry}}S_{\gamma_1}^{\mathtt{gz}}|\Omega\rangle, \\
        S_{\zeta_1}^{\mathtt{rz}} (S_{\gamma_1}^{\mathtt{gx}}S_{\gamma_3}^{\mathtt{by}}|\Omega\rangle)  =   -S_{\gamma_1}^{\mathtt{gx}}S_{\zeta_1}^{\mathtt{rz}}S_{\gamma_3}^{\mathtt{by}}|\Omega\rangle = S_{\gamma_1}^{\mathtt{gx}}S_{\gamma_3}^{\mathtt{by}}S_{\zeta_1}^{\mathtt{rz}}|\Omega\rangle= S_{\gamma_1}^{\mathtt{gx}}S_{\gamma_3}^{\mathtt{by}}|\Omega\rangle. 
    \end{cases}
\end{align*}
This means that the states $S_{\gamma_2}^{\mathtt{rx}}S_{\gamma_3}^{\mathtt{bz}}|\Omega\rangle$, $S_{\gamma_2}^{\mathtt{ry}}S_{\gamma_1}^{\mathtt{gz}}|\Omega\rangle$, and $S_{\gamma_1}^{\mathtt{gx}}S_{\gamma_3}^{\mathtt{by}}|\Omega\rangle$ satisfy the same relations under the action of $S_{\zeta_1}^{\mathtt{r}k_1}$ (for $k_1\in\{\mathtt{x},\mathtt{y},\mathtt{z}\}$). The same holds when acting with $S_{\zeta_i}^{c_ik_i}$ for $i=2,3$ and $k_i\in \{\mathtt{x},\mathtt{y},\mathtt{z}\}$. Consequently, these three states belong to the same topological excitation, denoted $\mathtt{f}_1$. In other words, $\mathtt{rx}\otimes \mathtt{bz}=\mathtt{ry}\otimes \mathtt{gz} = \mathtt{gx}\otimes \mathtt{by}=\mathtt{f_1}$ as observed in Table~\ref{tab:colorcode_fusion}. The same argument applies to the other five fermions.
\end{remark}

\subsection{Braiding} \label{subsec:braiding}

It remains to define the braiding. 
The approach for the toric code was developed in \cite{naaijkens2011localized}, following a method similar to the DHR program \cite{doplicher1971local}. 
We will follow the same argument. 

\begin{figure}[t]
    \centering
    \includegraphics[width=6cm]{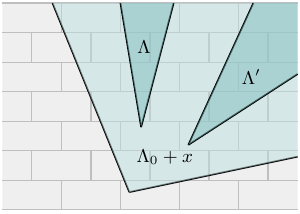}
    \caption{ The cone $\Lambda$ is to the left of $\Lambda'$. \label{fig:cone_left_right}}
\end{figure}

First we define a relation between two disjoint cones $\Lambda$ and $\Lambda'$ that are contained in $\Lambda_0+x$ for some $x$. Here $\Lambda_0$ is a fixed cone used to define the auxiliary algebra $\mathcal{A}^{au}$ given above. 
We define $\Lambda$ to be to the left of $\Lambda'$ if $\Lambda$ can be rotated counterclockwise around its apex until it intersects $(\Lambda_0+x)^c$ without meeting $\Lambda'$ during this process; otherwise, $\Lambda$ is to the right of $\Lambda'$. 
See Figure~\ref{fig:cone_left_right} for an illustration. 

Let $\rho$ and $\rho'$ be two transportable endomorphisms localized in cones $\Lambda$ and $\Lambda'$, respectively. 
Choose a cone $\tilde{\Lambda}'$ to the left of $\Lambda$. By transportability, there is a unitary $U$ such that $\tilde{\rho}'=U\rho'(\bullet)U^\dagger$ is localized in $\tilde{\Lambda}'$. The braiding is then given by 
\begin{equation*}
    \varepsilon_{\rho,\rho'} = (U\otimes I_\rho)^\dagger(I_\rho\otimes U) = U^\dagger \bar{\rho}(U),
\end{equation*}
where $\bar{\rho}$ denotes the extension of $\rho$ to
$\mathcal A^{{au}}$. Here $I_\rho=\mathds{1}\in \mathcal{A}$ is the identity matrix which can be regarded as an intertwiner from $\rho$ to itself when it has a subscript $\rho$.

This construction is a special case of the general braiding for
localized and transportable endomorphisms developed in
\cite{Ogata2022braided}. In particular,
$\varepsilon_{\rho,\rho'}$ is well defined and unitary, is natural in
both variables, and satisfies the braiding identities. Explicitly, for
intertwiners
\[
    S:\rho_1\to\rho_2, \quad T:\rho'_1\to\rho'_2,
\]
naturality gives
\begin{equation}
    \varepsilon_{\rho_2,\rho'}(S\otimes I_{\rho'})  = (I_{\rho'}\otimes S)\varepsilon_{\rho_1,\rho'}, \quad 
    \varepsilon_{\rho,\rho'_2}(I_\rho\otimes T)  = (T\otimes I_\rho)\varepsilon_{\rho,\rho'_1}. \label{eq:braid_nat} 
\end{equation}
Moreover,
\begin{align}
    \varepsilon_{\rho,\rho'\otimes \rho''}  = (I_{\rho'}\otimes \varepsilon_{\rho,\rho''})(\varepsilon_{\rho,\rho'}\otimes I_{\rho''}), \quad 
    \varepsilon_{\rho\otimes \rho',\rho''}  = (\varepsilon_{\rho,\rho''}\otimes I_{\rho'})(I_\rho\otimes\varepsilon_{\rho',\rho''}). \label{eq:braid_hexagon}
\end{align}

To compute the braiding explicitly, we will use the following
description of the charge transporters associated with half-infinite
strings.

\begin{proposition}[{\cite[Lemma~4.1, Theorem~4.1]{naaijkens2011localized}}] \label{prop:unitary}
    Let $\gamma$ and $\gamma'$ be two half-infinite strings of type $c$ in a cone $\Lambda$, starting at the faces $f$ and $f'$, respectively. Then there is a unitary intertwiner $U$ from $\rho_{\gamma}^{ck}$ to $\rho_{\gamma'}^{ck}$ such that $U\Omega=S_{\tilde{\gamma}}^{ck}\Omega$, where $\tilde{\gamma}$ is any finite string of type $c$ from $f$ to $f'$. Moreover, $U$ can be taken as the weak limit $U=\operatorname{\textit{w}\text{-}lim}_{n\to\infty} U_n$, where $U_n = S_{\gamma_n}^{ck}S_{\tilde{\gamma}_n}^{ck}S_{\gamma'_n}^{ck}$, and $\tilde{\gamma}_n$ is any string connecting the terminating faces of $\gamma_n$ and $\gamma'_n$ such that $\operatorname{dist}(f,\tilde{\gamma}_n)\to\infty$, $\operatorname{dist}(f',\tilde{\gamma}_n)\to\infty$ as $n\to\infty$.
\end{proposition}

\begin{proof}
    By Proposition~\ref{prop:local-transp}, $\pi_0\comp\rho_\gamma^{ck}$ and $\pi_0\comp\rho_{\gamma'}^{ck}$ are unitarily equivalent. We now construct explicitly a unitary intertwiner with the properties stated above, following \cite[Lemma~4.1]{naaijkens2011localized}.

    For simplicity, we omit the superscript $ck$. Let $\tilde{\gamma}$ be a finite string of type $c$ connecting $f$ to $f'$, and concatenate $\tilde{\gamma}$ with $\gamma'$ to obtain a half-infinite string $\widehat{\gamma}:=\tilde{\gamma}\gamma'$ starting at $f$. We first construct an intertwiner from $\rho_\gamma$ to $\rho_{\widehat{\gamma}}$. Since $\gamma$ and $\widehat{\gamma}$ have the same starting face, for each $n$ choose a finite string $\eta_n$ connecting the terminating faces of $\gamma_n$ and $\widehat{\gamma}_n$, such that $\operatorname{dist}(f,\eta_n)\to \infty$ as $n\to\infty$. Then $\gamma_n\eta_n\widehat{\gamma}_n$ is a closed string. Define $W_n := S_{\gamma_n} S_{\eta_n} S_{\widehat{\gamma}_n}$. 
    The operator $W_n$ is a product of stabilizer operators, and hence $W_n\Omega=\Omega$. Moreover, each $W_n$ is unitary, so $\|W_n\|=1$ for all $n$. We claim that $\{W_n\}_n$ converges in the weak operator topology. Since $\rho_\gamma(\mathcal A_{\rm loc})\Omega$ is dense in $\mathcal H$ and $\{W_n\}_n$ is uniformly norm bounded, it suffices to consider $x=\rho_\gamma(A)\Omega$, $y=\rho_\gamma(B)\Omega$, for $A,B\in\mathcal A_{\rm loc}$. Choose $N\geq1$ such that, for all $n\geq N$, the support of $B$ is disjoint from $\eta_n$, $\gamma_n\setminus\gamma_N$, and $\widehat{\gamma}_n\setminus\widehat{\gamma}_N$. By locality,
    \[
        W_n\rho_\gamma(B) = \rho_{\widehat{\gamma}}(B)W_n.
    \]
    Therefore,
    \begin{align*}
        \lim_{n\to\infty}\left\langle\rho_\gamma(A)\Omega,W_n\rho_\gamma(B)\Omega\right\rangle &= \lim_{n\to\infty}\left\langle\rho_\gamma(A)\Omega,\rho_{\widehat{\gamma}}(B)W_n\Omega\right\rangle
        \\
        &= \left\langle\rho_\gamma(A)\Omega,\rho_{\widehat{\gamma}}(B)\Omega\right\rangle .
    \end{align*}
    Hence $W_n$ converges weakly to an operator $W$ satisfying
    \[
        W\rho_\gamma(A) = \rho_{\widehat{\gamma}}(A)W, \quad W\Omega=\Omega.
    \]
    Since $\rho_\gamma$ and $\rho_{\widehat{\gamma}}$ are irreducible and unitarily equivalent, $W$ is unitary. By construction, $\rho_{\widehat{\gamma}} = \operatorname{Ad}(S_{\tilde{\gamma}}) \comp\rho_{\gamma'}$. Therefore, setting $U:=S_{\tilde{\gamma}}W$, we obtain, for every $A\in\mathcal A$,
    \[
        U\rho_\gamma(A) = S_{\tilde{\gamma}}W\rho_\gamma(A) = S_{\tilde{\gamma}} \rho_{\widehat{\gamma}}(A)W = \rho_{\gamma'}(A)S_{\tilde{\gamma}}W = \rho_{\gamma'}(A)U.
    \]
    Thus $U$ is a unitary intertwiner from $\rho_\gamma$ to $\rho_{\gamma'}$, and $U\Omega = S_{\tilde{\gamma}}W\Omega = S_{\tilde{\gamma}}\Omega$. 
    Finally, since $U=S_{\tilde{\gamma}}W$ and $W=\operatorname{\textit{w}\text{-}lim}_{n\to\infty}W_n$, we have $U=\operatorname{\textit{w}\text{-}lim}_{n\to\infty}S_{\tilde{\gamma}}W_n$. By combining the fixed string $\tilde{\gamma}$ with $\eta_n$ and the corresponding initial segment of $\widehat{\gamma}_n$, the operator $S_{\tilde{\gamma}}W_n$ can be written in the form $U_n=S_{\gamma_n}S_{\tilde{\gamma}_n}S_{\gamma'_n}$, where $\tilde{\gamma}_n$ connects the terminating faces of $\gamma_n$ and $\gamma'_n$ and can be chosen so that $\operatorname{dist}(f,\tilde{\gamma}_n)\to\infty$, $\operatorname{dist} (f',\tilde{\gamma}_n)\to\infty$. Hence $U=\operatorname{\textit{w}\text{-}lim}_{n\to\infty}U_n$, as expected. 
\end{proof}

\begin{figure}[t]
    \centering
    \includegraphics[width=5cm]{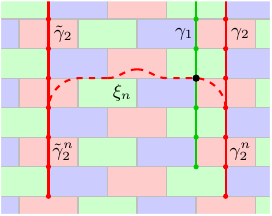}
    \caption{ To compute the braiding $\varepsilon_{\rho_1,\rho_2}$, one applies a charge transporter to make another endomorphism $\tilde{\rho}_2$, which is unitarily equivalent to $\rho_2$, such that it can be represented by $\tilde{\gamma}_2$ localized in a cone which is to the left of the cone in which $\rho_1$ is localized.
    \label{fig:braiding_strings}}
\end{figure}

Let us compute the braiding for all endomorphisms mentioned above. 
For simplicity, denote $\varepsilon_{\mathtt{rx},\mathtt{bz}}:=\varepsilon_{\rho^\mathtt{rx},\rho^\mathtt{bz}}$, $\varepsilon_{\mathtt{rx},\mathtt{f}_1} := \varepsilon_{\rho^{\mathtt{rx}},\rho^{\mathtt{f}_1}}$, etc. 

\begin{proposition}
The braiding $\varepsilon$ satisfies $\varepsilon_{ck,c'k'} = \mathds{1}$ if $c=c'$ or $k=k'$, and $\varepsilon_{ck,c'k'} = \pm \mathds{1}$ if $c\neq c'$ and $k\neq k'$. For the latter case, if $\varepsilon_{ck,c'k'} = - \mathds{1}$, then $\varepsilon_{c'k',ck} = \mathds{1}$, and vice versa.   
\end{proposition}

\begin{proof}
    Let $\rho_i$ be an endomorphism localized in cone $\Lambda_i$, $i=1,2$. To compute $\varepsilon_{\rho_1,\rho_2}$, choose a charge transporter $U$ such that $\tilde{\rho}_2:=U^\dagger \rho_2(\bullet) U$ is localized in a cone $\tilde{\Lambda}_2$ that is to the left of $\Lambda_1$. 
    Then $\varepsilon_{\rho_1,\rho_2} = U^\dagger \bar{\rho}_1(U)$. Suppose $\rho_1$, $\rho_2$, $\tilde{\rho}_2$ are represented by half-infinite strings $\gamma_1$, $\gamma_2$, $\tilde{\gamma}_2$, respectively. 
    Denote by $\gamma_2^n$ (resp.~$\tilde{\gamma}_2^n$) the finite string formed by first $n$ faces of $\gamma_2$ (resp.~$\tilde{\gamma}_2$).  Let $\xi_n$ be a string connecting the terminating faces of $\gamma_2^n$ and $\tilde{\gamma}_2^n$ such that $\lim_n\operatorname{dist}(\partial_0\gamma_2,\xi_n) = \lim_n\operatorname{dist}(\partial_0\tilde{\gamma}_2,\xi_n)=\infty$; see Figure~\ref{fig:braiding_strings}. Then, by Proposition~\ref{prop:unitary}, $U$ can be taken as the weak limit of $\{U_n\}_n$, where $U_n$ is the string operator associated with the finite string $\gamma_2^n\xi_n\tilde{\gamma}_2^n$. 

    If $k=k'$, the corresponding string operators commute at every intersection. If $c=c'$, the string $\xi_n$ may intersect $\gamma_1$, but the number of intersections is even. Hence, also in the case $k\neq k'$, the anticommutation signs occur an even number of times. Therefore, if either $c=c'$ or $k=k'$, we have $\bar{\rho}_1(U) = U$, which leads to $\varepsilon_{\rho_1,\rho_2} = \mathds{1}$.   
    Suppose now both $c \neq c'$ and $k \neq k'$.
    Let us take Figure~\ref{fig:braiding_strings} as an illustration.
    In this case, $\xi_n$ intersects $\gamma_1$ in an odd number of vertices.
    Since $\sigma^k$ and $\sigma^{k'}$ anticommute, it follows that $\bar{\rho}_1(U) = -U$.
    Thus, $\varepsilon_{\rho_1,\rho_2} = -U^\dagger U = -\mathds{1}$.
    Moreover, by the relative locations of $\gamma_1$ and $\gamma_2$, the corresponding $\xi_n'$ for string $\gamma_1$ will intersect $\gamma_2$ in an even number of vertices, which gives $\varepsilon_{\rho_2,\rho_1} = \mathds{1}$. This establishes the claim. 
\end{proof}

\begin{remark}
    To compute $\varepsilon_{ck,\mathtt{f}_i}$ and $\varepsilon_{\mathtt{f}_i,\mathtt{f}_j}$, we can appeal to braiding identities \eqref{eq:braid_hexagon}. 
    Hence we actually have all the braiding information among irreducible superselection sectors. 
\end{remark}

\begin{remark}\label{rem:gauge_braiding}
The braiding constructed here is slightly different from that presented in the appendix. 
The two constructions are related by a gauge transformation acting on the fusion/splitting space. 
Since all anyons in the color code are Abelian, these local gauge transformations reduce to complex phase factors. 
Both choices guarantee that the physical observable (the monodromy) is invariant, namely
$\varepsilon_{b,a}\varepsilon_{a,b} = M_{a,b} = R_{b,a} R_{a,b}$. 
In fact, if we choose the three colored strings in the relative positions shown below
\begin{equation*}
    \begin{tikzpicture}
        \draw[red, line width=1pt] (0,0) -- (0,1.2);  
        \path[fill = red] (0,0) circle (1.3pt);
        \draw[green!50!black, line width=1pt] (1,0) -- (1,1.2);  
        \path[fill = green!50!black] (1,0) circle (1.3pt);
        \draw[blue, line width=1pt] (2,0) -- (2,1.2);  
        \path[fill = blue] (2,0) circle (1.3pt);
    \end{tikzpicture}
\end{equation*}
then the braiding $\varepsilon_{a,b}$ defined in this section will satisfy the monodromy in Table~\ref{tab:colorcode_monodromy}. 
One can also show that the braiding $R_{a,b}$ given in the appendix also satisfies this table. 
\end{remark}

\subsection{Completeness of the superselection sectors}

We have constructed sixteen pairwise inequivalent irreducible superselection sectors. It remains to show that these exhaust all irreducible representations satisfying the superselection criterion. We adapt the strategy of \cite[Section~5]{bols2025classification} to the color code model. The argument simplifies considerably in the present setting because the color code is a commuting stabilizer model and all of its anyon charges are Abelian. Here is the main idea of the proof. Starting from an arbitrary irreducible superselection representation, we first reduce to a pure state with finitely many stabilizer violations, and then move all remaining violations into a fixed local region. These violations determine one of the sixteen anyon charges. Removing the corresponding charge yields the unique frustration-free ground state, which proves that the original representation is equivalent to one of the sixteen sectors constructed above.

For each face $f\in F(\Sigma)$, define
\[
    P_f^K:=\frac{\mathds{1}+K_f}{2}, \quad P_f^J:=\frac{\mathds{1}+J_f}{2}.
\]
These are mutually commuting orthogonal projections. For a bounded region $S\subset\mathbb R^2$, define
\[
    P_S := \prod_{\partial f\subset S}P_f^K P_f^J.
\]
Then $P_S\in\mathcal A(S)$ is the projection onto the simultaneous $+1$ eigenspace of all stabilizers whose supports are contained in $S$. (By abuse of notation, here we also write $\mathcal{A}(S)$ for $\mathcal{A}(S\cap\Sigma)$.) For a possibly unbounded region $S$, let $\{S_m\}_{m\geq1}$ be an increasing sequence of bounded regions with $\bigcup_m S_m=S$. Since the projections $\pi(P_{S_m})$ form a decreasing sequence, they converge in the strong operator topology to a projection
\[
    p_S:=\operatorname*{s-lim}_{m\to\infty}\pi(P_{S_m}).
\]
The resulting projection does not depend on the choice of the exhausting sequence $\{S_m\}$.

In this subsection, let $(\pi_0,\mathcal{H},\Omega)$ be a GNS representation for the ground state $\omega_0$. 

\begin{proposition}\label{prop:finite_stabilizer_violations}
    Let $\pi:\mathcal A\to\mathbf B(\mathcal H_\pi)$ be an irreducible representation satisfying the superselection criterion with respect to $\pi_0$. Then there exist a pure state $\omega$ on $\mathcal A$, whose GNS representation is unitarily equivalent to $\pi$, and a finite set of faces $F_0\subset F(\Sigma)$ such that $\omega(K_f)=\omega(J_f)=1$ for every $f\notin F_0$.
\end{proposition}

\begin{proof}
    We adapt the proof of \cite[Proposition~5.2]{bols2025classification} to the color code. Let $B_n\subset\mathbb R^2$ denote the closed ball of radius $n$. Since the supports of the stabilizers have uniformly bounded diameter, we may choose two cones $\Lambda_1,\Lambda_2$ such that every face $f$ satisfies $\partial f\subset\Lambda_1$ or $\partial f\subset\Lambda_2$. In particular, every $K_f$ and $J_f$ is supported in at least one of the two cones.

    Since $\pi$ satisfies the superselection criterion, for $i=1,2$ there is a unitary $U_i:\mathcal H\to\mathcal H_\pi$ such that $\pi(A)=U_i\pi_0(A)U_i^\dagger$, $A\in\mathcal A(\Lambda_i)$. Put $\xi_i:=U_i\Omega$ and let
    \[
        \omega_i(A):=\langle\xi_i,\pi(A)\xi_i\rangle.
    \]
    Then, for every $A\in\mathcal A(\Lambda_i)$,
    \[
        \omega_i(A) = \langle\Omega,U_i^\dagger\pi(A)U_i\Omega\rangle = \langle\Omega,\pi_0(A)\Omega\rangle
        = \omega_0(A).
    \]

    For $m,n\geq1$, set $\Lambda_i^{>n}:=\Lambda_i\setminus B_n$, $\Lambda_i^{n,n+m}:=\Lambda_i^{>n}\setminus\Lambda_i^{>n+m}$. The latter is bounded. Since
    $P_{\Lambda_i^{n,n+m}}\in\mathcal A(\Lambda_i)$ and $P_f^K\Omega=P_f^J\Omega=\Omega$ for every face $f$, we have
    \[
        \langle\xi_i,\pi(P_{\Lambda_i^{n,n+m}})\xi_i\rangle = \omega_i(P_{\Lambda_i^{n,n+m}})=\omega_0(P_{\Lambda_i^{n,n+m}})=1.
    \]
    Letting $m\to\infty$ gives $\langle\xi_i,p_{\Lambda_i^{>n}}\xi_i\rangle=1$, and therefore
    \begin{equation}\label{eq:cone_projector_fix}
        p_{\Lambda_i^{>n}}\xi_i=\xi_i
        \quad  (i=1,2).
    \end{equation}

    Since $\pi$ is irreducible, every nonzero vector is cyclic. Consequently, $\omega_1$ and $\omega_2$ are pure states whose GNS representations are both unitarily equivalent to $\pi$.
    By Proposition~\ref{prop:equiv_GNS}, for every $\varepsilon>0$ there exists $n(\varepsilon)$ such that
    \begin{equation}\label{eq:omega12_close}
        |\omega_1(A)-\omega_2(A)| < \varepsilon\|A\|
    \end{equation}
    for every local operator $A$ supported in $B_{n(\varepsilon)}^c$. Fix $0<\varepsilon<1$ and $n\geq n(\varepsilon)$. Applying \eqref{eq:omega12_close} to $P_{\Lambda_2^{n,n+m}}$, whose norm equals one, gives
    \[
        \left|\langle\xi_1,\pi(P_{\Lambda_2^{n,n+m}})\xi_1\rangle-\langle\xi_2,\pi(P_{\Lambda_2^{n,n+m}})\xi_2\rangle\right|<\varepsilon.
    \]
    The second expectation value is equal to one, and hence
    \[
        \langle\xi_1,\pi(P_{\Lambda_2^{n,n+m}})\xi_1\rangle>1-\varepsilon.
    \]
    Taking the strong limit as $m\to\infty$, we obtain 
    \begin{equation}\label{eq:second_cone_positive}
        \langle\xi_1,p_{\Lambda_2^{>n}}\xi_1\rangle
        \geq 1-\varepsilon>0.
    \end{equation}

    Let $r_0>0$ be larger than the diameter of the support of every face stabilizer. We claim that
    \begin{equation}\label{eq:projector_inclusion}
        p_{\Lambda_2^{>n}}p_{\Lambda_1^{>n}}
        \leq
        p_{B_{n+r_0}^c}.
    \end{equation}
    Indeed, let $f$ be a face such that $\partial f\subset B_{n+r_0}^c$. By the choice of $\Lambda_1,\Lambda_2$, the support $\partial f$ is contained in at least one $\Lambda_i$. Moreover, $\partial f\subset B_n^c$, and hence $\partial f\subset\Lambda_i^{>n}$ for some $i\in\{1,2\}$. Therefore $p_{\Lambda_i^{>n}}\pi(P_f^K)=p_{\Lambda_i^{>n}}$, $p_{\Lambda_i^{>n}}\pi(P_f^J)=p_{\Lambda_i^{>n}}$. All the stabilizer projections commute, so the strong limit projections above commute as well. Thus $p_{\Lambda_2^{>n}}p_{\Lambda_1^{>n}}\pi(P_f^K) = p_{\Lambda_2^{>n}}p_{\Lambda_1^{>n}}$, and similarly $p_{\Lambda_2^{>n}}p_{\Lambda_1^{>n}}\pi(P_f^J) = p_{\Lambda_2^{>n}}p_{\Lambda_1^{>n}}$. Since this holds for every face whose support is contained in $B_{n+r_0}^c$, the range of $p_{\Lambda_2^{>n}}p_{\Lambda_1^{>n}}$ is contained in the range of $p_{B_{n+r_0}^c}$, which proves
    \eqref{eq:projector_inclusion}.

    Using \eqref{eq:cone_projector_fix},
    \eqref{eq:second_cone_positive}, and
    \eqref{eq:projector_inclusion}, we obtain
    \begin{align*}
        \langle\xi_1,p_{B_{n+r_0}^c}\xi_1\rangle \geq \langle\xi_1, p_{\Lambda_2^{>n}}p_{\Lambda_1^{>n}}\xi_1\rangle = \langle\xi_1,p_{\Lambda_2^{>n}}\xi_1\rangle \geq 1-\varepsilon>0.
    \end{align*}
    Hence $p_{B_{n+r_0}^c}\xi_1\neq0$. Define
    \[
        \xi := \frac{p_{B_{n+r_0}^c}\xi_1}{\|p_{B_{n+r_0}^c}\xi_1\|},\quad \omega_\xi(A):=\langle \xi,\pi(A)\xi\rangle \;\; \text{for}\;\; A\in \mathcal{A}. 
    \]
    We now verify explicitly that all stabilizer constraints sufficiently far from the origin are satisfied. If
    $\partial f\subset B_{n+r_0}^c$, then by the definition of $p_{B_{n+r_0}^c}$,
    \[
        p_{B_{n+r_0}^c}\pi(P_f^K) = p_{B_{n+r_0}^c} = p_{B_{n+r_0}^c}\pi(P_f^J).
    \]
    Consequently,
    \[
        \omega_\xi(P_f^K) = \frac{\langle p_{B_{n+r_0}^c}\xi_1,\pi(P_f^K)p_{B_{n+r_0}^c}\xi_1\rangle}{\|p_{B_{n+r_0}^c}\xi_1\|^2} = \frac{\langle p_{B_{n+r_0}^c}\xi_1,p_{B_{n+r_0}^c}\xi_1\rangle}{\|p_{B_{n+r_0}^c}\xi_1\|^2}=1.
    \]
    Similarly, $\omega_\xi(P_f^J)=1$. Since $P_f^K={(\mathds{1}+K_f)}/{2}$, $P_f^J={(\mathds{1}+J_f)}/{2}$, it follows that $\omega_\xi(K_f)=\omega_\xi(J_f)=1$. 
    Finally, let
    \[
        F_0 := \{f\in F(\Sigma):\partial f\not\subset B_{n+r_0}^c\}.
    \]
    The lattice is locally finite, so $F_0$ is finite. Thus
    \[
        \omega_\xi(K_f)=\omega_\xi(J_f)=1
        \quad 
        \text{for all }f\notin F_0.
    \]
    Since $\pi$ is irreducible and $\xi\neq 0$, the vector $\xi$ is cyclic. Hence $\omega_\xi$ is pure, and its GNS representation is unitarily equivalent to $\pi$. This completes the proof.
\end{proof}

We next show that the finitely many remaining stabilizer violations can be moved into a fixed finite region. Fix a vertex $v_*$ and denote by $f_{\mathtt r}$, $f_{\mathtt g}$, and $f_{\mathtt b}$ the three faces incident to $v_*$, colored $\mathtt r$, $\mathtt g$, and $\mathtt b$, respectively.

\begin{lemma}\label{lem:sweep_violations}
    Let $\omega$ be a pure state on $\mathcal A$ such that $\omega(K_f)=\omega(J_f)=1$, for all but finitely many faces $f$. Then there exists a pure state $\psi$ whose GNS representation is unitarily equivalent to that of $\omega$ and such that $\psi(K_f)=\psi(J_f)=1$ for every $f\notin \{f_{\mathtt r},f_{\mathtt g},f_{\mathtt b}\}$. 
\end{lemma}

\begin{proof}
    Let $(\pi,\mathcal H_\pi,\xi)$ be the GNS triple of $\omega$. Then $\xi$ satisfies all but finitely many stabilizer constraints. We first remove the remaining $K$-violations away from $f_{\mathtt r},f_{\mathtt g},f_{\mathtt b}$. Let $f\neq f_{c}$ be a face of color $c\in\{\mathtt r,\mathtt g,\mathtt b\}$, and choose a finite $c$-colored string $\gamma$ joining $f$ to $f_{c}$. Let $P=P_f^K$, $S=S_\gamma^{cz}$. The string operator $S$ anticommutes with $K_f$ and $K_{f_{c}}$, and commutes with every other $K$-stabilizer and with every $J$-stabilizer. Hence $P S=S(\mathds{1}-P)$. One has 
    \begin{align*}
        \|\pi(P)\xi\|^2 + \|\pi(PS)\xi\|^2 & = \|\pi(P)\xi\|^2
        + \|\pi(S(\mathds{1}-P))\xi\|^2\\
        &= \|\pi(P)\xi\|^2 + \|\pi(\mathds{1}-P)\xi\|^2 =1.
    \end{align*}
    Therefore at least one of $\pi(P)\xi$, $\pi(PS)\xi$ is nonzero. Denote by $u$ a nonzero choice among them. Since $P$ is a projection, it follows that $\pi(P)u=u$. After normalization, this gives a unit vector $\xi'$ satisfying $\pi(K_f)\xi'=\xi'$. Since both $P$ and $S$ commute with all stabilizers except possibly $K_f$ and $K_{f_{c}}$, all stabilizer constraints already imposed away from $f_{c}$ remain unchanged. Applying this procedure finitely many times removes all $K$-violations except possibly those at $f_{\mathtt r},f_{\mathtt g},f_{\mathtt b}$. 
    
    We next remove the remaining $J$-violations in the same way. For a $c$-colored face $f\neq f_{c}$, choose a finite $c$-colored string $\gamma$ from $f$ to $f_{c}$ and use $P=P_f^J$, $S=S_\gamma^{cx}$. Now $S$ anticommutes with $J_f$ and $J_{f_{c}}$, while commuting with every other $J$-stabilizer and with every $K$-stabilizer. Repeating the preceding argument removes all $J$-violations except possibly at the three faces $f_{\mathtt r},f_{\mathtt g},f_{\mathtt b}$, without changing any of the $K$-constraints already imposed. Since there were only finitely many violations to begin with, this process terminates after finitely many steps. Let $\eta$ be the resulting unit vector and define $\psi(A):=\langle\eta,\pi(A)\eta\rangle$. Then $\psi(K_f)=\psi(J_f)=1$ for every $f\notin\{f_{\mathtt r},f_{\mathtt g},f_{\mathtt b}\}$. Finally, since $\pi$ is irreducible and $\eta\neq0$, the vector $\eta$ is cyclic. Hence $(\pi,\mathcal H_\pi,\eta)$ is a GNS representation of $\psi$. In particular, $\psi$ is pure and its GNS representation is unitarily equivalent to that of $\omega$.
\end{proof}

\begin{lemma}\label{lem:definite_local_syndrome}
    Let $\psi$ be a pure state on $\mathcal A$ satisfying $\psi(K_f)=\psi(J_f)=1$ for every $f\notin\{f_{\mathtt r},f_{\mathtt g},f_{\mathtt b}\}$. Then there exists a pure state $\varphi$ whose GNS representation is unitarily equivalent to that of $\psi$ such that $\varphi(K_f)=\varphi(J_f)=1$ for every $f\notin\{f_{\mathtt r},f_{\mathtt g}\}$, and there exist $\varepsilon_{\mathtt r}^K,\, \varepsilon_{\mathtt r}^J,\,\varepsilon_{\mathtt g}^K,\, \varepsilon_{\mathtt g}^J \in\{\pm1\}$ such that $\varphi(K_{f_{c}})=\varepsilon_{c}^K$, $\varphi(J_{f_{c}})=\varepsilon_{c}^J$ for $c\in\{\mathtt r,\mathtt g\}$. 
\end{lemma}

\begin{proof}
    Let $(\pi,\mathcal H,\eta)$ be the GNS triple of $\psi$. We first remove the possible $K$- and $J$-violations at $f_{\mathtt b}$. Indeed, since $v_*$ is incident to precisely the three faces $f_{\mathtt r},f_{\mathtt g},f_{\mathtt b}$, the operator $S=\sigma_{v_*}^z$ anticommutes with the three corresponding $K$-stabilizers and commutes with every other $K$-stabilizer and every $J$-stabilizer. Applying the same argument as in Lemma~\ref{lem:sweep_violations} with $P=P_{f_{\mathtt b}}^K$, $S=\sigma^z_{v_*}$ and $\eta$, we obtain a unit vector $\eta_1$ such that $\pi(K_{f_{\mathtt b}})\eta_1=\eta_1$,  while all stabilizer constraints away from $\{f_{\mathtt r},f_{\mathtt g},f_{\mathtt b}\}$ are preserved. Starting from $\eta_1$, we next remove the possible $J$-violation at $f_{\mathtt b}$. With $P=P_{f_{\mathtt b}}^J$ and $S=\sigma_{v_*}^x$, the same argument yields a unit vector $\eta_2$ satisfying $\pi(K_{f_{\mathtt b}})\eta_2 = \pi(J_{f_{\mathtt b}})\eta_2 = \eta_2$,  while all stabilizer constraints away from $f_{\mathtt r}$ and $f_{\mathtt g}$ remain unchanged.

    It remains to choose definite eigenvalues for the four remaining commuting stabilizers. For $\boldsymbol\varepsilon = (\varepsilon_{\mathtt r}^K,\varepsilon_{\mathtt r}^J,\varepsilon_{\mathtt g}^K,\varepsilon_{\mathtt g}^J)\in\{\pm1\}^4$, 
    define
    \[
        Q_{\boldsymbol\varepsilon} := \prod_{c\in\{\mathtt r,\mathtt g\}}\frac{\mathds{1}+\varepsilon_{c}^K K_{f_{c}}}{2}\frac{\mathds{1}+\varepsilon_{c}^J J_{f_{c}}}{2}.
    \]
    Since the four stabilizers commute, the $Q_{\boldsymbol\varepsilon}$ are mutually orthogonal projections satisfying $\sum_{\boldsymbol\varepsilon\in\{\pm1\}^4} Q_{\boldsymbol\varepsilon} = \mathds{1}$. Consequently, for at least one $\boldsymbol\varepsilon$, $\pi(Q_{\boldsymbol\varepsilon})\eta_2\neq0$. Fix such a choice and set
    \[
        \zeta := \frac{\pi(Q_{\boldsymbol\varepsilon})\eta_2}{\|\pi(Q_{\boldsymbol\varepsilon})\eta_2\|}.
    \]
    Then, for $c\in\{\mathtt r,\mathtt g\}$, $\pi(K_{f_{c}})\zeta = \varepsilon_{c}^K\zeta$, $\pi(J_{f_{c}})\zeta = \varepsilon_{c}^J\zeta$. Moreover, $Q_{\boldsymbol\varepsilon}$ commutes with every stabilizer, so all the previously imposed $+1$ constraints are preserved. Define $\varphi(A):=\langle\zeta,\pi(A)\zeta\rangle$. Since $\pi$ is irreducible and $\zeta\neq0$, the vector $\zeta$ is cyclic. Thus $\varphi$ is pure and its GNS representation is unitarily equivalent to that of $\psi$. The stated stabilizer eigenvalues follow immediately.
\end{proof}

We now identify the possible states obtained in Lemma~\ref{lem:definite_local_syndrome}. Fix half-infinite strings $\gamma_{\mathtt r}$ and $\gamma_{\mathtt g}$ starting at $f_{\mathtt r}$ and $f_{\mathtt g}$, respectively. For $\boldsymbol{\varepsilon}= (\varepsilon_{\mathtt r}^K,\varepsilon_{\mathtt r}^J,\varepsilon_{\mathtt g}^K,\varepsilon_{\mathtt g}^J)\in\{\pm1\}^4$, put
\[
    m_c^K:=\frac{1-\varepsilon_c^K}{2}, \quad m_c^J:=\frac{1-\varepsilon_c^J}{2}, \quad c\in\{\mathtt r,\mathtt g\},
\]
and define
\begin{equation*}
    \rho_{\boldsymbol{\varepsilon}} := \left(\rho_{\gamma_{\mathtt r}}^{\mathtt r\mathtt z}\right)^{m_{\mathtt r}^K} \comp \left(\rho_{\gamma_{\mathtt r}}^{\mathtt r\mathtt x}\right)^{m_{\mathtt r}^J} \comp \left(\rho_{\gamma_{\mathtt g}}^{\mathtt g\mathtt z}\right)^{m_{\mathtt g}^K} \comp \left(\rho_{\gamma_{\mathtt g}}^{\mathtt g\mathtt x}\right)^{m_{\mathtt g}^J},\quad \omega_{\boldsymbol{\varepsilon}} :=\omega_0\comp\rho_{\boldsymbol{\varepsilon}}.
\end{equation*}
Here an exponent equal to zero means that the corresponding factor is omitted. By the fusion rules obtained above, the sixteen $\rho_{\boldsymbol{\varepsilon}}$ represent precisely the vacuum, the nine bosonic sectors $\rho^{ck}$, and the six fermionic sectors
$\rho^{\mathtt f_i}$.

\begin{lemma}\label{lem:unique_anyon_syndrome}
    Let $\boldsymbol{\varepsilon}= (\varepsilon_{\mathtt r}^K,\varepsilon_{\mathtt r}^J,\varepsilon_{\mathtt g}^K,\varepsilon_{\mathtt g}^J)\in\{\pm1\}^4$ be fixed. Let $\omega$ be a state satisfying
    \[
        \omega(K_f)=\omega(J_f)=1, \quad f\notin\{f_{\mathtt r},f_{\mathtt g}\},
    \]
    and
    \[
        \omega(K_{f_c})=\varepsilon_c^K, \quad \omega(J_{f_c})=\varepsilon_c^J, \quad c\in\{\mathtt r,\mathtt g\}.
    \]
    Then $\omega=\omega_{\boldsymbol{\varepsilon}}$. 
\end{lemma}

\begin{proof}
    By construction, $\rho_{\boldsymbol{\varepsilon}}$ fixes every
    stabilizer away from $f_{\mathtt r}$ and $f_{\mathtt g}$, while
    \[
        \rho_{\boldsymbol{\varepsilon}}(K_{f_c}) = \varepsilon_c^K K_{f_c},
        \quad \rho_{\boldsymbol{\varepsilon}}(J_{f_c}) = \varepsilon_c^J J_{f_c}, \quad c\in\{\mathtt r,\mathtt g\}.
    \]
    Consequently, the state $\widetilde{\omega} := \omega\comp\rho_{\boldsymbol{\varepsilon}}^{-1}$ satisfies $\widetilde{\omega}(K_f) = \widetilde{\omega}(J_f) = 1$ for every face $f$. By Theorem~\ref{prop:GS}, $\widetilde{\omega}=\omega_0$. Therefore $\omega=\omega_0\comp\rho_\varepsilon=\omega_\varepsilon$.
\end{proof}

\begin{remark}
    In the proof of Lemma~\ref{lem:definite_local_syndrome}, there may a priori be more than one $\boldsymbol{\varepsilon}\in\{\pm1\}^4$ such that $\pi(Q_{\boldsymbol{\varepsilon}})\eta_2\neq0$. Choosing different nonzero components would then produce different vector states. However, since $\pi$ is irreducible, the GNS representation of each such vector state is unitarily equivalent to $\pi$. After identifying these states with the corresponding canonical anyon states, two distinct choices $\boldsymbol{\varepsilon}\neq\boldsymbol{\varepsilon}'$ would imply that the two corresponding canonical sectors are unitarily equivalent. This contradicts the pairwise inequivalence established in Theorems~\ref{thm:sectors} and \ref{thm:sectors_fermion}. Hence, in fact, at most one $\boldsymbol{\varepsilon}$ can occur.
\end{remark}

We can now prove completeness.

\begin{theorem}\label{thm:complete_sectors}
    Every irreducible representation satisfying the superselection
    criterion is unitarily equivalent to exactly one of $\pi_0$, $\pi^{ck}$, $\pi^{\mathtt f_i}$ where $c\in\{\mathtt r,\mathtt g,\mathtt b\}$, $k\in\{\mathtt x,\mathtt y,\mathtt z\}$, and $1\leq i\leq6$. 
\end{theorem}

\begin{proof}
    Let $\pi$ be an irreducible representation satisfying the superselection criterion. By Proposition~\ref{prop:finite_stabilizer_violations}, there exists a pure state $\omega$ whose GNS representation is unitarily equivalent to $\pi$ and such that $\omega(K_f)=\omega(J_f)=1$ for all but finitely many faces $f$. By Lemma~\ref{lem:sweep_violations}, there exists a pure state $\psi$ whose GNS representation is unitarily equivalent to that of $\omega$ and such that $\psi(K_f)=\psi(J_f)=1$ for every $f\notin\{f_{\mathtt r},f_{\mathtt g},f_{\mathtt b}\}$. Applying Lemma~\ref{lem:definite_local_syndrome}, we obtain a pure state $\varphi$ whose GNS representation is unitarily equivalent to that of $\psi$ such that $\varphi(K_f)=\varphi(J_f)=1$ for every $f\notin\{f_{\mathtt r},f_{\mathtt g}\}$, and $\varphi(K_{f_{\mathtt c}}) = \varepsilon_{\mathtt c}^K$, $\varphi(J_{f_{c}}) = \varepsilon_{c}^J$, $c\in\{\mathtt r,\mathtt g\}$, for some $\boldsymbol{\varepsilon}\in\{\pm1\}^4$. By Lemma~\ref{lem:unique_anyon_syndrome}, this implies $\varphi=\omega_{\boldsymbol{\varepsilon}}$. Hence the GNS representation of $\varphi$ is one of $\pi_0$, $\pi^{ck}$, $\pi^{\mathtt f_i}$ where $c\in\{\mathtt r,\mathtt g,\mathtt b\}$, $k\in\{\mathtt x,\mathtt y,\mathtt z\}$, and $1\leq i\leq6$. Since the GNS representations of $\omega$, $\psi$, and $\varphi$ are successively unitarily equivalent, the representation $\pi$ is unitarily equivalent to one of the sixteen representations listed above. Their pairwise inequivalence was established in Theorem~\ref{thm:sectors} and Theorem~\ref{thm:sectors_fermion}. Therefore this representation is unique.
\end{proof}

\begin{corollary}\label{coro:irreducible_objects_complete}
    Up to unitary equivalence, the irreducible objects of
    $\mathcal D_{\rm loc,tr}$ are precisely $\iota$, $\rho^{ck}$, $\rho^{\mathtt f_i}$, where $\iota$ is the identity endomorphism of $\mathcal A^{au}$, $\rho^{ck}$ denotes a representative of the sector $\pi^{ck}$ localized in the fixed cone $\Lambda_0$, with $c\in\{\mathtt r,\mathtt g,\mathtt b\}$, $k\in\{\mathtt x,\mathtt y,\mathtt z\}$, and $\rho^{\mathtt f_i}$ denotes a representative of the sector
    $\pi^{\mathtt f_i}$ localized in $\Lambda_0$, with
    $1\leq i\leq 6$.
\end{corollary}

\subsection{Equivalence of categories}

We now identify a full subcategory of $\mathcal D_{\rm loc,tr}$ with the representation category of the Drinfeld quantum double $D(\mathbb Z_2\times\mathbb Z_2)$. The whole category $\mathcal D_{\rm loc,tr}$ is larger than $\Rep(D(\mathbb Z_2\times\mathbb Z_2))$, since it admits countable direct sums, whereas $\Rep(D(\mathbb Z_2\times\mathbb Z_2))$ consists of finite-dimensional representations. To obtain a category that can be compared with $\Rep(D(\mathbb Z_2\times\mathbb Z_2))$, we therefore restrict to the full subcategory of dualizable objects in $\mathcal D_{\rm loc,tr}$. We will prove that this subcategory is braided tensor equivalent to $\Rep(D(\mathbb Z_2\times\mathbb Z_2))$. See the appendix for a review of the latter category.

\begin{definition}\label{def:dualizable_subcategory}
    Let $\mathcal D_{\rm loc,tr}^{\rm dual}$ denote the full subcategory of $\mathcal D_{\rm loc,tr}$ consisting of all dualizable objects. Namely, an object $\rho\in\mathcal D_{\rm loc,tr}$ belongs to $\mathcal D_{\rm loc,tr}^{\rm dual}$ if there exists an object $\bar\rho\in\mathcal D_{\rm loc,tr}$ and morphisms
    \[
        R:\iota\to \bar\rho\otimes\rho,
        \quad
        \bar R:\iota\to \rho\otimes\bar\rho
    \]
    satisfying the conjugate equations
    \[
        (\bar R^\dagger\otimes\mathds{1}_\rho) (\mathds{1}_\rho\otimes R) = \mathds{1}_\rho, \quad (R^\dagger\otimes\mathds{1}_{\bar\rho}) (\mathds{1}_{\bar\rho}\otimes\bar R) = \mathds{1}_{\bar\rho}.
    \]
    For $\rho,\sigma\in\mathcal D_{\rm loc,tr}^{\rm dual}$, the morphism space is inherited from $\mathcal D_{\rm loc,tr}$:
    \[
        \Hom_{\mathcal D_{\rm loc,tr}^{\rm dual}}(\rho,\sigma)
        :=
        \Hom_{\mathcal D_{\rm loc,tr}}(\rho,\sigma).
    \]
\end{definition}

By Proposition~\ref{prop:fusion_rules} and the fusion rules in Table~\ref{tab:colorcode_fusion}, each of the sixteen irreducible sectors constructed above is invertible, and in fact is its own tensor inverse. Hence all sixteen sectors are dualizable. Moreover, Theorem~\ref{thm:complete_sectors} shows that they exhaust all irreducible objects of $\mathcal D_{\rm loc,tr}$.

Finite direct sums and subobjects are available in $\mathcal D_{\rm loc,tr}$ by the standard construction; see \cite{Ogata2022braided}. Since $\omega_0$ is pure, the tensor unit $\iota$ is irreducible. By \cite[Lemma~3.2]{Longo1997dimension}, every dualizable object has a finite-dimensional endomorphism algebra and decomposes as a finite direct sum of irreducible objects. Theorem~\ref{thm:complete_sectors} therefore shows that every object of $\mathcal D_{\rm loc,tr}^{\rm dual}$ is a finite direct sum of the sixteen irreducible sectors constructed above. Consequently,
\[
    \operatorname{Irr}\bigl(\mathcal D_{\rm loc,tr}^{\rm dual}\bigr) = \bigl\{\iota,\rho^{ck},\rho^{\mathtt f_i}\bigr\},
\]
where $c\in\{\mathtt r,\mathtt g,\mathtt b\}$, $k\in\{\mathtt x,\mathtt y,\mathtt z\}$, and $1\leq i\leq6$.

The following is the main result of this paper.

\begin{theorem}\label{thm:category_equivalence}
    The category $\mathcal D_{\rm loc,tr}^{\rm dual}$ is braided tensor equivalent to the representation category $\Rep(D(\mathbb{Z}_2\times \mathbb{Z}_2))$.
\end{theorem}

\begin{proof}
    Denote $G := \mathbb{Z}_2\times \mathbb{Z}_2$. In the appendix, we have included a review of the representation category of the quantum double $D(G)$. Irreducible representations of $D(G)$ are labeled by pairs $(g,\chi)$, $g\in G, \chi\in\widehat{G}$, where $\widehat{G}$ denotes the character group of $G$. 
    Write $G = \{0,a,b,c\}$ where $a+b=c$. Similarly, write $\widehat{G} = \{1,\alpha,\beta,\gamma\}$ where $\alpha\beta=\gamma$. 
    Using these labels, we obtain a full list of $\Irr(\Rep(D(G)))$. Let us denote 
    \begin{align*}
    	& \Pi_\one = (0,1),\; \Pi_{\mathtt{rx}} = (0,\alpha),\; \Pi_{\mathtt{bx}} = (0,\beta),\; \Pi_{\mathtt{gx}} = (0,\gamma), \\
    	& \Pi_{\mathtt{bz}} = (a,1),\; \Pi_{\mathtt{f}_1} = (a,\alpha),\; \Pi_{\mathtt{by}} = (a,\beta),\; \Pi_{\mathtt{f}_4} = (a,\gamma), \\
    	& \Pi_{\mathtt{rz}} = (b,1),\; \Pi_{\mathtt{ry}} = (b,\alpha),\; \Pi_{\mathtt{f}_2} = (b,\beta),\; \Pi_{\mathtt{f}_3} = (b,\gamma), \\
    	& \Pi_{\mathtt{gz}} = (c,1),\; \Pi_{\mathtt{f}_6} = (c,\alpha),\; \Pi_{\mathtt{f}_5} = (c,\beta),\; \Pi_{\mathtt{gy}} = (c,\gamma). 
    \end{align*}
    With these new labels, one can verify that their fusion rules are exactly as shown in Table~\ref{tab:colorcode_fusion}. 
    For example, $\Pi_{\mathtt{rx}}\otimes\Pi_{\mathtt{ry}} = (0,\alpha)\otimes(b,\alpha) = (b,1) = \Pi_{\mathtt{rz}}$, $\Pi_{\mathtt{f}_1}\otimes \Pi_{\mathtt{f}_2} = (a,\alpha)\otimes (b,\beta)  = (c,\gamma) = \Pi_{\mathtt{gy}}$, $\Pi_{\mathtt{ry}}\otimes \Pi_{\mathtt{bz}} = (b,\alpha)\otimes (a,1) = (c,\alpha) = \Pi_{\mathtt{f}_6}$, etc. 
    Let $R_{i,j}$ denote the braiding $\Pi_i\otimes \Pi_j\to \Pi_j\otimes \Pi_i$. Up to a gauge transformation on the fusion/splitting spaces (cf.~Remark~\ref{rem:gauge_braiding}), one has the following nontrivial braiding among the nine bosons $R_{\mathtt{bz},\mathtt{rx}}$, $R_{\mathtt{by},\mathtt{rx}}$, $R_{\mathtt{gz},\mathtt{rx}}$, $ R_{\mathtt{gy},\mathtt{rx}}$, $R_{\mathtt{ry},\mathtt{bx}}$, $R_{\mathtt{bz},\mathtt{ry}}$, $R_{\mathtt{ry},\mathtt{gx}}$, $R_{\mathtt{gz},\mathtt{ry}}$, $R_{\mathtt{rz},\mathtt{bx}}$, $R_{\mathtt{rz},\mathtt{by}}$, $R_{\mathtt{rz},\mathtt{gx}}$, $R_{\mathtt{rz},\mathtt{gy}}$, all of which equal $-1$. 
    The braiding for which at least one of the indices equals $\mathtt{f}_i$ can be calculated using the braiding identities \eqref{eq:braid_hexagon}.

    On the side of $\mathcal{D}_{\rm loc, tr}^{\rm dual}$, choose three half-infinite colored strings with relative positions as shown in Remark~\ref{rem:gauge_braiding}. 
    In this case,  the braiding $\varepsilon_{\mathtt{bz},\mathtt{rx}}$, $\varepsilon_{\mathtt{by},\mathtt{rx}}$, $\varepsilon_{\mathtt{gz},\mathtt{rx}}$, $\varepsilon_{\mathtt{gy},\mathtt{rx}}$, $\varepsilon_{\mathtt{ry},\mathtt{bx}}$, $\varepsilon_{\mathtt{bz},\mathtt{ry}}$, $\varepsilon_{\mathtt{ry},\mathtt{gx}}$, $\varepsilon_{\mathtt{gz},\mathtt{ry}}$, $\varepsilon_{\mathtt{rz},\mathtt{bx}}$, $\varepsilon_{\mathtt{rz},\mathtt{by}}$, $\varepsilon_{\mathtt{rz},\mathtt{gx}}$, 
    $\varepsilon_{\mathtt{rz},\mathtt{gy}}$ for corresponding endomorphisms are all equal to $-\mathds{1}$. 
    Here, for example, $\varepsilon_{\mathtt{bz},\mathtt{rx}}$ denotes the intertwiner $\rho^{\mathtt{bz}}\otimes \rho^{\mathtt{rx}} \to \rho^{\mathtt{rx}}\otimes \rho^{\mathtt{bz}}$.
    Let us define a functor 
    \[
        \Phi:\Rep(D(G)) \to \mathcal{D}_{\rm loc, tr}^{\rm dual}. 
    \]
    On the level of irreducible objects, define $\Phi(\Pi_k) = \rho^k$, where $k$ runs over all 16 labels, and $\rho^k$ corresponds to half-strings of type $k$. 
    Since each $\Pi_k$ is of dimension $1$, $\Hom(\Pi_k,\Pi_j) = \delta_{k,j}\mathbb{C}$. 
    Then on the level of morphisms, define $\Phi(\lambda) = \lambda\mathds{1}_k$ for $\lambda\in \Hom(\Pi_k,\Pi_k)$. 
    Since $\Rep(D(G))$ is semisimple, every object is a finite direct sum of the sixteen irreducible representations above. We therefore extend $\Phi$ additively. 
    For a representation $X = \oplus_k m_k\Pi_k$, define $\Phi(X) = \oplus_k m_k\rho^k$. Here all multiplicities $m_k$ are finite. For two representations $X=\oplus_k m_k\Pi_k$ and $Y= \oplus_k n_k\Pi_k$, we have $\Hom(X,Y) \cong \oplus_k M_{n_k\times m_k}(\mathbb C)$. 
    For $T=(T_k)_k\in\Hom(X,Y)$, 
    define $\Phi(T)=\oplus_k\bigl(T_k\otimes\mathds{1}_{\rho^k}\bigr)$.  
    This defines a functor of Abelian categories. 
    For representations $X,Y\in \Rep(D(G))$, define $J_{X,Y}:\Phi(X\otimes Y) \to \Phi(X)\otimes \Phi(Y)$ by $J_{X,Y}=\mathds{1}$; and define $\varphi:\iota \to \Phi(\Pi_{\one})$ by $\varphi = \mathds{1}$. 
    These data make $\Phi$ a monoidal functor, and it is direct to check that $\Phi$ preserves the relevant tensor category structures. 
    Moreover, it is straightforward but tedious to verify that $\Phi$ preserves the braiding; for example, the diagram
    \begin{equation*}
        \begin{tikzcd}
        	{\Phi(\Pi_{\mathtt{ry}}\otimes\Pi_{\mathtt{bx}})} & {\Phi(\Pi_{\mathtt{bx}}\otimes\Pi_{\mathtt{ry}})} \\
        	{\Phi(\Pi_{\mathtt{ry}})\otimes \Phi(\Pi_{\mathtt{bx}})} & {\Phi(\Pi_{\mathtt{bx}})\otimes\Phi(\Pi_{\mathtt{ry}})}
        	\arrow["{\Phi(R_{\mathtt{ry},\mathtt{bx}})}", from=1-1, to=1-2]
        	\arrow["{J_{\mathtt{ry},\mathtt{bx}}}"', from=1-1, to=2-1]
        	\arrow["{J_{\mathtt{bx},\mathtt{ry}}}", from=1-2, to=2-2]
        	\arrow["{\varepsilon_{\mathtt{ry},\mathtt{bx}}}", from=2-1, to=2-2]
        \end{tikzcd}
    \end{equation*}
    commutes, since $J_{\mathtt{ry},\mathtt{bx}} = J_{\mathtt{bx},\mathtt{ry}} = \mathds{1}$, $\varepsilon_{\mathtt{ry},\mathtt{bx}} = -\mathds{1}$, and $\Phi(R_{\mathtt{ry},\mathtt{bx}}) = R_{\mathtt{ry},\mathtt{bx}}\mathds{1} = -\mathds{1}$. 
    Thus $\Phi$ is a braided tensor functor $\Phi:\Rep(D(G)) \to \mathcal D_{\rm loc,tr}^{\rm dual}$. It is fully faithful, since the sixteen objects $\rho^k$ are pairwise inequivalent irreducible objects and $\Hom(\rho^k,\rho^\ell) = \delta_{k,\ell}\mathbb C$. It remains to verify essential surjectivity. By Theorem~\ref{thm:complete_sectors} (or equivalently Corollary~\ref{coro:irreducible_objects_complete}), the sixteen objects $\rho^k$ form the complete list of irreducible objects of $\mathcal D_{\rm loc,tr}^{\rm dual}$. As observed above, every object of $\mathcal D_{\rm loc,tr}^{\rm dual}$ is a finite direct sum of these irreducible objects. Hence every object of $\mathcal D_{\rm loc,tr}^{\rm dual}$ is isomorphic to an object in the image of $\Phi$. Therefore $\Phi$ is a braided tensor equivalence
    \[
        \Rep(D(G)) \simeq_{\rm br,\otimes} \mathcal D_{\rm loc,tr}^{\rm dual}.
    \]
    This completes the proof. 
\end{proof}

\section{Haag duality} \label{sec:haag_duality}

Haag duality is a fundamental locality property in algebraic quantum field theory and plays an important role in the Doplicher--Haag--Roberts analysis of superselection sectors; see the original works \cite{doplicher1971local,doplicher1974local}. In two-dimensional quantum lattice systems, cone Haag duality has been established for the toric code \cite{naaijkens2012haag}, for Kitaev quantum double models with finite Abelian gauge groups \cite{fiedler2015haag}, and, more recently, for broad classes of two-dimensional quantum spin systems \cite{ogata2025haag}. See also \cite{naaijkens2026local} for a recent approach based on local topological order.

For a cone $\Lambda$, let
\[
    \mathcal R_\Lambda := \pi_0(\mathcal A(\Lambda))'', \quad \mathcal{R}_{\Lambda^c} := \pi_0(\mathcal A(\Lambda^c))''.
\]
Haag duality for the ground state representation is the equality
\[
    \mathcal R_\Lambda = \mathcal R_{\Lambda^c}'.
\]
The inclusion $\mathcal R_\Lambda \subseteq \mathcal R_{\Lambda^c}'$ follows immediately from locality. The nontrivial part is the reverse inclusion.

The operator-algebraic reduction used below is the one developed by Naaijkens for the toric code \cite[Section~3]{naaijkens2012haag}. We do not reproduce the model-independent parts of that argument. Instead, we isolate the property that must be verified for the color code: colored string configurations carrying no stabilizer violation in the interior of one side of the boundary can be replaced, on the vacuum vector, by colored strings lying on the other side, while preserving the relevant commutation parities. Once this property is established, the remainder of the proof follows from the standard reduction of
\cite{naaijkens2012haag}.

Let 
\[
    \mathcal S_\Lambda := \Bigl\{S_\gamma^{ck}\,:\,\gamma \text{ is a finite string contained in }\Lambda,\;c\in\{\mathtt r,\mathtt g,\mathtt b\},\;k\in\{\mathtt x,\mathtt y,\mathtt z\}\Bigr\},
\]
and define $\mathcal S_{\Lambda^c}$ analogously. As before, a single vertex is also regarded as a string, so that the corresponding string operators include the single site Pauli operators. Consequently, products of operators from $\mathcal S_\Lambda$ and $\mathcal S_{\Lambda^c}$ generate the local algebra. Since $\mathcal A_{\rm loc}$ is norm dense in $\mathcal A$ and $\Omega$ is cyclic, the linear span
\[
    \operatorname{span}\left\{SS'\Omega\,:\,S\in\langle\mathcal S_\Lambda\rangle,\;S'\in\langle\mathcal S_{\Lambda^c}\rangle\right\}
\]
is dense in $\mathcal H$, where $\langle\mathcal S_\Lambda\rangle$ denotes the set of finite products of elements of $\mathcal S_\Lambda$.
Define
\[
    \mathcal H_\Lambda:=\overline{\operatorname{span}\left\{S\Omega\,:\,S\in\langle\mathcal S_\Lambda\rangle\right\}} \subseteq \mathcal H,
\]
and let $P_\Lambda$ denote the orthogonal projection onto $\mathcal H_\Lambda$.

For a face $f$, we say that $f$ is a boundary face of $\Lambda$ if its support contains vertices in both $\Lambda$ and $\Lambda^c$. The following lemma provides the key ingredient needed to establish Haag duality for the color code.

\begin{lemma}\label{lem:boundary_string_replacement}
    Let $X$ be either $\Lambda$ or $\Lambda^c$, and let $X^{\rm op}$ denote the other region. Suppose that $S\in\langle\mathcal S_X\rangle$ commutes with every stabilizer $K_f$ and $J_f$ whose support is contained in $X$. Then there exists $S^{\rm op} \in \langle\mathcal S_{X^{\rm op}}\rangle$ such that $S^{\rm op}\Omega = \pm S\Omega$. Moreover, $S^{\rm op}$ can be chosen so that $S^\dagger=\varepsilon  S$ if and only if $(S^{\rm op})^\dagger=\varepsilon  S^{\rm op}$, where $\varepsilon\in \{\pm 1\}$. 
\end{lemma}

\begin{proof}
    Suppose
    \[
        S = \prod_{i=1}^N S_{\gamma_i}^{c_i k_i}, \quad c_i\in\{\mathtt r,\mathtt g,\mathtt b\}, \quad k_i\in\{\mathtt x,\mathtt y,\mathtt z\},
    \]
    where each $\gamma_i$ is contained in $X$. By the property of string operators,
    \[
        [S_{\gamma_i}^{c_i k_i},K_f]\neq0
        \quad\text{or}\quad
        [S_{\gamma_i}^{c_i k_i},J_f]\neq0
    \]
    can occur only when $f\in \{\partial_0\gamma_i,\partial_1\gamma_i\}$. Using Lemma~\ref{lem:defom}, we may combine components with common endpoints and remove closed components. Indeed, if $\gamma$ is closed, then $S_\gamma^{ck}$ is a product of some $K_f$ and $J_f$, and hence $S_\gamma^{ck}\Omega=\Omega$. We may therefore assume that every remaining endpoint $f$ is a face at which $S$ produces a nontrivial stabilizer violation. Suppose that one such endpoint $f$ lies in the interior of $X$. Then there exists $Q_f\in\{K_f,J_f\}$ such that $Q_fS=-SQ_f$. Since $\operatorname{supp}(Q_f)\subset X$, this contradicts the assumption that $S$ commutes with every stabilizer supported in $X$. Consequently, $\partial_0\gamma_i,\, \partial_1\gamma_i \subset\partial\Lambda$ for every remaining component $\gamma_i$. For each $i$, choose a $c_i$-colored string $\gamma_i^{\rm op}\subset X^{\rm op}$ connecting the same two boundary faces as $\gamma_i$, and define
    \[
        S_i:=S_{\gamma_i}^{c_i k_i}, \quad S_i^{\rm op} := S_{\gamma_i^{\rm op}}^{c_i k_i}, \quad S^{\rm op} := \prod_{i=1}^N S_i^{\rm op}.
    \]
    For each $i$, the concatenation $\gamma_i\gamma_i^{\rm op}$ is a closed $c_i$-colored string. Hence, $S_iS_i^{\rm op}\Omega=\Omega$ up to the irrelevant ordering convention for the Pauli string operators. Reordering the factors therefore gives $S^{\rm op}\Omega = \pm S\Omega$. 

    It remains to compare the adjoints of $S$ and $S^{\rm op}$. For any $i\neq j$, consider the two closed strings obtained by joining $\gamma_i$ with $\gamma_i^{\rm op}$ and $\gamma_j$ with $\gamma_j^{\rm op}$, respectively. Any two closed strings in the plane intersect an even number of times: whenever one closed string crosses the other, it must cross back in order to return to its original side. Since $\gamma_i,\gamma_j\subset X$, while $\gamma_i^{\rm op},\gamma_j^{\rm op}\subset X^{\rm op}$, the intersections of the two closed strings consist of the intersections between $\gamma_i$ and $\gamma_j$ on one side of the boundary and those between $\gamma_i^{\rm op}$ and $\gamma_j^{\rm op}$ on the other side. Consequently, $\gamma_i$ and $\gamma_j$ intersect an even number of times if and only if $\gamma_i^{\rm op}$ and $\gamma_j^{\rm op}$ do.

    Since the Pauli labels $k_i$ and $k_j$ are unchanged under the replacement, the corresponding pairs of string operators have the same commutation relation. More precisely,
    \[
        S_iS_j=S_jS_i \quad\Leftrightarrow\quad S_i^{\rm op}S_j^{\rm op} = S_j^{\rm op}S_i^{\rm op},
    \]
    and
    \[
        S_iS_j=-S_jS_i \quad\Leftrightarrow\quad S_i^{\rm op}S_j^{\rm op} = -S_j^{\rm op}S_i^{\rm op}.
    \] 
    Each $S_i$ and $S_i^{\rm op}$ is self-adjoint. Hence taking the adjoint of $S=S_1\cdots S_N$ reverses the order of the factors $S^\dagger=S_N\cdots S_1$, and similarly $(S^{\rm op})^\dagger = S_N^{\rm op}\cdots S_1^{\rm op}$. Restoring the original order in each expression produces a sign determined entirely by the pairwise commutation relations of the factors. Since these commutation relations are preserved under the replacement, the two signs are identical. Therefore, for every
    $\varepsilon\in\{\pm1\}$,
    \[
        S^\dagger=\varepsilon S \quad\Leftrightarrow\quad (S^{\rm op})^\dagger = \varepsilon S^{\rm op}.
    \]
    This proves the lemma.
\end{proof}

We now prove Haag duality. We follow the operator-algebraic reduction developed by Naaijkens for the toric code in \cite[Lemmas~3.5--3.8 and Theorem~3.1]{naaijkens2012haag}, and spell out the argument in the present notation. The basic idea is to restrict the two von Neumann algebras to the subspace $\mathcal H_\Lambda$, on which the vectors obtained by acting with string operators supported in $\Lambda$ are dense, establish the commutant relation there by the theorem of Rieffel and van Daele, and then lift this relation back to $\mathcal H$. Lemma~\ref{lem:boundary_string_replacement} provides the key ingredient needed to adapt this reduction to the color code.

\begin{theorem}\label{thm:haag_duality}
The ground state representation of the color code model satisfies Haag duality. Namely, for every cone $\Lambda$,
\[
    \pi_0(\mathcal A(\Lambda))''=\pi_0(\mathcal A(\Lambda^c))'.
\]
\end{theorem}

\begin{proof}
As observed above, locality gives $\mathcal R_\Lambda\subseteq\mathcal R_{\Lambda^c}'$. We prove the reverse inclusion.

We first note that an operator in $\mathcal R_{\Lambda^c}'$ is uniquely determined by its restriction to $\mathcal H_\Lambda$. Indeed, suppose that $A,B\in\mathcal R_{\Lambda^c}'$ agree on $\mathcal H_\Lambda$. For $S\in\langle\mathcal S_\Lambda\rangle$ and $S'\in\langle\mathcal S_{\Lambda^c}\rangle$, we have
\[
    AS'S\Omega=S'AS\Omega=S'BS\Omega=BS'S\Omega.
\]
Since the vectors of the form $S'S\Omega$ span a dense subspace of $\mathcal H$, it follows that $A=B$. This is the restriction argument used in \cite[Lemma~3.5]{naaijkens2012haag}.

We next show that $\mathcal R_{\Lambda^c}'\mathcal H_\Lambda\subseteq\mathcal H_\Lambda$ by following the orthogonality argument of \cite[Lemma~3.6]{naaijkens2012haag}. Let $A\in\mathcal R_{\Lambda^c}'$ and $\zeta=S\Omega$ with $S\in\langle\mathcal S_\Lambda\rangle$. Consider a vector $\eta=S'T\Omega$, where $S'\in\langle\mathcal S_{\Lambda^c}\rangle$ and $T\in\langle\mathcal S_\Lambda\rangle$. Such vectors span a dense subspace of $\mathcal H$. Suppose first that there is a stabilizer $Q_f\in\{K_f,J_f\}$ supported in $\Lambda^c$ such that $Q_fS'=-S'Q_f$. Since $Q_f$ commutes with every string operator supported in $\Lambda$ and satisfies $Q_f\Omega=\Omega$, we have $Q_f\eta=-\eta$, whereas $Q_fU\Omega=U\Omega$ for every $U\in\langle\mathcal S_\Lambda\rangle$. Hence $\eta\perp\mathcal H_\Lambda$. Moreover, $Q_f\zeta=\zeta$ and $[A,Q_f]=0$, so
\[
    \langle\eta,A\zeta\rangle=\langle Q_f\eta,Q_fA\zeta\rangle=-\langle\eta,A\zeta\rangle,
\]
and therefore $\langle\eta,A\zeta\rangle=0$. Suppose now that no stabilizer supported in $\Lambda^c$ anticommutes with $S'$. By Lemma~\ref{lem:boundary_string_replacement}, applied with $X=\Lambda^c$, there exists $S^{\rm op}\in\langle\mathcal S_\Lambda\rangle$ such that $S^{\rm op}\Omega=\pm S'\Omega$. Since $S'$ and $T$ have disjoint supports, they commute, and hence $\eta=S'T\Omega=TS'\Omega=\pm TS^{\rm op}\Omega\in\mathcal H_\Lambda$. Thus the spanning vectors above either belong to $\mathcal H_\Lambda$ or are orthogonal to it and, in the latter case, have zero inner product with $A\zeta$. It follows that $A\zeta\in\mathcal H_\Lambda$. Since such vectors $\zeta=S\Omega$ are dense in $\mathcal H_\Lambda$, we conclude that $\mathcal R_{\Lambda^c}'\mathcal H_\Lambda\subseteq\mathcal H_\Lambda$.

Since $\mathcal H_\Lambda$ is invariant under the von Neumann algebra $\mathcal R_{\Lambda^c}'$, the orthogonal projection $P_\Lambda$ onto $\mathcal H_\Lambda$ belongs to $(\mathcal R_{\Lambda^c}')'=\mathcal R_{\Lambda^c}$. Since $\mathcal H_\Lambda$ is also invariant under $\mathcal R_\Lambda$, we may consider the reduced von Neumann algebras on $\mathcal H_\Lambda$,
\[
    \mathcal A_\Lambda:=\mathcal R_\Lambda P_\Lambda,\qquad \mathcal B_\Lambda:=P_\Lambda\mathcal R_{\Lambda^c}P_\Lambda.
\]
Let $\mathcal A_{\Lambda,\mathrm{sa}}$ and $\mathcal B_{\Lambda,\mathrm{sa}}$ denote their self-adjoint parts. We claim that
\[
    \overline{\mathcal A_{\Lambda,\mathrm{sa}}\Omega+i\mathcal B_{\Lambda,\mathrm{sa}}\Omega}=\mathcal H_\Lambda.
\]
This is the analogue of \cite[Lemma~3.8]{naaijkens2012haag}. Since the vectors $S\Omega$, with $S\in\langle\mathcal S_\Lambda\rangle$, span a dense subspace of $\mathcal H_\Lambda$, it is enough to show that both $S\Omega$ and $iS\Omega$ belong to $\mathcal A_{\Lambda,\mathrm{sa}}\Omega+i\mathcal B_{\Lambda,\mathrm{sa}}\Omega$. Since $S$ is a finite product of string operators, one has $S^\dagger=\varepsilon S$ for some $\varepsilon\in\{\pm1\}$. Thus, if $S^\dagger=S$, then $S\Omega\in\mathcal A_{\Lambda,\mathrm{sa}}\Omega$, while if $S^\dagger=-S$, then $iS\Omega\in\mathcal A_{\Lambda,\mathrm{sa}}\Omega$.
Suppose first that some stabilizer $Q_f\in\{K_f,J_f\}$ supported in $\Lambda$ anticommutes with $S$. If $S^\dagger=S$, then $iQ_fS$ is self-adjoint and $iQ_fS\Omega=-iS\Omega$, so $iS\Omega\in\mathcal A_{\Lambda,\mathrm{sa}}\Omega$. If $S^\dagger=-S$, then $Q_fS$ is self-adjoint and $Q_fS\Omega=-S\Omega$, so $S\Omega\in\mathcal A_{\Lambda,\mathrm{sa}}\Omega$. Hence both required vectors are obtained in this case.
It remains to consider the case in which $S$ commutes with every stabilizer supported in $\Lambda$. Lemma~\ref{lem:boundary_string_replacement}, applied with $X=\Lambda$, gives $S^{\rm op}\in\langle\mathcal S_{\Lambda^c}\rangle$ such that $S^{\rm op}\Omega=\pm S\Omega$ and $S^\dagger=\varepsilon S$ if and only if $(S^{\rm op})^\dagger=\varepsilon S^{\rm op}$ with $\varepsilon =\pm 1$. If $S^\dagger=S$, then $P_\Lambda S^{\rm op}P_\Lambda\in\mathcal B_{\Lambda,\mathrm{sa}}$ and therefore $iS\Omega\in i\mathcal B_{\Lambda,\mathrm{sa}}\Omega$. If $S^\dagger=-S$, then $P_\Lambda(iS^{\rm op})P_\Lambda\in\mathcal B_{\Lambda,\mathrm{sa}}$ and therefore $S\Omega\in i\mathcal B_{\Lambda,\mathrm{sa}}\Omega$. This proves the claimed density.

By the commutation theorem of Rieffel and van Daele \cite[Theorem~2]{rieffel1975commutation}, this density implies $\mathcal A_\Lambda=\mathcal B_\Lambda'$ as von Neumann algebras on $\mathcal H_\Lambda$. Moreover, the standard compression result for von Neumann algebras gives $\mathcal B_\Lambda'=\mathcal R_{\Lambda^c}'P_\Lambda$; see, for example, \cite[Proposition~II.3.10]{takesaki1979theory}. Hence
\[
    \mathcal R_\Lambda P_\Lambda=\mathcal R_{\Lambda^c}'P_\Lambda.
\]
Let $A\in\mathcal R_{\Lambda^c}'$. There exists $\widehat A\in\mathcal R_\Lambda$ such that $AP_\Lambda=\widehat A P_\Lambda$, so $A$ and $\widehat A$ agree on $\mathcal H_\Lambda$. By locality, $\widehat A\in\mathcal R_\Lambda\subseteq\mathcal R_{\Lambda^c}'$. Since elements of $\mathcal R_{\Lambda^c}'$ are uniquely determined by their restrictions to $\mathcal H_\Lambda$, we obtain $A=\widehat A\in\mathcal R_\Lambda$. Therefore $\mathcal R_{\Lambda^c}'\subseteq\mathcal R_\Lambda$. Together with locality, this gives $\mathcal R_\Lambda=\mathcal R_{\Lambda^c}'$, as required.
\end{proof}

\section{KMS states}\label{sec:KMS}

We now discuss the thermal equilibrium states of the color code model. For an infinite quantum spin system, thermal equilibrium at inverse temperature $\beta>0$ is characterized by the KMS condition \cite{kubo1957statistical,martin1959KMStheory,Haag1967KMS,bratteli1997operator}. We follow the strategy used for the toric code in \cite{alicki2007statistical} and adapt it to the color code setting.

\subsection{The KMS condition and locality of the dynamics}

\begin{definition}\label{def:KMS}
    Let $\beta>0$. A state $\omega$ on $\mathcal A$ is called a $\beta$-KMS state for the dynamics $\alpha$ if, for every $A,B\in\mathcal A$, there exists a bounded continuous function
    \[
        F_{A,B}:\{z\in\mathbb C:0\leq\operatorname{Im}z\leq\beta\}\to \mathbb C
    \]
    which is analytic in the interior of the strip and satisfies
    \[
        F_{A,B}(t) =\omega\bigl(A\alpha_t(B)\bigr),\quad F_{A,B}(t+i\beta)=\omega\bigl(\alpha_t(B)A\bigr)
    \]
    for every $t\in\mathbb R$.
\end{definition}

An observable $A\in \mathcal{A}$ is called analytic if the function $\alpha_t(A):\mathbb{R} \to \mathcal{A}$ can be extended to an entire function $\mathbb{C}\to \mathcal{A}$. 
For entire analytic elements, the KMS condition is equivalently characterized by
\begin{equation}\label{eq:KMS_analytic}
    \omega\bigl(A\alpha_{i\beta}(B)\bigr)=\omega(BA).
\end{equation}
Indeed, the forward implication follows by setting $t=0$ in the KMS boundary condition. Conversely, if the above identity holds for analytic $A$ and $B$, then
\[
    F_{A,B}(z):=\omega\bigl(A\alpha_z(B)\bigr)
\]
has the required analyticity, boundedness, and boundary values.
In the present model, every local observable is an entire analytic element as a consequence of the commuting stabilizer structure, as shown in Proposition~\ref{prop:exact_locality} below.

For a finite subset $\Lambda\subset V(\Sigma)$, recall its one-step enlargement by
\begin{equation*}
    \Lambda(1) := \bigcup_{f:\,\partial f\cap\Lambda\neq\emptyset}\partial f.
\end{equation*}
The following locality property is the color code analogue of the corresponding argument for the toric code in \cite[Section~2.2.2, (30)--(41)]{alicki2007statistical}.

\begin{proposition}\label{prop:exact_locality}
Let $A\in\mathcal A(\Lambda)$ be a local observable. Then, for every $z\in\mathbb C$,
\begin{equation}\label{eq:complex_dynamics_local}
\alpha_z(A)=e^{izH_{\Lambda(1)}}Ae^{-izH_{\Lambda(1)}}.
\end{equation}
In particular, $\alpha_z(A)$ is localized in the fixed finite region $\Lambda(1)$ for all $z\in\mathbb C$, and every element of $\mathcal A_{\rm loc}$ is entire analytic for $\alpha$.
\end{proposition}

\begin{proof}
It suffices to consider stabilizer terms in the local Hamiltonian whose supports intersect $\Lambda$. By the definition of $\Lambda(1)$, the support of every such stabilizer is contained in $\Lambda(1)$. Hence $\delta(A)=i[H_{\Lambda(1)},A]$. We claim more generally that, for every $n\geq1$,
\begin{equation}\label{eq:delta_iterated_local}
\delta^n(A)=i^n\operatorname{ad}_{H_{\Lambda(1)}}^n(A),
\end{equation}
where $\operatorname{ad}_{H_{\Lambda(1)}}^n(A)=[H_{\Lambda(1)},[H_{\Lambda(1)},\ldots,[H_{\Lambda(1)},A]\ldots]]$ ($n$ nested commutators). Indeed, let $Q\in\{K_f,J_f\}$ be a stabilizer whose support is not contained in $\Lambda(1)$. Then $Q$ cannot intersect $\Lambda$, since otherwise its support would be contained in $\Lambda(1)$ by definition. Thus $[Q,A]=0$. Moreover, all stabilizers commute with one another, and therefore $[Q,H_{\Lambda(1)}]=0$. It follows inductively from the Jacobi identity that $[Q,\operatorname{ad}_{H_{\Lambda(1)}}^n(A)]=0$ for every $n\geq0$. Consequently, when $\delta$ is applied repeatedly to $A$, no stabilizer outside $\Lambda(1)$ contributes, which proves \eqref{eq:delta_iterated_local}.

The real-time dynamics therefore has the norm-convergent expansion
\[
\alpha_t(A)
=
\sum_{n=0}^{\infty}\frac{t^n}{n!}\delta^n(A)
=
\sum_{n=0}^{\infty}\frac{(it)^n}{n!}\operatorname{ad}_{H_{\Lambda(1)}}^n(A)
=
e^{itH_{\Lambda(1)}}Ae^{-itH_{\Lambda(1)}}.
\]
Since $H_{\Lambda(1)}$ belongs to the finite-dimensional algebra $\mathcal A(\Lambda(1))$, it is bounded. Hence, for every $z\in\mathbb C$, the series
\[
e^{izH_{\Lambda(1)}}=\sum_{n=0}^{\infty}\frac{(iz)^n}{n!}H_{\Lambda(1)}^n
\]
converges in norm, and the map
\[
z\mapsto e^{izH_{\Lambda(1)}}Ae^{-izH_{\Lambda(1)}}
\]
is entire. This gives the extension \eqref{eq:complex_dynamics_local} and proves that every local observable is an entire analytic element for $\alpha$.
\end{proof}

\subsection{The finite-volume Gibbs states}

Recall that $\mathcal A_{\rm stab}$ denotes the Abelian subalgebra generated by the stabilizers
\[
    \{K_f,J_f:f\in F(\Sigma)\}.
\]
We first record a property of the color code stabilizers that will be useful below.

\begin{lemma}\label{lem:finite_stab_indep}
    There are no nontrivial finite relations among the stabilizers. More precisely, if $F_K,F_J\subset F(\Sigma)$ are finite and
    \begin{equation} \label{eq:finite_stab_rel}
        \prod_{f\in F_K}K_f\prod_{g\in F_J}J_g=\mathds{1},
    \end{equation}
    then $F_K=F_J=\emptyset$. 
\end{lemma}

\begin{proof}
    Suppose that \eqref{eq:finite_stab_rel} holds. At every vertex, the total Pauli operator must therefore be the identity. Since the $K$-stabilizers contribute only $\sigma^{\mathtt{x}}$ operators and the $J$-stabilizers contribute only $\sigma^{\mathtt{z}}$ operators, the number of selected $K$-stabilizers and the number of selected $J$-stabilizers incident at each vertex must both be even. It is therefore enough to consider the two families separately.
    
    Suppose, for example, that $F_K$ is nonempty. The same north-east boundary argument used in the stabilizer elimination procedure in the proof of Theorem~\ref{prop:GS} shows that a nonempty finite collection of faces has an outermost vertex at which an odd number of the selected face stabilizers acts. Hence $\prod_{f\in F_K}K_f$ acts nontrivially at this vertex, contradicting the assumption that it is the identity. Thus $F_K=\emptyset$. The argument for $F_J$ is identical.
\end{proof}

Consequently, the stabilizer algebra can be identified with the algebra of continuous functions on the classical configuration space
\begin{equation}\label{eq:stabilizer_configuration_space}
    \mathcal{X}_{\rm stab} := \{\pm1\}^{F(\Sigma)\times\{K,J\}},
\end{equation}
where $K_f$ and $J_f$ correspond to the coordinate functions on the two copies of $\{\pm1\}$ associated with the face $f$. This is analogous to the classical configuration-space description of the toric code stabilizer algebra in \cite[Section~2.2.2]{alicki2007statistical}.

Let $\Lambda\subset V(\Sigma)$ be finite and set
\[
    F_\Lambda := \{f\in F(\Sigma):\partial f\subset\Lambda\}, \quad N_\Lambda:=|F_\Lambda|.
\]
The finite-volume Gibbs state at inverse temperature $\beta$ is
\begin{equation}\label{eq:finite_Gibbs}
    \omega_{\beta,\Lambda}(A) := \frac{\operatorname{Tr}_{\Lambda}\bigl(e^{-\beta H_\Lambda}A\bigr)}{\operatorname{Tr}_{\Lambda}\bigl(e^{-\beta H_\Lambda}\bigr)},
    \quad A\in\mathcal A(\Lambda).
\end{equation}
For convenience, put
\begin{equation}\label{eq:m_beta}
    m_\beta:=\tanh\beta.
\end{equation}

Since every stabilizer is a self-adjoint involution,
\[
    e^{\beta K_f} =\cosh\beta\,\bigl(\mathds{1}+m_\beta K_f\bigr),\quad e^{\beta J_f}=\cosh\beta\,\bigl(\mathds{1}+m_\beta J_f\bigr).
\]
Therefore,
\begin{equation}\label{eq:Gibbs_factorization}
    e^{-\beta H_\Lambda}=(\cosh\beta)^{2N_\Lambda}\prod_{f\in F_\Lambda}\bigl(\mathds{1}+m_\beta K_f\bigr)\bigl(\mathds{1}+m_\beta J_f\bigr).
\end{equation}

\begin{proposition}\label{prop:finite_Gibbs_expectation}
    Let $A\in\mathcal A(\Lambda)$ be a Pauli monomial. Then $\omega_{\beta,\Lambda}(A) = m_\beta^n$ if $A=Q_1\cdots Q_n$ for distinct $Q_i\in \{K_f,J_f:f\in F_\Lambda\}$, and $\omega_{\beta,\Lambda}(A) = 0$ otherwise. 
\end{proposition}

\begin{proof}
    First notice that when expanding \eqref{eq:Gibbs_factorization}, every term is a product of distinct stabilizers. By Lemma~\ref{lem:finite_stab_indep}, no nonempty such product is equal to $\mathds{1}$. Since every Pauli monomial which is not a scalar multiple of $\mathds{1}$ has vanishing trace, we obtain
    \[
        \operatorname{Tr}_{\Lambda} \bigl(e^{-\beta H_\Lambda}\bigr) = 2^{|\Lambda|} (\cosh\beta)^{2N_\Lambda}.
    \]
    By expanding \eqref{eq:Gibbs_factorization}, the numerator of the Gibbs expectation takes the form
    \begin{equation} \label{eq:expansion-num}
        \operatorname{Tr}_\Lambda\bigl(e^{-\beta H_\Lambda}A\bigr)
        =
        (\cosh\beta)^{2N_\Lambda}
        \sum_{S}m_\beta^{|S|}
        \operatorname{Tr}_\Lambda(Q_SA),
    \end{equation}
    where the sum runs over all subsets of the stabilizers contained in $F_\Lambda$, and $Q_S$ denotes the product of the stabilizers in $S$. Suppose now that $A=Q_1\cdots Q_n$, where $Q_1,\ldots,Q_n\in\{K_f,J_f:f\in F_\Lambda\}$ are distinct stabilizers. Then by Lemma~\ref{lem:finite_stab_indep}, $\operatorname{Tr}_\Lambda(Q_SA) \neq 0$ only when $Q_S=A$.  Its contribution is therefore $m_\beta^n\operatorname{Tr}_\Lambda(\mathds{1})=m_\beta^n2^{|\Lambda|}$. For every other choice of $S$, the operator $Q_SA$ is a Pauli monomial and hence has vanishing trace. Consequently,
    \[
        \operatorname{Tr}_\Lambda\bigl(e^{-\beta H_\Lambda}A\bigr) = (\cosh\beta)^{2N_\Lambda}m_\beta^n2^{|\Lambda|}= m_\beta^n
        \operatorname{Tr}_\Lambda\bigl(e^{-\beta H_\Lambda}\bigr).
    \]
    Then the assertion for $A=Q_1\cdots Q_n$ follows by the definition of $\omega_{\beta,\Lambda}$. 
    If $A$ is not a product of stabilizers, then for every $S$ appearing in the expansion \eqref{eq:expansion-num}, the operator $Q_SA$ is not the identity and hence has vanishing trace. Therefore,
    \[
        \operatorname{Tr}_\Lambda\bigl(e^{-\beta H_\Lambda}A\bigr)=0,
    \]
    and consequently $\omega_{\beta,\Lambda}(A)=0$.
\end{proof}

As in \cite[Section~2]{alicki2007statistical}, we verify directly that the finite-volume Gibbs state $\omega_{\beta,\Lambda}$ satisfies the KMS condition in Definition~\ref{def:KMS} for the finite-volume dynamics $\alpha_z^\Lambda(A):=e^{izH_\Lambda}Ae^{-izH_\Lambda}$. Indeed, for $A,B\in\mathcal A(\Lambda)$, define $ F_{A,B}(z):=\omega_{\beta,\Lambda}\bigl(A\alpha_z^\Lambda(B)\bigr)$, which is analytic in the interior of the strip and bounded and continuous on its closure $\{z\in\mathbb C:0\leq\operatorname{Im}z\leq\beta\}$. Moreover,
\[
\begin{aligned}
\operatorname{Tr}_\Lambda\!\bigl(e^{-\beta H_\Lambda}\bigr)F_{A,B}(t+i\beta)
&=\operatorname{Tr}_\Lambda\!\bigl(e^{-\beta H_\Lambda}A e^{itH_\Lambda}e^{-\beta H_\Lambda}B e^{\beta H_\Lambda}e^{-itH_\Lambda}\bigr)
\\
&= \operatorname{Tr}_\Lambda\!\bigl(e^{-\beta H_\Lambda}e^{itH_\Lambda}B e^{-itH_\Lambda}A\bigr),
\end{aligned}
\]
where we used the cyclicity of the trace and the fact that $e^{-\beta H_\Lambda}$ commutes with $e^{itH_\Lambda}$. Hence $F_{A,B}(t+i\beta) = \omega_{\beta,\Lambda}\bigl(\alpha_t^\Lambda(B)A\bigr)$, and $\omega_{\beta,\Lambda}$ satisfies the KMS condition.

\subsection{Existence and uniqueness of the KMS state}

We now pass to the infinite-volume system. The following is the main result of this section. 

\begin{theorem}\label{thm:unique_KMS}
    For every $\beta>0$, the color code model on the infinite plane admits a unique $\beta$-KMS state $\omega_\beta$. This state is completely determined on Pauli monomials as follows. For any finite subsets $F_K,F_J\subset F(\Sigma)$,
    \begin{equation}\label{eq:KMS_stabilizer_expectation}
        \omega_\beta\left(\prod_{f\in F_K}K_f\prod_{g\in F_J}J_g\right)=m_\beta^{|F_K|+|F_J|},\quad \text{with}\;\;  m_\beta=\tanh\beta.
    \end{equation}
    If a Pauli monomial $A$ is not a product of stabilizers, then 
    \begin{equation}\label{eq:KMS_nonstabilizer_zero}
        \omega_\beta(A)=0.
    \end{equation}
\end{theorem}

\begin{proof}
    We first prove existence. Let
    \[
        \Lambda_1\subset\Lambda_2\subset\cdots,\quad V(\Sigma) = \bigcup_n \Lambda_n
    \]
    be an increasing sequence of finite regions exhausting $V(\Sigma)$. Consider the finite-volume Gibbs states $\omega_{\beta,\Lambda_n}$ defined in
    \eqref{eq:finite_Gibbs}.

    Let $A$ be a fixed local Pauli monomial. If $A$ is a product of stabilizers, we can write 
    \[
        A= \prod_{f\in F_K}K_f \prod_{g\in F_J}J_g
    \]
    with $F_K$ and $F_J$ finite. For all sufficiently large $n$, the supports of all these stabilizers are contained in $\Lambda_n$. Proposition~\ref{prop:finite_Gibbs_expectation} then gives
    \[
        \omega_{\beta,\Lambda_n}(A) = m_\beta^{|F_K|+|F_J|}.
    \]
    If $A$ is not a product of stabilizers, then
    \[
        \omega_{\beta,\Lambda_n}(A)=0
    \]
    for all sufficiently large $n$. Hence, for every local Pauli monomial $A$, the sequence $\omega_{\beta,\Lambda_n}(A)$ is eventually constant, and therefore convergent. By linearity, the limit
    \[
        \omega_\beta(A):=\lim_{n\to\infty}\omega_{\beta,\Lambda_n}(A)
    \]
    exists for every $A\in\mathcal A_{\rm loc}$. Since each $\omega_{\beta,\Lambda_n}$ is a state, we have
    \[
        |\omega_\beta(A)|\leq\|A\|,\quad
        \omega_\beta(A^\dagger A)\geq0,\quad
        \omega_\beta(\mathds{1})=1.
    \]
    Thus $\omega_\beta$ extends uniquely by continuity to a state on $\mathcal A$, and the norm density of $\mathcal A_{\rm loc}$ together with $\|\omega_{\beta,\Lambda_n}\|=1$ implies that $\omega_{\beta,\Lambda_n}\xrightarrow{w^*}\omega_\beta$ on $\mathcal A$. The limit is independent of the chosen exhaustion and satisfies \eqref{eq:KMS_stabilizer_expectation} and \eqref{eq:KMS_nonstabilizer_zero}.

    We next verify the KMS condition. Let $A,B\in\mathcal A_{\rm loc}$. By Proposition~\ref{prop:exact_locality}, there is a finite region $\Lambda_0$ such that $A$, $B$, and $\alpha_z(B)$ are supported in $\Lambda_0$ for every $z\in\mathbb C$. For all sufficiently large $n$, the finite-volume and infinite-volume evolutions of $B$ agree:
    \[
        \alpha_z^{\Lambda_n}(B) =\alpha_z(B).
    \]
    Since $\omega_{\beta,\Lambda_n}$ is the Gibbs state for the finite-dimensional Hamiltonian $H_{\Lambda_n}$, it satisfies
    \[
        \omega_{\beta,\Lambda_n}\bigl(A\alpha_{i\beta}^{\Lambda_n}(B)\bigr)=\omega_{\beta,\Lambda_n}(BA).
    \]
    Taking $n\to\infty$ gives
    \[
        \omega_\beta\bigl(A\alpha_{i\beta}(B)\bigr)=\omega_\beta(BA).
    \]
    Since $\mathcal A_{\rm loc}$ is norm dense in $\mathcal A$ and consists of entire analytic elements, this proves that $\omega_\beta$ is a $\beta$-KMS state.

    It remains to prove uniqueness. Let $\omega$ be an arbitrary $\beta$-KMS state. As in \cite[Section~2.2.2]{alicki2007statistical}, we first use the stabilizer symmetries to reduce the problem to the restriction of $\omega$ to $\mathcal A_{\rm stab}$. Every stabilizer $Q\in\{K_f,J_f:f\in F(\Sigma)\}$ is fixed by the dynamics. Applying \eqref{eq:KMS_analytic} to $QA$ and $Q$, for $A\in\mathcal A_{\rm loc}$, gives 
    \[
        \omega(QAQ)=\omega((QA)\alpha_{i\beta}(Q))=\omega(QQA)=\omega(A).
    \]
    If a local Pauli monomial $A$ anticommutes with $Q$, then $\omega(A)=-\omega(A)$, and hence $\omega(A)=0$. The stabilizer elimination argument in the proof of Theorem~\ref{prop:GS} shows that every local Pauli monomial commuting with all stabilizers is a product of stabilizers. Consequently, $\omega$ is completely determined by its restriction to $\mathcal A_{\rm stab}$.

    We now determine this restriction using finite string operators. Fix a color $c\in\{\mathtt r,\mathtt g,\mathtt b\}$ and two distinct faces $f,g$ of color $c$. Choose a finite string $\gamma$ of color $c$ joining $f$ to $g$, and set $S:=S_\gamma^{c\mathtt z}$. Then $S^\dagger=S$ and $S^2=\mathds{1}$. By the endpoint properties of the string operators, $S$ anticommutes with $K_f$ and $K_g$ and commutes with every other stabilizer. Proposition~\ref{prop:exact_locality} therefore gives 
    \[
        \alpha_{i\beta}(S)=e^{\beta(K_f+K_g)}Se^{-\beta(K_f+K_g)}=Se^{-2\beta(K_f+K_g)}.
    \]
    
    Let $B$ be any finite product of stabilizers in which neither $K_f$ nor $K_g$ occurs. The empty product is allowed. Then $B$ commutes with $S$. For $\varepsilon,\eta\in\{\pm1\}$, define 
    \[
        P_{\varepsilon,\eta}:=\frac14(\mathds{1}+\varepsilon K_f)(\mathds{1}+\eta K_g).
    \]
    These projections satisfy 
    \[
        SP_{\varepsilon,\eta}S=P_{-\varepsilon,-\eta}, \quad (K_f+K_g)P_{\varepsilon,\eta}=(\varepsilon+\eta)P_{\varepsilon,\eta}.
    \]
    Applying \eqref{eq:KMS_analytic} to $BP_{\varepsilon,\eta}S$ and $S$, we obtain
    \[
        \omega(BP_{-\varepsilon,-\eta})=\omega(SBP_{\varepsilon,\eta}S)=\omega(BP_{\varepsilon,\eta}S\alpha_{i\beta}(S))=e^{-2\beta(\varepsilon+\eta)}\omega(BP_{\varepsilon,\eta}).
    \]
    Taking $(\varepsilon,\eta)=(1,-1)$ gives $\omega(BP_{-1,1})=\omega(BP_{1,-1})$. Since 
    \[
        P_{1,-1}-P_{-1,1}=\frac{1}{2}(K_f-K_g),
    \]
    it follows that $\omega(BK_f)=\omega(BK_g)$. Taking $(\varepsilon,\eta)=(1,1)$ gives $\omega(BP_{-1,-1})=e^{-4\beta}\omega(BP_{1,1})$. Expanding these projections and using $\omega(BK_f)=\omega(BK_g)$, we obtain 
    \begin{equation} \label{eq:induction-step}
        \omega(B)+\omega(BK_fK_g)=2\coth(2\beta)\omega(BK_f).
    \end{equation}
    Both identities hold for every such product $B$ and every pair of distinct faces $f,g$ of color $c$.

    Fix a finite product $C$ of stabilizers. For $n\geq1$, choose distinct faces $f_1,\ldots,f_n$ of color $c$ such that none of the stabilizers $K_{f_1},\ldots,K_{f_n}$ occurs in $C$, and put
    \[
        a_n:=\omega(CK_{f_1}\cdots K_{f_n}), \quad a_0:=\omega(C). 
    \]
    The identity $\omega(BK_f)=\omega(BK_g)$ allows us to replace any one of the selected faces by another face of color $c$ not already selected. By successive replacements, $a_n$ is independent of the choice of $f_1,\ldots,f_n$. Since $C$ contains only finitely many factors, we may choose two further distinct faces $f,g$ of color $c$ outside $\{f_1,\ldots,f_n\}$ such that neither $K_f$ nor $K_g$ occurs in $C$. Applying the identity \eqref{eq:induction-step} with $B=CK_{f_1}\cdots K_{f_n}$ yields 
    \[
        a_{n+2}-2\coth(2\beta)a_{n+1}+a_n=0
    \]
    for every $n\geq0$. Recall that $m_\beta=\tanh\beta$. Since $2\coth(2\beta)=m_\beta+m_\beta^{-1}$, the characteristic roots of this recurrence are $m_\beta$ and $m_\beta^{-1}$. Thus 
    \[
        a_n=um_\beta^n+vm_\beta^{-n}
    \]
    for some constants $u,v\in\mathbb C$. Every product $CK_{f_1}\cdots K_{f_n}$ is unitary, so $|a_n|\leq1$ for all $n$. Since $0<m_\beta<1$ for every finite $\beta>0$, boundedness forces $v=0$. Evaluating at $n=0$ gives $u=\omega(C)$. Taking $n=1$, we obtain 
    \[
        \omega(CK_f)=m_\beta\omega(C),
    \]
    whenever $K_f$ does not occur in $C$. Using $S_\gamma^{c\mathtt x}$ instead of $S_\gamma^{c\mathtt z}$ gives the same identity for any $J_f$ that does not occur in $C$. Since the color $c$ was arbitrary, 
    \[
        \omega(CQ)=m_\beta\omega(C)
    \]
    for every stabilizer $Q$ not occurring in $C$. Applying this identity repeatedly, we obtain
    \[
        \omega\left(\prod_{f\in F_K}K_f\prod_{g\in F_J}J_g\right)=m_\beta^{|F_K|+|F_J|}
    \]
    for all finite subsets $F_K,F_J\subset F(\Sigma)$, since $\omega(\mathds{1})=1$.
    
    Thus $\omega$ and $\omega_\beta$ agree on every finite product of stabilizers and, by linearity, on every scalar multiple of such a product. The preceding vanishing argument shows that they also agree on all remaining local Pauli monomials. Since these monomials linearly span $\mathcal A_{\rm loc}$, which is norm dense in $\mathcal A$, we conclude that $\omega=\omega_\beta$.
\end{proof}

\begin{remark}
    In \cite[Section~2.2.2]{alicki2007statistical}, the uniqueness of the KMS state for the toric code is obtained by identifying the restriction to the stabilizer algebra with the unique classical equilibrium state of a free Ising system. In the proof above, we make the corresponding step explicit for the color code by deriving the stabilizer expectations directly from the KMS condition and finite string operators.
\end{remark}

\begin{remark}\label{rem:KMS_normalization}
    The factor $m_\beta=\tanh\beta$ in \eqref{eq:KMS_stabilizer_expectation} depends on the normalization of the Hamiltonian. With the convention used throughout this paper, a single stabilizer contribution is $-Q_f$, where $Q_f\in\{K_f,J_f\}$, and therefore
    \[
        \frac{\operatorname{Tr}(Q_f e^{\beta Q_f})}{\operatorname{Tr}(e^{\beta Q_f})} = \tanh\beta.
    \]
    If instead one uses the positive frustration-free normalization \eqref{eq:Hamiltonian-1}, i.e., 
    \[
        \widetilde H_\Lambda = \sum_{f:\,\partial f\subset\Lambda}\frac{\mathds{1}-K_f}{2}+\sum_{f:\,\partial f\subset\Lambda}\frac{\mathds{1}-J_f}{2},
    \]
    then at inverse temperature $\beta$ for $\widetilde H_\Lambda$, the corresponding thermal parameter is $\tanh(\beta/2)$. The two expressions differ only by this normalization of the Hamiltonian and the corresponding rescaling of inverse temperature.
\end{remark}

The zero and infinite temperature limits follow immediately from Theorem~\ref{thm:unique_KMS}.

\begin{corollary}\label{coro:KMS_limits}
    In the weak-$*$ topology,
    \begin{equation}\label{eq:KMS_zero_temperature}
        \lim_{\beta\to\infty}\omega_\beta  = \omega_0.
    \end{equation}
    Moreover,
    \begin{equation}\label{eq:KMS_infinite_temperature}
        \lim_{\beta\to0}\omega_\beta = \tau,
    \end{equation}
    where $\tau$ is the tracial state on the quasi-local algebra $\mathcal A$.
\end{corollary}

\begin{proof}
    As $\beta\to\infty$, one has $\omega_\beta(K_f)=\omega_\beta(J_f)=\tanh\beta\to1$ for every face $f$. Hence every weak-$*$ limit point satisfies all stabilizer conditions and equals $\omega_0$ by Theorem~\ref{prop:GS}. Compactness of the state space therefore gives $\omega_\beta\to\omega_0$.
    
    As $\beta\to0$, the explicit expectation formulas show that the expectation of every Pauli monomial that is not a scalar multiple of $\mathds{1}$ tends to zero. On scalar multiples of $\mathds{1}$, every state agrees with $\tau$. Thus $\omega_\beta(A)\to\tau(A)$ for every local observable $A$, and the norm density of $\mathcal A_{\rm loc}$ gives weak-$*$ convergence on $\mathcal A$.
\end{proof}

\subsection{Thermal distribution of anyonic excitations}

The explicit form of $\omega_\beta$ also gives a simple description of the thermal distribution of the elementary color code excitations. Define
\begin{equation}\label{eq:p_pm}
    p_+ := \frac{1+m_\beta}{2} = \frac{e^\beta}{2\cosh\beta}, \quad p_- := \frac{1-m_\beta}{2} = \frac{e^{-\beta}}{2\cosh\beta}.
\end{equation}
Thus, for every face $f$,
\[
    \operatorname{Prob}_\beta(K_f=+1) = \operatorname{Prob}_\beta(J_f=+1) = p_+,
\]
and
\[
    \operatorname{Prob}_\beta(K_f=-1) = \operatorname{Prob}_\beta(J_f=-1) = p_-.
\]
In particular,
\begin{equation}\label{eq:thermal_violation_density}
    p_- = \frac{1}{1+e^{2\beta}}.
\end{equation}

Let $f$ be a face of color $c\in\{\mathtt r,\mathtt g,\mathtt b\}$. By the string operator convention of Section~\ref{sec:string-ope}, the four possible values of $(K_f,J_f)$ correspond to the vacuum and the three bosonic charges $c\mathtt x$, $c\mathtt z$, and $c\mathtt y$ as follows:
\begin{center}
\begin{tabular}{c|c|c|c}
    \toprule
    $K_f$ & $J_f$ & charge at $f$ & probability \\ \hline
    $+1$ & $+1$ & $\mathds{1}$ & $p_+^2$ \\
    $+1$ & $-1$ & $c\mathtt x$ & $p_+p_-$ \\
    $-1$ & $+1$ & $c\mathtt z$ & $p_-p_+$ \\
    $-1$ & $-1$ & $c\mathtt y$ & $p_-^2$ \\
    \bottomrule
\end{tabular}
\end{center}
Indeed, an $\mathtt{x}$-type string operator anticommutes with the $J$-stabilizer at its endpoint, a $\mathtt{z}$-type string operator anticommutes with the $K$-stabilizer, and a $\mathtt{y}$-type string operator changes both stabilizer eigenvalues.

The probability that a given face carries a nontrivial elementary charge is therefore
\begin{equation}\label{eq:thermal_anyon_probability}
    1-p_+^2.
\end{equation}
The mean energy contribution per face is
\begin{equation}\label{eq:thermal_energy_face}
    \omega_\beta(-K_f-J_f) = -2\tanh\beta.
\end{equation}
At low temperature,
\[
    p_- = \frac{1}{1+e^{2\beta}} \sim e^{-2\beta}, \quad \beta\to\infty,
\]
so stabilizer violations are exponentially suppressed, but occur with strictly positive density at every finite inverse temperature.

The uniqueness of $\omega_\beta$ for every finite $\beta$ shows that the infinite color code has no finite-temperature phase coexistence in the equilibrium KMS sense. This is analogous to the two-dimensional toric code discussed in \cite{alicki2007statistical}. We emphasize, however, that uniqueness of the equilibrium state by itself does not determine the lifetime of encoded information at finite temperature. Such a question requires a separate analysis of the open-system dynamics, for example through a Davies weak-coupling generator.

\section{Conclusion and outlooks}

In this work, we rigorously define the anyon superselection sectors of the color code model and show that the resulting anyon superselection category is equivalent to $\mathsf{Rep}(D(\mathbb{Z}_2 \times \mathbb{Z}_2))$, in agreement with analyses on finite lattices.  
We also prove Haag duality for the color code model, and investigate the existence and uniqueness of the KMS state at every finite inverse temperature.  
These results have potential applications in the study of topological quantum phases as well as in quantum error correction.

The following are several interesting directions for future research:  

(1) Studying the finite-group-valued color code on an infinite lattice remains an interesting and largely open problem.  
Extending the color code model to general finite groups~\cite{Brell2015colorcode}, or even to (weak) Hopf algebra input, is still unresolved.  
Such generalizations are expected to have close connections to Kitaev's quantum double models~\cite{Kitaev2003,Buerschaper2013a,jia2023boundary,Jia2023weak,jia2022electricmagnetic,jia2024generalized,jia2024weakhopf,bols2025classification,Cha2018infinite} and to string-net models with (multi)fusion category input~\cite{Levin2005,Hu2018full,jia2024weakTube,jia2025tube,bols2025sector}.  
A systematic investigation along these lines, for both finite and infinite lattices, will be crucial for developing a comprehensive understanding of generalized color codes.  

(2) Another intriguing direction concerns topological boundary conditions and bulk-to-boundary anyon condensation~\cite{wallick2023algebraic,jones2025local,Ogata2024boundary}.  
The six possible topological boundary conditions for the color code are classified in~\cite{kesselring2024anyon}; developing an infinite-lattice boundary theory will be essential for understanding the stability of the code and the mechanisms of anyon condensation.  
Our framework can be naturally extended to incorporate boundaries.  

(3) The color code model also admits higher-dimensional generalizations \cite{Kubica2015color}.  
Using tools from algebraic quantum field theory to study these higher-dimensional models would be particularly valuable for obtaining a deeper understanding of their quantum phases.

These problems will be left for future research.

\subsection*{Acknowledgements}

We thank the anonymous referee for many insightful comments and suggestions that substantially improved both the content and the
presentation of this paper.
Z.J. is supported by the National Natural Science Foundation of China (Grant
No. 12605015), the National Research Foundation in Singapore, the A*STAR under its CQT Bridging Grant, CQT-Return of PIs EOM YR1-10 Funding,  CQT Young Researcher Career Development Grant, the start-up grant of Central South University (Grant No. 502045031) and the Frontier Interdisciplinary Direction ``Quantum Technologies'' Project of Central South University (Grant No. 506010805).
S.T. is supported by the National Natural Science Foundation of China (Grant No.~12601116), the Beijing Natural Science Foundation (Grant No.~1264052), a Beijing municipal research program for returned overseas scholars, and start-up fund from Capital Normal University.

\appendix

\section{Representation theory of $D(\mathbb{Z}_2\times \mathbb{Z}_2)$} \label{app:rep_doubld}


In this appendix, we recall the basic structure of the representation category of the quantum double $D(\mathbb Z_2\times\mathbb Z_2)$. See \cite{bakalov2001lectures,gould1993quantum} for more details on quantum doubles of finite groups and their representations.
We write
\begin{equation*}
G=\mathbb{Z}_2 \times \mathbb{Z}_2=\{0,a,b,c\},
\quad  
a+a=b+b=c+c=0,\quad a+b=c.
\end{equation*}
Here $a=(1,0)$, $b=(0,1)$ and $c=(1,1)$.
The character group $\widehat{G}$ is isomorphic to $G$, and we adopt the standard labels
\begin{equation*}
\widehat{G}=\{1,\alpha,\beta,\gamma\},
\quad  
\quad 
\alpha^2=\beta^2=\gamma^2=1,\quad \alpha\beta=\gamma.
\end{equation*}
Let $\chi_{x,y}$ denote the character
\begin{equation*}
\chi_{x,y}(u,v)
=\exp\!\left(2\pi i\,\frac{x u}{2}\right)
 \exp\!\left(2\pi i\,\frac{y v}{2}\right),
\end{equation*}
so that $\alpha=\chi_{1,0}$, $\beta=\chi_{0,1}$ and $\gamma=\chi_{1,1}$.
This convention is standard and convenient.
The explicit values of these characters are listed below:
\begin{center}
\begin{tabular}{c|cccc}
\toprule
\(\chi\) & \(\chi(0)\) & \(\chi(a)\) & \(\chi(b)\) & \(\chi(c)\) \\
\midrule
\(1\) & \(1\) & \(1\) & \(1\) & \(1\) \\
\(\alpha\) & \(1\) & \(-1\) & \(1\) & \(-1\) \\
\(\beta\) & \(1\) & \(1\) & \(-1\) & \(-1\) \\
\(\gamma\) & \(1\) & \(-1\) & \(-1\) & \(1\) \\
\bottomrule
\end{tabular}
\end{center}

\subsection*{Simple objects}

Since $G$ is Abelian, every conjugacy class consists of a single element and the centralizer of each element is all of $G$.  
Consequently, the irreducible representations (simple objects) of the Drinfeld double $D(G)$ are labeled by pairs
\begin{equation*}
(g,\chi),\quad  g\in G,\ \chi\in\widehat{G}.
\end{equation*}
There are $4\times 4 = 16$ simple objects, each of quantum dimension $1$; hence, the category is pointed.  
We list them in lexicographic order of the pairs $(g,\chi)$ with $g\in\{0,a,b,c\}$ and $\chi\in\{1,\alpha,\beta,\gamma\}$. The quantum dimension of $(g,\chi)$ is $\operatorname{qdim}(g,\chi)=1$, so all excitations are Abelian anyons.
Total quantum dimension:
  \begin{equation}
      \operatorname{qdim}(\Rep(D(\Zbb_2 \times \Zbb_2))) = \sqrt{\sum_i d_i^2} = \sqrt{16} = 4.
  \end{equation}

\subsection*{Fusion rules}

Fusion is componentwise (the pointed fusion inherited from $G\times\widehat{G}$):
\begin{equation*}
    (g,\chi)\otimes(g',\chi') = (g+g',\ \chi\chi'),
\end{equation*}
where $(\chi\chi')(h)=\chi(h)\chi'(h)$ for $h\in G$.  
In particular, the fusion rules realize the Abelian group
\begin{equation*}
K(\Rep(D(\mathbb{Z}_2\times\mathbb{Z}_2))) 
= G \times \widehat{G}
\cong (\mathbb{Z}_2\times\mathbb{Z}_2)\times(\mathbb{Z}_2\times\mathbb{Z}_2).
\end{equation*}
Thus every simple object is invertible, and all fusion coefficients are either $0$ or $1$.

\subsection*{Twists (topological spins) and bosons/fermions}

The twist (topological spin) of a simple object $(g,\chi)$ in the Drinfeld double of an Abelian group is given by
\begin{equation*}
    \theta_{(g,\chi)} = \chi(g) \in \{\pm 1\}.
\end{equation*}
We call $(g,\chi)$ a \emph{boson} if $\theta_{(g,\chi)}=1$ and a \emph{fermion} if $\theta_{(g,\chi)}=-1$.  
Using the explicit character values given above, we obtain the following table of twists and statistics:

\begin{longtable}{llcc}
\toprule
Label & Pair & Twist $\theta$ & Type \\
\midrule
\endhead
$(0,1)$  & $g=0$, $\chi=1$  & $1$ & boson \\
$(0,\alpha)$ & $g=0, \ \chi=\alpha$ & $1$ & boson \\
$(0,\beta)$  & $g=0, \ \chi=\beta$ & $1$ & boson \\
$(0,\gamma)$ & $g=0, \ \chi=\gamma$ & $1$ & boson \\
\midrule
$(a,1)$ & $g=a$, $\chi=1$ & $1$ & boson \\
$(a,\alpha)$ & $g=a, \ \chi=\alpha$ & $-1$ & fermion \\
$(a,\beta)$  & $g=a, \ \chi=\beta$ & $1$ & boson \\
$(a,\gamma)$ & $g=a, \ \chi=\gamma$ & $-1$ & fermion \\
\midrule
$(b,1)$ & $g=b$, $\chi=1$ & $1$ & boson \\
$(b,\alpha)$ & $g=b, \ \chi=\alpha$ & $1$ & boson \\
$(b,\beta)$ & $g=b, \ \chi=\beta$ & $-1$ & fermion \\
$(b,\gamma)$ & $g=b, \ \chi=\gamma$ & $-1$ & fermion \\
\midrule
$(c,1)$ & $g=c$,  $\chi=1$ & $1$ & boson \\
$(c,\alpha)$ & $g=c, \ \chi=\alpha$ & $-1$ & fermion \\
$(c,\beta)$ & $g=c, \ \chi=\beta$ & $-1$ & fermion \\
$(c,\gamma)$ & $g=c, \ \chi=\gamma$ & $1$ & boson \\
\bottomrule
\end{longtable}

Counting fermions from the table yields $6$ fermions total (two for each nontrivial flux $a,b,c$) and $10$ bosons.

\subsection*{Braiding}

For two simple objects $(g,\chi),(g',\chi')$, the braiding (monodromy factor) is given by the bicharacter
\begin{equation*}
    R_{(g,\chi),(g',\chi')} = \chi'(g).
\end{equation*}
The double-braiding (monodromy) is
\begin{equation*}
    M_{(g,\chi),(g',\chi')} = R_{(g,\chi),(g',\chi')} R_{(g',\chi'),(g,\chi)} 
    = \chi'(g)\chi(g').
\end{equation*}

\subsection*{Modular $S$-matrix}

The modular $S$-matrix for the Drinfeld double of an Abelian group has the closed form
\begin{equation}
    \label{eq:S-general}
    S_{(g,\chi),(g',\chi')} = \frac{1}{\operatorname{qdim} \Rep(D(G))} \, 
    M_{(g,\chi),(g',\chi')} 
    = \frac{1}{|G|} \, \chi(g') \, \chi'(g).
\end{equation}
Since $|G| = 4$, every entry of $S$ is $\pm \tfrac{1}{4}$.  
An explicit expression is
\begin{equation}
    S = S^{(1)} \otimes S^{(2)},
\end{equation}
where
\begin{equation*}
S^{(1)} = \frac{1}{2}
\begin{pmatrix}
1 & 1 & 1 & 1\\
1 & 1 & -1 & -1\\
1 & -1 & 1 & -1\\
1 & -1 & -1 & 1
\end{pmatrix},
\quad 
S^{(2)} = \frac{1}{2}
\begin{pmatrix}
1 & 1 & 1 & 1\\
1 & 1 & -1 & -1\\
1 & -1 & 1 & -1\\
1 & -1 & -1 & 1
\end{pmatrix}.
\end{equation*}
Thus $S$ is a $16\times 16$ matrix whose entries are all $\pm \tfrac{1}{4}$.  
The formula~\eqref{eq:S-general} directly explains this Kronecker-product structure: the factor $\chi(g')$ depends only on the character and the target flux, while $\chi'(g)$ depends only on the other character and flux.

\subsection*{Correspondence to double-layer toric code}

Because the Drinfeld double is multiplicative over direct products,
\begin{equation*}
D(\mathbb{Z}_2 \times \mathbb{Z}_2) \cong D(\mathbb{Z}_2) \boxtimes D(\mathbb{Z}_2),
\end{equation*}
it follows that as modular tensor categories (representation categories),
\begin{equation*}
\Rep(D(\mathbb{Z}_2 \times \mathbb{Z}_2)) \simeq \Rep(D(\mathbb{Z}_2)) \boxtimes \Rep(D(\mathbb{Z}_2)).
\end{equation*}
Each factor $\Rep(D(\mathbb{Z}_2))$ is the toric code category, which is pointed with four invertible objects and $S$-matrix entries $\pm 1/2$. Consequently, the color code category $\Rep(D(\mathbb{Z}_2 \times \mathbb{Z}_2))$ is pointed (all simple objects have dimension $1$) and does not contain non-Abelian Ising anyons (which require quantum dimension $\sqrt{2}$). 
Thus we have the factorization 
\begin{equation*}
 \mathsf{CC}_{\Zbb_2}\simeq \mathsf{TC}_{\Zbb_2} \boxtimes  \mathsf{TC}_{\Zbb_2}.
\end{equation*}

To make the correspondence explicit, recall that toric code anyons in $\Rep(D(\Zbb_2))$ are labeled by
\begin{equation*}
(k,\chi_x) ,\quad \chi_x = \exp\Big(2\pi i \frac{x\bullet}{2}\Big), \quad k,x \in \Zbb_2,
\end{equation*}
giving $1=(0,\chi_0)$, $e=(0,\chi_1)$, $m=(1,\chi_0)$, $f=(1,\chi_1)$.  
For a color code anyon $(g,\chi)$ with
\begin{equation*}
g=(k,l), \quad  \chi_{x,y} = \chi_x \chi_y,
\end{equation*}
we then have
\begin{equation*}
((k,l), \chi_{x,y}) = (k,\chi_x) \boxtimes (l,\chi_y), \quad k,l,x,y \in \Zbb_2.
\end{equation*}
To summarize, we have the following:
\begin{center}
\begin{longtable}{ccccc}
\toprule
$(g,\chi)$ Label & $g=(k,l)$ (Flux) & $\chi=\chi_{x,y}$ (Charge) & Toric Code & Color Code \\
\midrule
\multicolumn{5}{c}{\text{Trivial Flux $g=0 \ (k=0, l=0)$}} \\
\midrule
$(0, 1)$ & $(0, 0)$ & $\chi_{0, 0}$ & $1 \boxtimes 1$ & $\mathbbm{1}$ \\
$(0, \alpha)$ & $(0, 0)$ & $\chi_{1, 0}$ & $e \boxtimes 1$ & $\mathtt{rx}$ \\
$(0, \beta)$ & $(0, 0)$ & $\chi_{0, 1}$ & $1 \boxtimes e$ & $\mathtt{bx}$ \\
$(0, \gamma)$ & $(0, 0)$ & $\chi_{1, 1}$ & $e \boxtimes e$ & $\mathtt{gx}$ \\
\midrule
\multicolumn{5}{c}{\text{Flux $g=a \ (k=1, l=0)$}} \\
\midrule
$(a, 1)$ & $(1, 0)$ & $\chi_{0, 0}$ & $m \boxtimes 1$ & $\mathtt{bz}$ \\
$(a, \alpha)$ & $(1, 0)$ & $\chi_{1, 0}$ & $f \boxtimes 1$ & $\mathtt{f}_1$ \\
$(a, \beta)$ & $(1, 0)$ & $\chi_{0, 1}$ & $m \boxtimes e$ & $\mathtt{by}$ \\
$(a, \gamma)$ & $(1, 0)$ & $\chi_{1, 1}$ & $f \boxtimes e$ & $\mathtt{f}_4$ \\
\midrule
\multicolumn{5}{c}{\text{Flux $g=b \ (k=0, l=1)$}} \\
\midrule
$(b, 1)$ & $(0, 1)$ & $\chi_{0, 0}$ & $1 \boxtimes m$ & $\mathtt{rz}$ \\
$(b, \alpha)$ & $(0, 1)$ & $\chi_{1, 0}$ & $e \boxtimes m$ & $\mathtt{ry}$ \\
$(b, \beta)$ & $(0, 1)$ & $\chi_{0, 1}$ & $1 \boxtimes f$ & $\mathtt{f}_2$ \\
$(b, \gamma)$ & $(0, 1)$ & $\chi_{1, 1}$ & $e \boxtimes f$ & $\mathtt{f}_3$ \\
\midrule
\multicolumn{5}{c}{\text{Flux $g=c \ (k=1, l=1)$}} \\
\midrule
$(c, 1)$ & $(1, 1)$ & $\chi_{0, 0}$ & $m \boxtimes m$ & $\mathtt{gz}$ \\
$(c, \alpha)$ & $(1, 1)$ & $\chi_{1, 0}$ & $f \boxtimes m$ & $\mathtt{f}_6$ \\
$(c, \beta)$ & $(1, 1)$ & $\chi_{0, 1}$ & $m \boxtimes f$ & $\mathtt{f}_5$ \\
$(c, \gamma)$ & $(1, 1)$ & $\chi_{1, 1}$ & $f \boxtimes f$ & $\mathtt{gy}$ \\
\bottomrule
\end{longtable}
\end{center}

\subsection*{Declarations}
\paragraph{Conflict of interest:} 
The authors certify that there are no conflicts of interest for this work.

\paragraph{Data Availability Statement:} 
The authors declare that the data supporting the findings of this study are available within the paper.


\bibliographystyle{apsrev4-1-title}
\bibliography{mybib}

\end{document}